\documentclass[12pt]{article}
\usepackage[latin1]{inputenc}
\usepackage{natbib}
\usepackage[a4paper,left=2.5cm,right=2.5cm,top=2cm,bottom=1.5cm,footnotesep=.75cm,headheight=13.6pt]{geometry}
\usepackage{setspace}
\usepackage{amsmath}
\usepackage{graphicx}
\usepackage[small]{caption}
\usepackage{subfigure}
\usepackage{authblk}
\usepackage{rotating}
\usepackage{float}
\usepackage{pdflscape}
\usepackage{newtxmath}
\usepackage{appendix}

\title{International vulnerability of inflation}

\author[1]{\large Ignacio Garr\'on}
\author[2]{\large Vladimir Rodr\'iguez-Caballero}
\author[1]{\large Esther Ruiz\thanks{Corresponding author.\\ E-mail addresses: igarron@est-econ.uc3m.es (I. Garr\'on), vladimir.rodriguez@itam.mx (V. Rodr\'iguez-Caballero), ortega@est-econ.uc3m.es (E. Ruiz)}}
\affil[1]{\small{Department of Statistics, Universidad Carlos III de Madrid}\\
\small{C/ Madrid 126, 28903 Getafe (Spain)}}
\affil[2]{\small{Department of Statistics, ITAM, Rio Hondo 1, Col. Progreso Tizap\'an, Alvaro Obreg\'on, CDMX 01080 (Mexico)}}

\date{\today}

\begin{document}

\maketitle

\begin{abstract}

In a globalised world, inflation in a given country  may be becoming less responsive to domestic economic activity, while being increasingly determined by international conditions. Consequently, understanding the international sources of vulnerability of domestic inflation is turning fundamental for policy makers. In this paper, we propose the construction of  Inflation-at-risk and Deflation-at-risk measures of vulnerability obtained using factor-augmented quantile regressions estimated with international factors extracted from a multi-level Dynamic Factor Model with overlapping blocks of inflations corresponding to economies grouped either in a given geographical region or according to their development level. The methodology is implemented to inflation observed monthly from 1999 to 2022 for over 115 countries. We conclude that,  in a large number of developed countries, international factors are relevant to explain the right tail of the distribution of inflation, and, consequently, they are more relevant for the vulnerability related to high inflation than for average or low inflation. However, while inflation of developing low-income countries is hardly affected by international conditions, the results for middle-income countries are mixed.  Finally, based on a rolling-window out-of-sample forecasting exercise, we show that the predictive power of international factors has increased in the most recent years of high inflation.

\end{abstract}



\setcounter{page}{1}

\newpage

\doublespacing

\section{Introduction}

The interest on modelling and forecasting inflation is without discussion. Low and stable inflation is usually associated with economic growth and employment while both very low and very high inflation are macroeconomic challenges. Inflation forecasts have implications for policy makers, central banks, and business decisions. Furthermore, due to recent geopolitical events, and, in particular, the Russian invasion of Ukraine in February 2022, inflation has a synchronized increase above inflation targets in many advanced economies with its forecasts moving to the forefront of the policy debate.

Although the literature on modelling and forecasting inflation is large and important, the debate on the best model to represent the dynamic evolution of inflation is still open.\footnote{This debate has been mainly focused on US inflation.} According to the Fed's view, three components matter for inflation dynamics: i) short-term fluctuations via the Phillips curve; ii) long term inflation shaped by monetary policy, via expectations; and iii) oil price that explain the volatile component of inflation; see \cite{hasenzagl2022}.  %To start with, the empirical evidence on domestic Phillips curve-based models is still under discussion; see, for example, \cite{atkeson2001}, \cite{stock2007}, \cite{dotsey2018}, and \cite{forbes2020}, for an examination of the forecasting ability of Phillips curves, and \cite{mankiw2024} for some concerns about them.\footnote{\cite{jorda2020} remark that the Phillips curve is not a forecasting model.} Monetary aggregates also have a poor record track in forecasting inflation; see \cite{mankiw2024}. Several authors analyze whether disaggregated prices help unpacking the effect of monetary policy on inflation; see \cite{baumeister2013}, \cite{dupor2023}, and \cite{aruoba2024}. the recent work by \cite{joseph2024}, who propose forecasting inflation using disaggregate item series comprising the consumer price index and novel machine learning tools to improve forecastability.  Recently, \cite{kilian2024} also analyse the inflationary impact of gasoline prices and different policy measures, and \cite{hasenzagl2022} consider a semi-structural model with all three components of inflation and conclude that the Phillips curve is alive and has been fairly stable since the early 1980's.
Most models for inflation, based on one or several of these components, focus mainly on domestic factors to explain domestic inflation. However, there is evidence of an increasing role of globalization, defined broadly as increased integration between individual countries and the rest of the world, on domestic inflation. Quoting \cite{draghi2015}, "Over the last decade there has been a growing interest in the concept of global inflation. This is the notion that, in a globalised world, inflation is becoming less responsive to domestic conditions, and is instead increasingly determined by global factors"; see \cite{auer2019}. Consequently, it is important to understand the factors underlying international movements in inflation and the way domestic inflation in a given country reacts to them. Ignoring these factors may overrate domestic inflation and mislead macroeconomic policies.

The debate on whether and how globalization affects domestic inflation is not new, going back to the 70's; see \cite{forbes2020} and \cite{ha2019a} for descriptions of this debate. However, there is a recent even stronger interest on worldwide fluctuations of inflation, with a generalized consensus about inflation being largely a global phenomenon; see, among others, %\cite{cecchetti2007}, \cite{wang2007}, \cite{monacelli2009}, \cite{ciccarelli2010}, %analyses year-on-year (\textit{yoy}) quarterly inflation in 22 OECD countries between 1960 and 2008, and show that inflation in industrialized countries is largely a global phenomenon with a global common factor accounting for nearly 70\% of their variance.\cite{neely2011}, \cite{muntaz2012}, \cite{eickmeier2013}, \cite{ahmad2017}, \cite{kabukcuoglu2018}, \cite{forbes2020}, \cite{kamber2020}, 
\cite{MuntazMusso}, \cite{arango2023}, \cite{ha2019a},  \cite{liu2021}, \cite{Carriero2022JAE}, \cite{ha2023}, \cite{pitarakis2023}, and \cite{medeiros2024}, for some recent references. % also observe global synchronization of inflation analysing \textit{yoy} inflation observed quarterly for a large set of economies that can be classified as advanced (38 countries), emerging and developing (171 countries) and low-income (27 countries).
Several drivers for international inflation synchronization have been proposed; see \cite{gali2010} for a theory-based discussion of the global drivers and \cite{ha2019a}, \cite{ha2019b},  \cite{jorda2020}, and \cite{forbes2020} for detailed descriptions and references. Among the main channels, one can find the growing trade integration, %; see, for example, \cite{frankel2006}, \cite{bianchi2015} and \cite{medeiros2024}.
 the role of emerging markets and their impact on commodity prices and, in particular, on oil prices, the greater similarity of monetary policy that trigger similar policy responses, % as drivers of global inflation; see, for example, \cite{ha2022}, \cite{hasenzagl2022}, and \cite{ha2023b}. \cite{auer2019} argue that
and the increasing use of supply chains to shift parts of the production to cheaper locations, contributing to the cross-border propagation of cost shocks through input-output linkages.

In parallel to the interest on the influence of international factors on domestic inflation, policy-makers also need to assess the risk of having either very low or very high inflation; see \cite{KilianManganelli2008}, who introduce a model in which the policy-maker acts as a risk manager trying to balance the risks of inflation and output stability. Consequently, central banks and financial institutions often discuss and monitor both upside and downside risks to inflation. Taking into account these risks, \cite{mankiw2024} recommends that central bankers offer as target for inflation a range instead of a single figure, recogniting how imprecise their ability to control inflation is. There is a nascent although growing fast literature on measuring risk to inflation. Early on, \cite{manzan2013} and \cite{xiu2015} propose modelling the inflation distribution by a quantile regression model with macroeconomic indicators as regressors. The term inflation at risk (IaR) followed soon after the proposal of Growth at Risk (GaR) by \cite{ABG2019}; see, for example, \cite{tabliabracci2020}, \cite{adam2021}, \cite{korobilis2021} and \cite{lopez2024}.\footnote{Predictive densities to asses the tail risk of inflation have also been proposed by, for example, \cite{andrade2014}, \cite{kilian2007}, \cite{korobilis2017}, \cite{ghysels2018}, and \cite{pfarrhofer2022}. These proposals are univariate in the sense that the distribution of inflation was not related to economic or financial indicators.} In these papers, the predictors for the distribution of inflation are domestic and, consequently, they do not take into account the international fluctuations of inflation. Only very recently, \cite{zheng2023} propose characterizing the dynamic distribution of inflation using quantile regression with the global energy connectedness index (GECI) as regressor, while \cite{banerjee2024} use quantile regressions including oil prices, among other variables, in a large panel of advanced and emerging market economies.\footnote{The existence of global and specific factors related with the level of economic development has been previously analysed in other contexts; see, for example, \cite{kose2012} and \cite{djogbenou2020} analysing the real economic activity, or \cite{djogbenou2024} for oil prices. \cite{aastveit2015} also estimate the developed economic activity factor and the emerging economy real factor to explain oil prices.} 

The novelty of this paper lies on bringing together two previously unrelated strands of the literature, namely, the impact of international inflation factors on domestic inflation and inflation vulnerability. We propose modelling domestic inflation density, and, consequently, inflation risk, in a given country, taking into account its dependence on international movements of inflation. Furthermore, while most extant works usually focus on a relatively reduced number of countries, in this paper, we analyse how inflation density in a given country depends on international factors by analysing a very large number of countries covering not only advanced (ADV) economies but also middle and high income emerging and developping (MHI-EMD) economies and low income emerging (LI-EMD) economies. A final novelty of our work is that we not only consider global international factors, but also group-specific international factors linked to countries in the same geographical regions, as in  \cite{medeiros2024}, or to countries classified as either ADV, MHI-EMD or LI-EMD, as in \cite{ha2019a},  \cite{jorda2020} and \cite{ha2023}.

The methodology used to estimate the inflation density in each country is based on factor-augmented quantile regressions (FA-QR) with global and group-specific international factors as regressors.\footnote{\cite{busetti2021OBES} propose using expectile regression instead of quantile regression to model the distribution of core inflation in the euro area as a function of domestic and foreign output gap.} The factors are extracted using a Multi-Level Dynamic Factor Model (ML-DFM) with overlapping blocks of variables; see \cite{gonzalez2024} for the same methodology in the context of modelling the distribution of growth.\footnote{A similar ML-DFM is proposed by \cite{beck2015} in the context of Euro area inflation considering overlapping factors related to countries and sectors.} It is important to point out that, by using extracted latent factors to explain the quantiles of domestic inflation, we are not able to address the question about what the drivers of international inflation are. However, in doing so, we are able to assess the relevance of different international factors on explaining different quantiles of the distribution of inflation in different countries. We show that international factors may have different impact on different quantiles, being, in general stronger in the tails than in the center of the domestic inflation distribution. In particular, the impact of international factors is generally stronger in the right quantiles and, consequently, on the risk of high inflation. Furthermore, we also find interesting heterogeneous patterns across countries. In concordance with the related literature, the impact of international factors is often stronger in ADV economies than in EMD economies.
 %Second, we measure inflation risks taking into account that inflation is a global phenomenon and construct scenarios for domestic inflation, which are generated by taking into account the risk associated with stressed international and specific underlying factors. We extend the usual scenario analysis by obtaining scenarios for the full distribution of inflation in each country. The methodology implemented to stress the factors is similar to that proposed by \cite{gonzalez2024} for growth in stress in the US.
Finally, we carry out an out-of-sample exercise, obtaining one-step-ahead forecasts of the quantiles of domestic inflation in each of the economies considered. We show that the forecasting power of international factors is relevant when forecasting domestic inflation. 

The rest of the paper is organized as follows. Section \ref{section:methodology} briefly describes the methodology used to estimate the domestic inflation densities based on FA-QR with the factors extracted using ML-DFMs. We also describe the evaluation measures for the forecasts of the quantiles of inflation. Section \ref{section:data} describes the main empirical stylized facts of the inflation data  analyzed in this paper. Section \ref{section:Distribution} is devoted to the estimation results of the ML-DFM and of the corresponding FA-QRs. Section 5 describes the main results of the out-of-sample exercise. Finally, Section \ref{section:final} summarizes the main takeaways. 

\section{Inflation densities: factor-augmented quantile regression and multi-level dynamic factor model}
\label{section:methodology}

In this section, we offer a brief overview of the econometric FA-QR model used to estimate and predict the probability distribution of domestic inflation in a given country, with the international factors extracted using a ML-DFM with overlapping blocks of variables.

\subsection{Factor-augmented quantile regression}

Let $y_{it}$ be the inflation observed in country $i$ at time $t$, for $i=1,...,N$, and $t=1,...,T$. The one-step-ahead $\tau$-quantile of the conditional distribution of $y_{it}$ is obtained by estimating the following FA-QR model\footnote{FA-QRs are standard in modelling quantiles of macroeconomic variables. \cite{manzan2005}, \cite{giglio2016}, \cite{ABG2019}, \cite{GMR2019}, \cite{AGLM2022}, and \cite{gonzalez2024}, among others, use FA-QRs to model the density of growth while \cite{nicolo2017} fit them to measure the tail risk of industrial production and employment in the US. \cite{banerjee2024}, and \cite{lopez2024} use quantile regressions for the distribution of inflation with the quantiles depending on a set of macroeconomic and financial variables.  Alternatively, in the case of the euro area inflation, \cite{tabliabracci2020} consider $F_t$ being a business cycle factor with the factor being the Eurocoin indicator of \cite{altisimo2010}. They show that there are asymmetries across quantiles of the distribution of inflation with downside risks related to the business cycle while upside risks are instead not affected by the sate of the economy, being  relatively stable over time.}
\begin{equation}
\label{eq:Quantile_regression}
q_{\tau}\left(y_{it+1}|y_{it}, F_{t} \right) = \mu(\tau,i) + \phi(\tau,i) y_{it} + \sum_{k=1}^r \beta_{k}(\tau,i) F_{kt},
\end{equation}
where $\theta(\tau,i)=\left( \mu(\tau,i), \phi(\tau,i), \beta_{1}(\tau,i),...,\beta_r(\tau,i)\right) ^{\prime}$, is the $(r+2) \times 1$ vector of parameters and $F_t=\left(F_{1t},...,F_{rt} \right) ^{\prime}$ is the $r\times 1$ vector of underlying unobserved international factors at time $t$, extracted from $Y_t=\left(y_{1t},...,y_{Nt} \right)$. Finally, note that, in practice, the underlying factors in (\ref{eq:Quantile_regression}) are replaced by estimated factors, $\hat{F}_t$, which are extracted from the ML-DFM as described below.

The parameters in equation (\ref{eq:Quantile_regression}) are estimated using the algorithms by \cite{koenker1987}, who implement the estimator proposed by \cite{koenker1978}. This estimator is a Maximum Likelihood (ML) estimator assuming independently distributed Laplace densities; see, for example, \cite{ando2011}. Asymptotic normality of the ML estimator of the parameters is proven by \cite{BaiNg2008} for the case of the FA-QR model; see also \cite{koenker1999} for the asymptotic distribution of the ML estimator when the predictors are observed variables instead of unobserved estimated factors. In practice, standard deviations of the estimated parameters can be computed using the sandwich formula proposed by \cite{powell1989}. The significance of individual parameters or subsets of parameters can then be tested using the Wald test. In particular, we may want to test for the joint significance of the parameters associated with the factors, i.e. test the benchmark quantile autoregressive regression (AR-QR) model against the FA-QR model. %Then, one can compute the following Likelihood Ratio (LR) statistic
%\begin{equation}
%\label{eq:LR}
%LR= 2 T ln\left( \frac{\sum_{t=h+1}^T \tilde{\nu}_{it} \left[\tau^* I(\tilde{\nu}_{it}\geq 0) + (\tau^*-1) I(\tilde{\nu}_{it}<0) \right] }{\sum_{t=h+1}^T \hat{\nu}_{it}\left[\tau^* (I(\hat{\nu}_{it} \geq 0)+(\tau^*-1) I(\hat{\nu}_{it}<0) \right] } \right) ,
%\end{equation}
%where $\hat{\nu}_{it}$ are defined as in (\ref{eq:R2}) and $\tilde{\nu}_{it}=y_{it}-\tilde{\mu}-\tilde{\phi}y_{it-h}$, with $\tilde{\mu}$ and $\tilde{\phi}$ being the parameters estimated in the restricted model. 
Under the null hypothesis, $H_0:\beta_{1}(\tau,i)=...=\beta_r(\tau,i)=0$, the Wald statistic is asymptotically distributed as an $F$ distribution with $T$ and $r$ degrees of freedom.

Furthermore, \cite{giglio2016} derive the consistency of the estimated quantiles in the presence of estimated factors. However, it is important to mention that, as far as we know, there are not results available for their asymptotic distribution. In the context of quantile regressions without estimated factors as predictors, \cite{gregory2018} and \cite{chernozhukov2022} propose using bootstrap. However, there is not any proposal to obtain the distribution of the quantiles in the presence of estimated factors; see the discussion by \cite{gonzalez2024}.

The in-sample predictive accuracy of the fitted model (\ref{eq:Quantile_regression}) can be evaluated via a quantile $R^1$ based on the loss function; see, for example, \cite{giglio2016}. For a given quantile $\tau$ and country $i$, \cite{koenker1999} propose measuring the goodness of fit of the estimated FA-QR by\footnote{From now on, we simplify the notation by dropping the indexes $\tau$ and $i$ from the parameters.}
\begin{equation}
\label{eq:R2}
R^1(i,\tau)=1-\frac{\sum_{t=2}^T \hat{\nu}_{it} \left[\tau \vmathbb{1} (\hat{\nu}_{it}\geq 0) + (\tau-1) \vmathbb{1} (\hat{\nu}_{it}<0) \right] }{\sum_{t=2}^T \tilde{\nu}_{it}\left[\tau \vmathbb{1} (\tilde{\nu}_{it} \geq 0)+(\tau-1) \vmathbb{1} (\tilde{\nu}_{it}<0) \right] },
\end{equation}
where $\hat{\nu}_{it}=y_{it} - \hat{\theta}(i,\tau)^{\prime} X_{it-1}$, with $X_{it}=\left(1, y_{it}, F_{1t},...,F_{rt} \right) ^{\prime}$, while $\tilde{\nu}_{it}=y_{it}-\tilde{\mu}_{i\tau}$ with $\tilde{\mu}_{i\tau}$ being the empirical $\tau$ quantile of $y_{it}$, and $\vmathbb{1} (\cdot)$ being an indicator function that takes value 1 if the argument is true and zero otherwise. Note that $R^1$ is the natural analogue of the $R^2$ coefficient in a regression model.
%More recently, \cite{troster2018} extends \cite{koenker1999} by providing an omnibus test for Granger causality in several quantiles simultaneously, while \cite{mayer2024} extend further this test to be robust against time-varying parameters and/or structural breaks. 

Following \cite{ando2011}, we also use the following Information Criteria, which takes into account the presence of estimated factors in the FA-QR
\begin{equation}
\label{eq:AIC}
AIC(i,\tau)= 2 T ln \left[ \frac{1}{T} \sum_{t=2}^T \hat{\nu}_{it} \left[\tau \vmathbb{1} (\hat{\nu}_{it}\geq 0) + (\tau-1) \vmathbb{1} (\hat{\nu}_{it}<0) \right] \right] + 2 T b,
\end{equation}
where $b$ is a correction bias that depends on the uncertainty of the estimated factors; see \cite{ando2011} for the expression $b$. Note that, for quantile regressions, the selection based on minimising AIC overestimates the model dimension with positive probability but may have superior performance for prediction; see \cite{koenker2005}.

Finally, after estimating model (\ref{eq:Quantile_regression}) for different quantiles $\tau$, we follow \cite{ABG2019} and obtain the conditional distribution of inflation by fitting the Skewed-t distribution of \cite{azzalinicapitanio2003} to the estimated quantiles, $\hat{q}_{\tau}\left(y_{it+1}|y_{it}, F_{t} \right)$. At each moment of time $t$, the four parameters that define the Skewed-t distribution are estimated by minimizing the squared distance between the estimated quantiles and the corresponding quantiles of the Skewed-t distribution. Denote this density by $\tilde{k}(y_{it+1})$.

As mentioned above, policy-makers are interested in evaluating the probability of deflation and/or high inflation rates. In doing so, they can gauge possible risks associated to future inflation, which are a major threat to policy effectiveness. Inflation-at-Risk (IaR) is the probability that inflation is larger than a given threshold $\pi$, and, therefore, measures the probability of high-inflation. The IaR of country $i$ at time $t+1$  for threshold $\pi$ is given by
\begin{equation}
\label{eq_IaR}
IaR_{it+1}(\pi)=\int_{\pi}^{\infty} \tilde{k}(y_{it+1}) dy_{it+1}.
\end{equation}
Alternatively, we can measure the probability of deflation, using Deflation-at-Risk (DaR), which is defined as the probability that inflation falls below a certain threshold, $\pi$. The DaR is given by
\begin{equation}
\label{eq_DaR}
DaR_{it+1}(\pi)=\int_{-\infty}^{\pi} \tilde{k}(y_{it+1}) dy_{it+1}.
\end{equation} 
\cite{tabliabracci2020} propose measuring DaR and IaR with $\pi=0$ and 2, respectively.\footnote{\cite{kilian2007} propose an alternative way of measuring the risk of inflation linked to the concept of expected loss in the context of financial risk. They define the deflation risk (DR) and the risk of excessive inflation (EIR) as probability weighted functions of the deviations of inflation from the threshold $\pi$. These measures include the IaR and DaR as special cases. Furthermore, they propose a Balance of risks measure based on DR and EIR. They implement these measures to estimate risk of inflation based on an AR-GARCH model using bootstrap to obtain the density; see \cite{wang2024} for an application using Bayesian procedures to estimate the inflation density depending on selected indicators and with time-varying parameters modelled as random walks. Alternatively, \cite{banerjee2024} propose measuring inflation risk using entropy measures.}

\subsection{Out-of-sample performance}

We assess the real-time forecasting content of international factors by obtaining out-of-sample forecasts of the quantiles $q_{\tau}(y_{it+1}|y_{it}, F_t)$. For this goal, the sample size $T$ is divided into an in-sample period of size $R$ and an out-of-sample period of size $P=T-R$. The parameters of the FA-QR model are estimated based on a rolling window scheme with the factors extracted in each rolling sample; see \cite{amburgey2022} for the relevance of real-time factor extraction in the assessment of the forecast performance of FA-QR models. Using a rolling window scheme attenuates potential issues associated to parameter instabilities during the forecasting period; see the discussions by \cite{giacomini2010} and \cite{inoue2017}.

When it comes to assess the out-of-sample accuracy of forecasts of a particular quantile, say $q_{\tau}(y_{it+1}|y_{it}, F_t)$, obtained using the estimated FA-QR model, $\hat{q}_{\tau}(y_{it+1})=\hat{\theta}^{\prime} X_{it}$, the literature usually recommends using the following consistent asymmetric piecewise linear tick loss function, denoted as quantile score,
\begin{equation}
\label{eq:QS}
QS(i,\tau)= \frac{1}{P} \sum_{t=R}^{P} ( \vmathbb{1} (y_{it+1} \leq \hat{q}_{\tau}(y_{it+1}) ) - \tau ) \left( \hat{q}_{\tau}(y_{it+1}) - y_{it+1} \right); %= \left\lbrace \begin{matrix}
%\tau \mid y_{it+1}- \tilde{q}_{\tau}(y_{it+1}) \mid, & y_{t+1} \geq \tilde{q}_{\tau}(y_{it+1}) \\
%(1-\tau)  \mid y_{it+1}- \tilde{q}_{\tau}(y_{it+1}) \mid, & yit+11-{} \leq \tilde{q}_{\tau}(y_{it+1})
%\end{matrix}
%\right
\end{equation}
see, for example, \cite{gneiting2011b}, \cite{amburgey2022}, and \cite{kleen2024}. Alternatively, one can use scoring rules specified in terms of the full density function; see \cite{gneiting2014} for an excellent review of scoring rules. In particular, we consider the following popular continuous ranked probability score (CRPS) proposed by \cite{gneiting2011}%\footnote{Alternatively, the densities can be evaluated using the PITs as proposed by \cite{rossi2019}; see \cite{wang2024} for an application.}
\begin{equation}
\label{eq:CRPS}
CRPS(i)= \frac{1}{J}\sum_{j=1}^J \omega(\tau_j) QS(i,\tau_j),
\end{equation}
where $J=5$, $\tau_1=0.05,..., \tau_5=0.95$, and $\omega(\tau)$ is one of the following weighting schemes: i) $\omega(\tau)= (\tau -1)^2$, denoted as left tail (CRPS-L); ii) $\omega(\tau)=\tau^2$, denoted as right tail (CRPS-R); and iii) $\omega(\tau)=1$, denoted as equal (CRPS-E). Recently, \cite{kleen2024} shows that QS and CRPS are not only invariant to transformations in the data but also insensitive to the fact that the observed value of the target variable is not necessarily equal to the value of the true predictand.  

Testing for equal predictive ability (based on either QS or CRPS) of a given FA-QR model with respect to a benchmark model of interest is carried out using the DM test proposed by \cite{diebold1995}.\footnote{Issues associated to testing for equal predictive ability among nested models are avoided by using a rolling window scheme. In fact, in this context, the test of \cite{giacomini2006} is equivalent to the traditional \cite{diebold1995} test; see, for example, \cite{medeiros2021}. 
} In this paper, the benchmark model considered is the quantile regression model in (\ref{eq:Quantile_regression}) without factors, i.e. only with a constant and the autoregressive term, which is denoted as AR-QR model. Although the DM test is robust to the presence of revisions, it does not take into account parameter uncertainty. Consequently, we also implement the AM test of equal predictive ability between two nested quantile regression models when the factors are subject to revisions proposed by \cite{amburgey2022}. It is important to note that the AM test can be implemented to the $QS(i, \tau)$ statistic but not to the CRPS($i$) statistics in (\ref{eq:CRPS}). %When comparing the predictive power of several restricted versions of the FA-QR model, we also use the Model Confidence Set (MCS) of \cite{hansen2011}. 
Finally, we test for equal predictive ability using the fluctuations test of \cite{giacomini2010}, which compares the predictive performance of the FA-QR and the benchmark AR-QR models in each moment of time.
%The significance of these tests is assessed using the test by \cite{diebold1995} corrected as suggested by \cite{harvey1997}.

%Finally, we compare the predictive performance of the FA-QR with that of the quantile autoregressive (QAR) model using the test recently proposed \cite{corradi2023}, which is based on comparing the distances between the actual and nominal coverages in both models. Using the quadratic loss function, for a given quantile, $\tau$, the null and alternative hypothesis are given by
%\begin{equation}
%\label{eq:null}
%H_0: E[ (c_1-\tau)^2- (c_2-\tau)^2=0
%\end{equation}
%\begin{equation}
%\label{eq:alternative}
%H_0: E[ (c_1-\tau)^2- (c_2-\tau)^2 \neq 0
%\end{equation}
%where $c_j$ is the coverage of the one-sided intervals constructed with model $j$, for $j=1$ for the FA-QR  and $j=2$ for the QAR model. Under the null the expected coverage error is the same in both models. One should choose the model that has more accurate coverage on average, placing less weight onto extreme and rare events.

%\cite{corradi2023} stablish the asymptotic validity of critical values based on a wild bootstrap.

%The test can be extended to compare simultaneously several quantiles.   

\subsection{Multi-level dynamic factor model}

The international factors needed to estimate the FA-QR in (\ref{eq:Quantile_regression}) can be extracted by Principal Components (PC) from the following static DFM fitted to $Y_t$
\begin{equation}
\label{eq:DFM}
Y_t=\Lambda F_t+\varepsilon_t,
\end{equation}
where $\Lambda$ is the $N\times r$ matrix of factor loadings, $F_t$ is defined as in (\ref{eq:Quantile_regression}), and $\varepsilon_t$ is the $N \times 1$ vector of idiosyncratic components, which are allowed to be weakly cross-sectionally correlated but uncorrelated with $F_t$; see, for example, \cite{ciccarelli2010}, who propose extracting the common international factor of inflation in OECD countries by PC fitting a DFM as (\ref{eq:DFM}) with a single factor, $r=1$.\footnote{Note that they use the factor to estimate a factor-augmented predictive regression for inflation instead of quantile regressions to estimate the density of inflation.}

However, PC does not take full advantage of the block structure of the system of inflation series. Consequently, PC factors are not able to separately identify specific factors for different blocks of variables and, they are not optimal; see \cite{BreitungEickmeier2015}, \cite{han2021}, and \cite{barigozzi2024} for discussions. Furthermore, the usual criteria for the determination of the number of factors are not conclusive when the eigenvalues of the sample covariance matrix of $Y_t$ have not a clear break, as it is often the case when there are specific factors that only load in subsets of variables. As a consequence, the estimated DFM in (\ref{eq:DFM}) could appear as either having specific weak common factors with many zeros in the loadings or with cross-sectionally correlated idiosyncratic errors when some specific factors are not considered as such; see, for example, the discussions by \cite{moench2013} and \cite{luciani2014}. In the first case, the presence of zeros in the loadings may bias the estimates of the underlying factors; see \cite{boivin2006} and \cite{BreitungEickmeier2015}. In the second case, weak cross-correlated idiosyncratic noises may have consequences for the measurement of the uncertainty of the factors; see \cite{fresoli2023}. Furthermore, it may also have consequences for the estimation of the FA-QR models in case of very persistent specific factors: see \cite{luciani2014}.

Alternatively, when the variables in $Y_t$ have a block structure, the factors are easily interpretable if they are extracted from a ML-DFM, which is obtained after imposing the adequate zero restrictions on the matrix of loadings, $\Lambda$; see, for example, \cite{ha2019a} and \cite{ha2023}, who propose a ML-DFM for inflation with a global common factor and, in recognition of differences in economic strcutures and policy frameworks, specific factors corresponding to blocks of Advanced, Emerging and Low Income economies; see also the analysis by  \cite{jorda2020}, who show that advanced economies differ considerably from emerging ones . Alternatively, several authors consider specific factors associated to the geographical area; see, for example, \cite{medeiros2024}. Consequently, we extract the factors based on a ML-DFM that decomposes the factor structure into different levels such that some factors are associated with the full cross-section of inflations while some others either impact on economies in a given geographical region or on blocks of economies classified according to their economic development level. %\footnote{This ML-DFM is closely related to the three-level model proposed by \cite{BreitungEickmeier2015}.}
%In order to specify the factor structure of the ML-DFM, i.e. to determine the zeros in the loading matrix $\Lambda$, we determine the number of factors within each block by visual inspection of the scree plot; see \cite{hindrayanto2016}, who also use the scree plot.

Estimation of the ML-DFM with overlapping blocks of variables is challenging as the factor structure does not allow estimating one level after another. In this paper, we estimate the parameters using the sequential procedure proposed by \cite{BreitungEickmeier2015}; see \cite{rodriguez2019} for details about the estimation algorithm, \cite{djogbenou2024}, who uses the same estimator in the context of oil prices, and \cite{gonzalez2024} for an application to extracting domestic and international factors underlying economic growth. First, initial estimates of the factors are obtained using canonical correlations and PC. Second, a sequential Least Squares (LS) procedure is implemented to estimate the loadings and factors. %\footnote{See \cite{choi2018} for a similar estimation procedure and \cite{aastveit2016} for an alternative estimation procedure for ML-DFMs and a bootstrap procedure to construct confidence bounds for the factors.}
 It is important to note that, although the ML-DFM could be identified because of the zeros imposed on the matrix of loadings $\Lambda$, identification restrictions are still needed in order to extract the factors in each step of the estimation procedure; see the discussions by \cite{choi2018} and \cite{lin2023} on identification of ML-DFMs. Consequently, all factors need to have norm one and the regional factors to be orthogonal with respect to the development level factors, and all block factors to be orthogonal with the global factor. Even though there is not yet a formal result on the asymptotic distribution of the factors extracted from ML-DFMs, in this paper, confidence regions for the factors are constructed using the asymptotic distribution derived by \cite{choi2018} for the pervasive factor, which is extracted in the first step and has the same asymptotic distribution derived by \cite{bai2003}. For the rest of the factors, which are extracted based on the residuals from the previous step, we also assume asymptotic normality. Since they are based on residuals, their asymptotic MSE will be affected by parameter estimation uncertainty but this problem should be mitigated by extending the subsampling procedure of \cite{MR2017}. which is designed to incorporate the uncertainty due to the estimation of the loading, to the ML-DFM framework.

\section{Data description}
\label{section:data}

\begin{figure}[]
\caption{Maps of countries classified according to their income (top) and countries classified according to the region (bottom).}
\label{fig:maps}
\begin{center}
\includegraphics[width=0.75\textwidth]{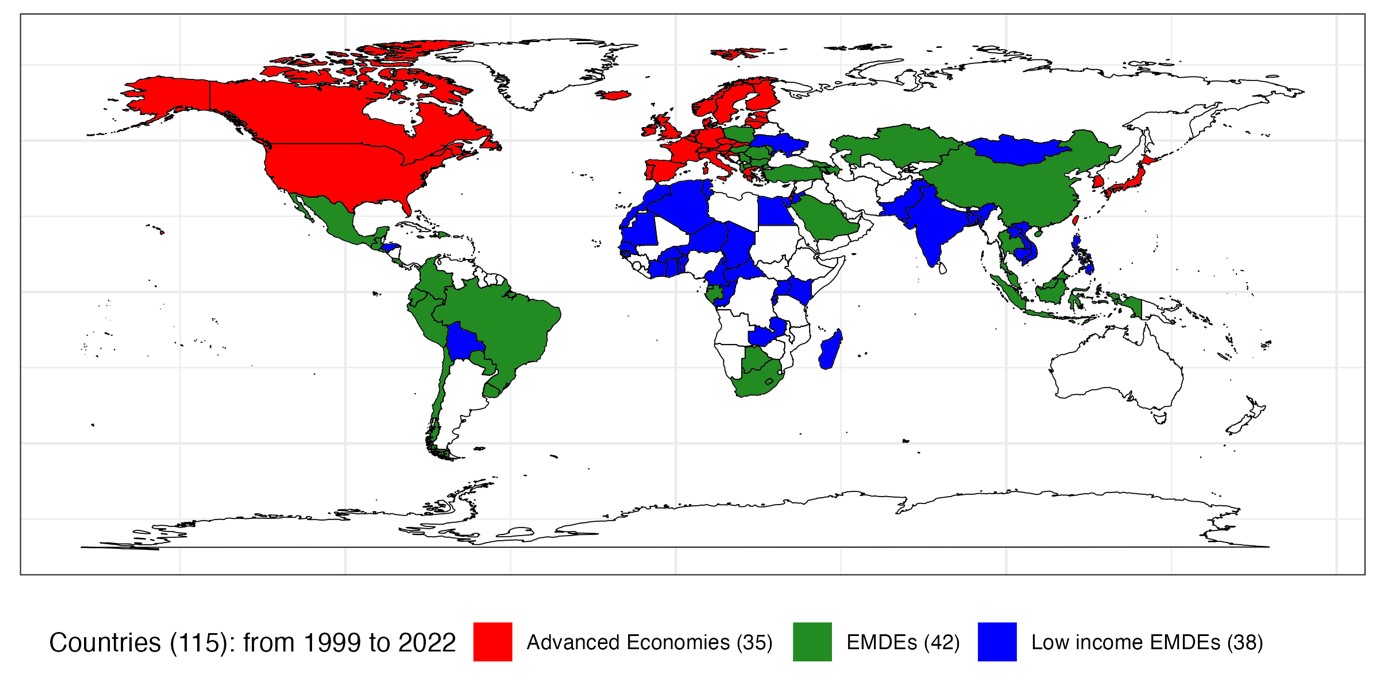}
\includegraphics[width=0.75\textwidth]{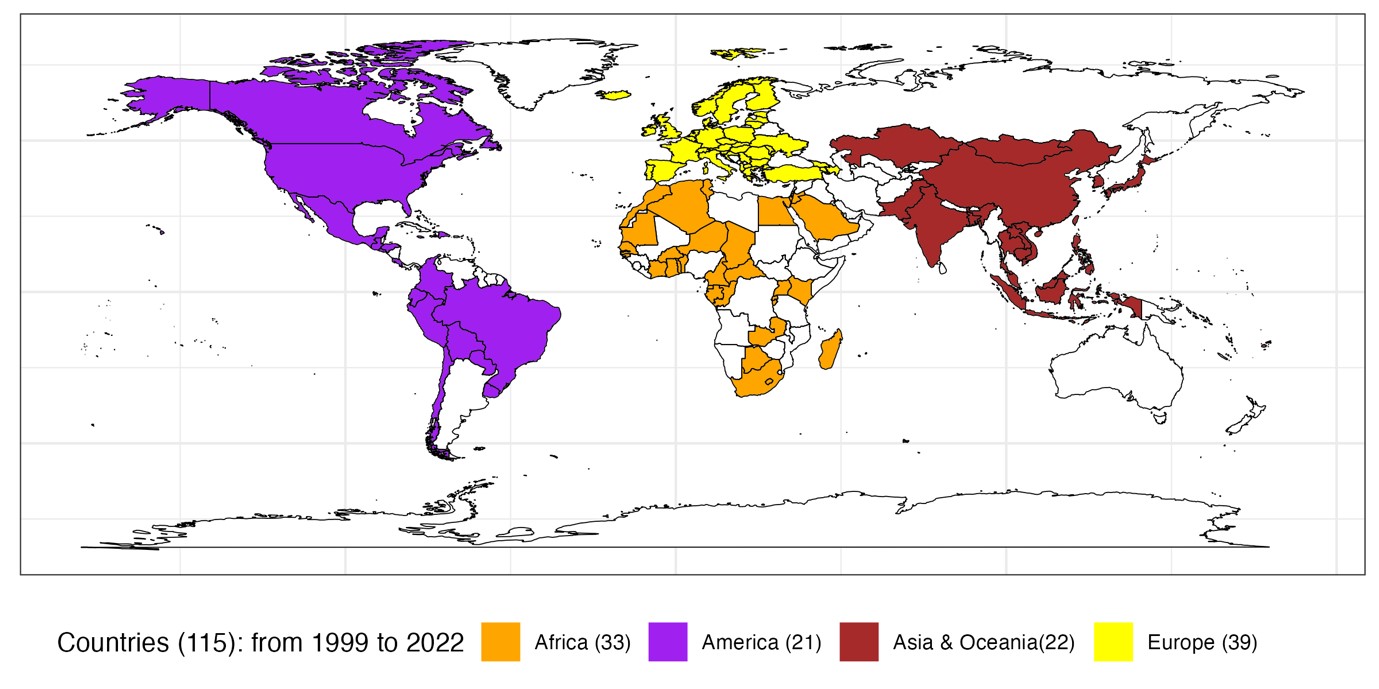}
\end{center}
\end{figure}

We obtain monthly observations of headline CPI from January 1999 to December 2022 ($T=288$) for a set of $N=115$ countries from the novel \textit{Global Database on Inflation} (GDI) constructed by \cite{ha2023}, who compile data on prices of a large set of countries from multiple sources, including various international institutions as well as large number of country-specific sources; see also \cite{medeiros2024} and \cite{pitarakis2023}, who consider GDI.\footnote{Headline prices are defined by \cite{ha2023} as \textit{ prices of all goods and services in a basket that is representative of consumer expenditures} and, consequently, it has the broadest coverage in terms of the basket of goods and services included.} We should note that inflation data in the GDI is available before 1999 for $209$ countries. However, we select $N=115$ countries in order to have a balanced data set since 1999, which corresponds with the introduction of the Euro and the pegging of European currencies. Furthermore, several countries, namely Belarus, Haiti, Sudan, Malawi, Tanzania, United Republic of Gambia, The Maldives and Hong Kong SAR (China), have been excluded from our analysis due to the very atypical behaviour of their inflation series. Table \ref{tab:countries} lists all countries considered in this paper, which are classified according to two different criteria. First, according to the International Monetary Found (IMF), they can be classified into 35 ADV economies and 80 emerging and developing economies. Furthermore, within this last group,  according to the Word Economic Outlook 2022 of the World Bank, we consider the group of 42 MHI-EMD economies and those of 38 LI-EMD economies. The top panel of Figure \ref{fig:maps} represents the world map with the distribution of countries classified according to their income. Second, the countries in our data base can be classified according to their geographical area as follows: Asia and Pacific (21), Europe and Central Asia (40), America (21), Middle East and Africa (33). %; see \cite{medeiros2024} for a similar classification of countries according to geographical area although they divide American countries into countries in North America and in South America. 
The bottom panel of Figure \ref{fig:maps} represents the world map with the distribution of countries classified according to their geographical region. Note that there is an important overlapping between countries classified according to their geographical region and according to the development level.

\begin{table}[ht!]
\caption{List of countries classified by income together with their codes and geographical regions.}
\label{tab:countries}
\begin{center}
\resizebox{450pt}{!}{
\begin{tabular}{c c c c c c c c c c c c c c c c c}
\multicolumn{3}{c}{Developed} && \multicolumn{3}{c}{Middle-High Emerging} && \multicolumn{3}{c}{Low Emerging} \\
\hline
Country & Code & Region && Country & Code & Region && Country & Code & Region\\
\hline
Austria & AUT & Europe && Albania & ALB	 & Europe && Algeria & DZA	& Africa\\
Belgium & BEL & Europe&& Antigua and Barbuda & ATG	& America && Bangladesh & BGD&	Asia\\
Canada & CAN & America&& Armenia & ARM & Europe && Benin & BEN	& Africa\\
Croatia& HRV & Europe&& Azerbaijan & AZE&	Europe && Bolivia & BOL	& America\\
Cyprus& CYP & Europe&& Bahamas & BHS&	America&& Burkina Faso & BFA &	Africa\\
Czech Republic& CZE & Europe&& Botswana & BWA &	Africa && Burundi & BDI	& Africa\\
Denmark&DNK & Europe&& Brazil & BRA	& America&& Cambodia & KHM &	Asia\\
Estonia& EST & Europe&& Bulgaria & BGR	&Europe && Cameroon & CMR &	Africa\\
Finland& FIN &  Europe&& Chile & CHL &	America && Central African Republic & CAF &	Africa\\
France&FRA & Europe&& China & CHN &	Asia && Chad & TCD	 & Africa\\
Germany& DEU & Europe&& Colombia & COL	& America && Congo, Rep. & COG &	Africa\\
Greece& GRC & Europe&& Costa Rica & CRI	& America && Cote d'Ivoire & CIV &	Africa\\
Iceland& ISL & Europe&& Dominican Republic & DOM &	America && Egypt, Arab Rep. & EGY &	Africa\\
Ireland& IRL & Europe&& Ecuador & ECU &	America && Ghana & GHA	& Africa\\
Israel& ISR & Africa && El Salvador & SLV &	America && Guinea-Bissau & GNB &	Africa\\
Italy& ITA & Europe&& Equatorial Guinea & GNQ	& Africa && Honduras & HND &	America\\
Japan&JPN & Asia&& Fiji & FJI &	Asia && India & IND	 & Asia\\
Korea, Rep.& KOR & Asia&& Gabon & GAB &	Africa && Jordan & JOR &	Africa\\
Latvia& LVA & Europe&& Georgia & GEO &	Europe && Kenya & KEN &	Africa\\
Lithuania& LTU & Europe&& Guatemala & GTM	& America && Lao, PDR & LAO	& Asia\\
Luxembourg& LUX & Europe&& Hungary & HUN &	Europe && Madagascar & MDG	& Africa\\
Macao SAR, China& MAC & Asia&& Indonesia & IDN &	Asia && Mauritania & MRT	& Africa\\
Malta& MLT & Africa&& Jamaica & JAM	& America && Mongolia & MNG	& Asia\\ 
Netherlands& NLD & Europe&& Kazakhstan & KAZ&	Asia && Morocco & MAR &	Africa\\
Norway& NOR & Europe&& Malaysia & MYS &	Asia && Niger & NER	& Africa\\
Portugal& PRT & Europe&& Mauritius & MUS &	Africa && Pakistan & PAK &	Asia\\
Singapore& SGP & Asia&& Mexico & MEX & 	America && Philippines & PHL &	Asia\\
Slovakia& SVK & Europe&& Moldova, Rep. & MDA &	Europe && Rwanda & RWA	& Africa\\
Slovenia& SVN & Europe&& North Macedonia & MKD	& Europe && Samoa & WSM	& Asia\\
Spain& ESP & Europe&& Paraguay & PRY &	America && Sao Tome and Principe & STP	& Africa\\
Sweden& SWE & Europe&& Peru & PER & America && Senegal & SEN &	Africa\\
Switzerland& CHE &Europe&& Poland & POL	& Europe && Solomon Islands & SLB &	Asia\\
Taiwan, China& TWN & Asia&& Romania & ROU &	Europe && Togo & TGO &	Africa\\
United Kingdom& GBR & Europe&& Saudi Arabia & SAU &	Africa && Tunisia & TUN	& Africa\\
United States & USA & America&& Serbia & SRB &	Europe && Uganda & UGA	& Africa\\
&& && South Africa & ZAF &	Africa && Ukraine & UKR	 & Europe\\
&& && St. Kitts and Nevis & KNA	& America && Vietnam & VNM	& Asia\\
&& &&St. Lucia & LCA	& America&& Zambia & ZMB	& Africa\\
&& && Thailand & THA & Asia\\
 && && Tonga & TON &	Asia\\
&& && Turkey & TUR	& Europe\\
 && && Uruguay & URY&	America
\end{tabular}
}
\end{center}
\end{table}

For each country considered, CPI prices are transformed to annualized month-on-month (\textit{mom}) inflation with each series of inflation sequentially cleaned of seasonal effects and outliers.%; see \cite{watson2002}, \cite{hansen2012}, \cite{tu2019} and \cite{bae2024}, among others, for \textit{mom} inflation.%, and \cite{eickmeier2013}, \cite{inoue2017}, and \cite{pitarakis2023}, for quarter-on-quarter (\textit{qoq}) inflation based on quarterly data.
\footnote{By considering \textit{mom} instead of year-over-year (\textit{yoy}) inflation, we avoid issues associated with potential non-stationarity; see \cite{watson1999}, who found little difference in $I(1)$ and $I(2)$ factor model forecasts for prices.} To deseasonalise, we follow \cite{harvey2006} and fit a structural time series model with the seasonal component estimated using the Kalman filter and subtracted from the original observations of inflation.\footnote{According to \cite{camacho2015}, factor extraction is not affected by deseasonalization.}$^{,}$\footnote{All computations in this paper have been carried out by the first two authors using their own Matlab programs.} After deseasonalizing each series, we follow \cite{mccracken2016} and identify outliers as observations that deviate from the sample median more than 10 interquartile ranges (IR), substituting them by $10 \times IR$ multiplied by the sign.

\begin{figure}[]
\caption{Deseasonalized annualized monthly inflations grouped by development level. Grey shaded areas represent recession periods as identified by the US National Bureau of Economic Research (NBER) and the read shaded area represents the Russia-Ukraine war period.}
\label{fig:inflations_1}
\begin{center}
\includegraphics[width=0.75\textwidth]{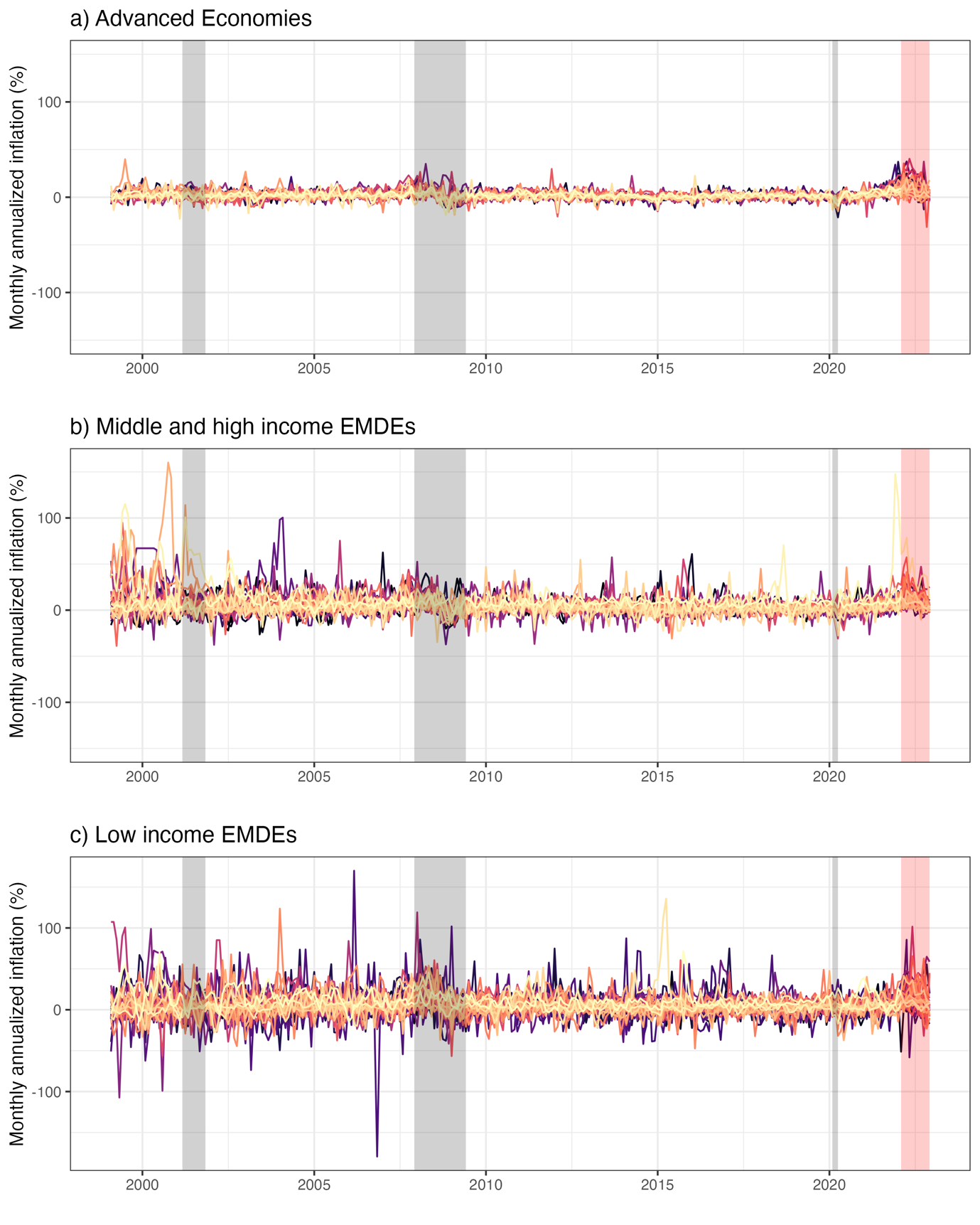}
\end{center}
\end{figure}

Figure \ref{fig:inflations_1} plots the corrected series of inflation separately in the ADV, MHI-EMD and LI-EMD economies. %while Figure \ref{fig:inflations_2} plots them grouped by geographical area.
It is remarkable that, with the exception of the ADV economies, we observe episodes of what can be called as hyperinflation in a large proportion of economies. %\footnote{\cite{cagan1956} defines hyperinflation as a monthly inflation greater than 50\% in at least one month.}
However, in the last few years since the Russian invasion of Ukraine, ADV economies are amid their worst inflation pressure over the sample period due to a plethora of unprecedented developments. In some ADV countries with a long history of low and stable inflation, monthly inflation has soared to 40\% also showing episodes of -30\% deflation. Furthermore, the extent and pace of this surge are not just a recurring problem in ADV economies, but it is also becoming an entrenched phenomenon across the world. This unprecedented extreme movements in inflation could be attributed, among other causes, to the strong rebound in aggregate demand caused by the extraordinary policy response to the COVID-19 pandemic and to global supply constraints and shock waves through international commodity markets triggered by the Russian invasion of Ukraine.

\begin{landscape}
\begin{figure}[]
\begin{center}
\subfigure[All economies]{\includegraphics[width=0.2\linewidth]{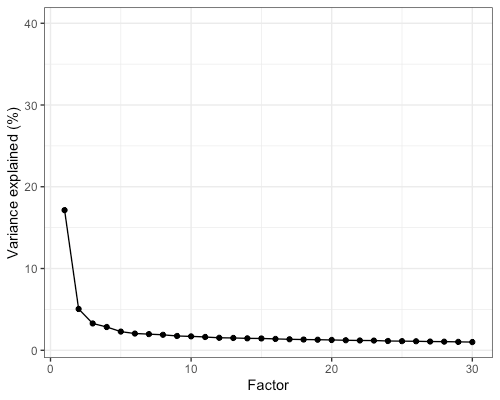}}\\
\subfigure[Advanced]{\includegraphics[width=0.2\linewidth]{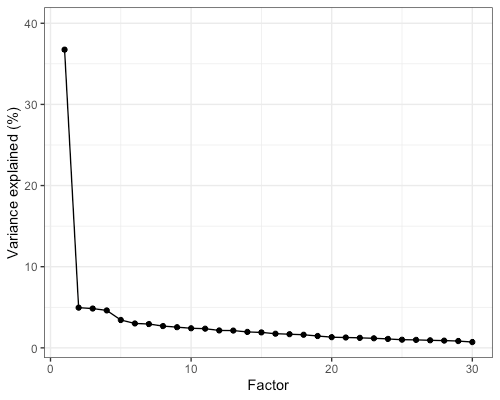}}
\subfigure[HMI]{\includegraphics[width=0.2\linewidth]{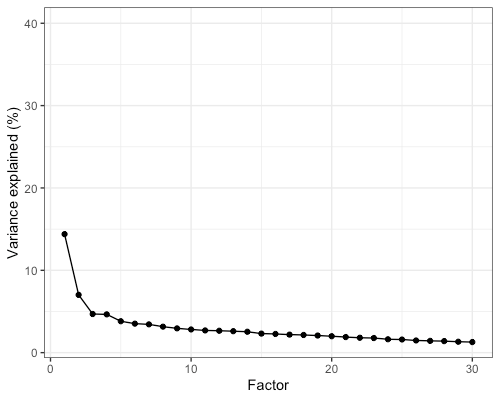}}
\subfigure[LIE]{\includegraphics[width=0.2\linewidth]{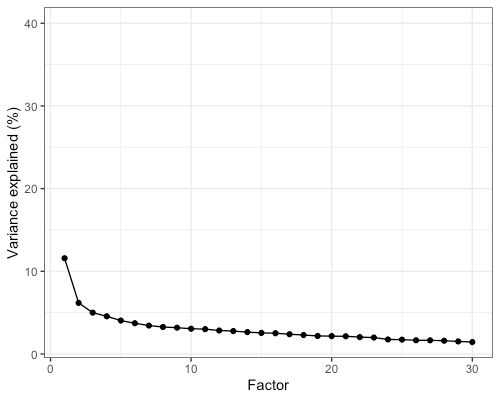}}\\
\subfigure[Europe]{\includegraphics[width=0.2\linewidth]{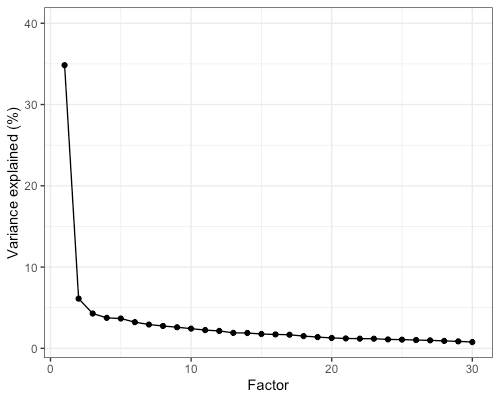}}
\subfigure[America]{\includegraphics[width=0.2\linewidth]{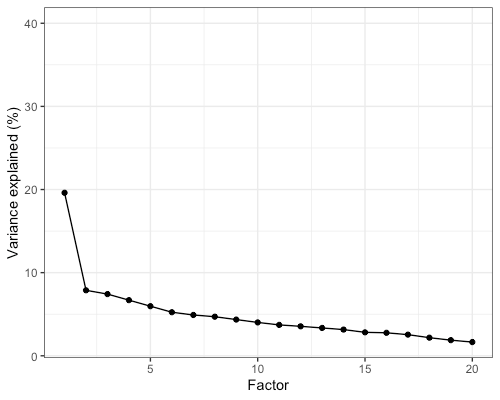}}
\subfigure[Africa]{\includegraphics[width=0.2\linewidth]{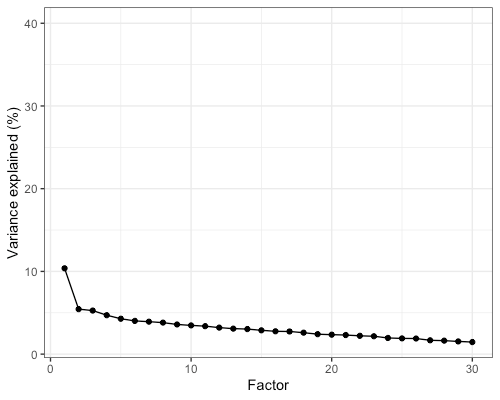}}
\subfigure[Asia]{\includegraphics[width=0.2\linewidth]{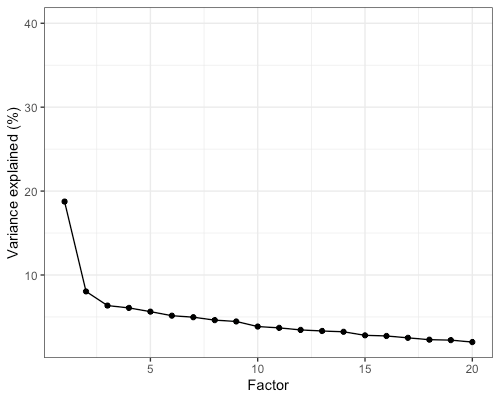}}
\caption{Scree plots: All inflations (top panel); inflations grouped according to income (second row); and to region (third row).}
\label{fig:Scree}
\end{center}
\end{figure}
\end{landscape}

In order to determine the structure of the ML-DFM used to extract the international common factors of inflations around the world, Figure \ref{fig:Scree} plots scree-plots for all inflations in the data set (top panel) together with scree-plots for inflations grouped according to their income level (second row) and to their geographical region (third row).\footnote{Due to the large number of subgroups of inflations and the small number of series within some of them, it is not possible to determine the structure of the ML-DFM by using the methodology proposed by \cite{hallin2011}, who propose analysing the pairwise correlations among the factors separately extracted from each block of variables.} Two main conclusions are clear. First, when looking at the scree-plot of all inflations in Panel (a) of Figure \ref{fig:Scree}, we observe that there is not a clear cut-off point, and therefore, the number of underlying common factors cannot be sharply determined. As mentioned above, this could be a consequence of the presence of factors affecting only some blocks of variables. Second, although there is not a clear cut-off point, we can see that, regardless of whether we look at the blocks classified according to geographical regions or to economic development, the number of factors within each block that explains more than 5\% of the total variability varies between 2 factors in ADV or European economies and 6 in economies in Africa or in Asia.

\section{The distribution of inflation: International underlying factors}
\label{section:Distribution}

In this section, we estimate the conditional quantiles of inflation in each country using FA-QRs, with the international factors extracted from the system of inflations using a ML-DFM with overlapping factors.

\subsection{The international factors underlying inflation}

In concordance with the results in the previous section, we represent the complex structure of common factors in monthly inflations by fitting a ML-DFM with one global factor, $F_{gt}$, which loads in all inflations in the system; one factor for each regional block, $F_{Aft}$, $F_{Amt}$, $F_{Ast}$, and $F_{Eut}$, which loads in inflations in Africa, America, Asia and Europe, respectively; and one factor for each level of development block, $F_{Adt}$, $F_{Mit}$, and $F_{Lit}$, which loads in inflations of ADV, MHI-EMD and LI-EMD economies, respectively. Denote by $Y_{1t}$, $Y_{2t}$ and $Y_{3t}$ the blocks of inflations in ADV, MHI-EMD and LI-EMD economies in Africa. Similarly, $Y_{4t}$, $Y_{5t}$ and $Y_{6t}$ denote the blocks of inflations in economies in America. Finally, $Y_{7t}, Y_{8t}$ and $Y_{9t}$ are the blocks of inflations in Asia while $Y_{10t}, Y_{11t}$ and $Y_{12t}$ are the blocks of inflations in Europe. The ML-DFM considered is given by
\begin{equation}
\label{eq:ML-DFM}
\left[ \begin{matrix}
Y_{1t} \\
Y_{2t} \\
Y_{3t} \\
Y_{4t} \\
Y_{5t} \\
Y_{6t} \\
Y_{7t} \\
Y_{8t} \\
Y_{9t} \\
y_{10t} \\
Y_{11t} \\
Y_{12t}
\end{matrix} \right] = 
\left[ \begin{matrix}
\lambda_{11} & \lambda_{12} & 0 & 0 & 0 & \lambda_{16} & 0 & 0 \\
\lambda_{21} & \lambda_{22} & 0 & 0 & 0 & 0 & \lambda_{27} & 0 \\
\lambda_{31} & \lambda_{32} & 0 & 0 & 0 & 0 & 0 & \lambda_{38} \\
\lambda_{41} & 0 & \lambda_{43} & 0 & 0 & \lambda_{46} & 0 & 0 \\
\lambda_{51} & 0 & \lambda_{53} & 0 & 0 & 0 & \lambda_{57} & 0\\
\lambda_{61} & 0 & \lambda_{63} & 0 & 0 & 0 & 0 & \lambda_{68} \\
\lambda_{71} & 0 & 0 & \lambda_{74} & 0 & \lambda_{76} & 0 & 0\\
\lambda_{81} & 0 & 0 & \lambda_{84} & 0 & 0 & \lambda_{87} & 0 \\
\lambda_{91} & 0 & 0 & \lambda_{94} & 0 & 0 & 0 & \lambda_{98} \\
\lambda_{10,1} & 0 & 0 & 0 & \lambda_{10,5} & \lambda_{10,6} & 0 & 0 \\
\lambda_{11,1} & 0 & 0 & 0 & \lambda_{11,5} & 0 & \lambda_{11,7} & 0 \\
\lambda_{12,1} & 0 & 0 & 0 & \lambda_{12,5} & 0 & 0 & \lambda_{12,8} 
\end{matrix} \right]
\left[ \begin{matrix}
F_{gt} \\
F_{AFt} \\
F_{Amt} \\
F_{Ast} \\
F_{Eut} \\
F_{Adt} \\
F_{Mit} \\
F_{Lit}
\end{matrix} \right] + \varepsilon_t.
\end{equation}

\begin{figure}[]
\caption{Estimated factors (blue lines) of ML-DFM together with $95\%$ confidence bounds (grey areas). Grey shaded areas represent recession periods as identified by the US National Bureau of Economic Research (NBER) and the read shaded area represents the Russia-Ukraine war period.}
\label{fig:Estimated_factors}
\begin{center}
\includegraphics[width=1\textwidth]{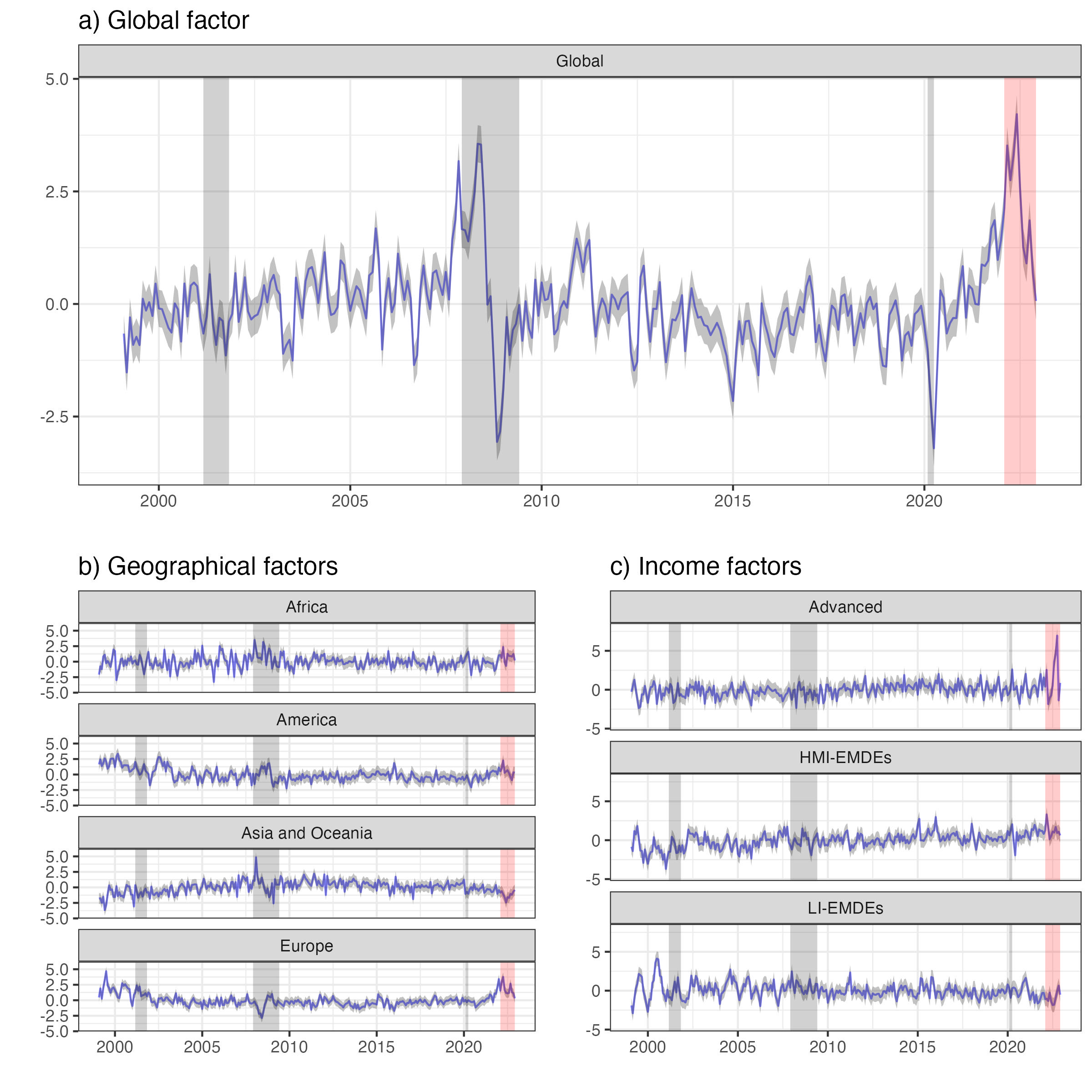}
\end{center}
\end{figure}

\begin{figure}[]
\caption{Estimated factor loadings of ML-DFM together with $95\%$ confidence bounds.}
\label{fig:Estimated_loadings}
\begin{center}
\includegraphics[width=1\textwidth]{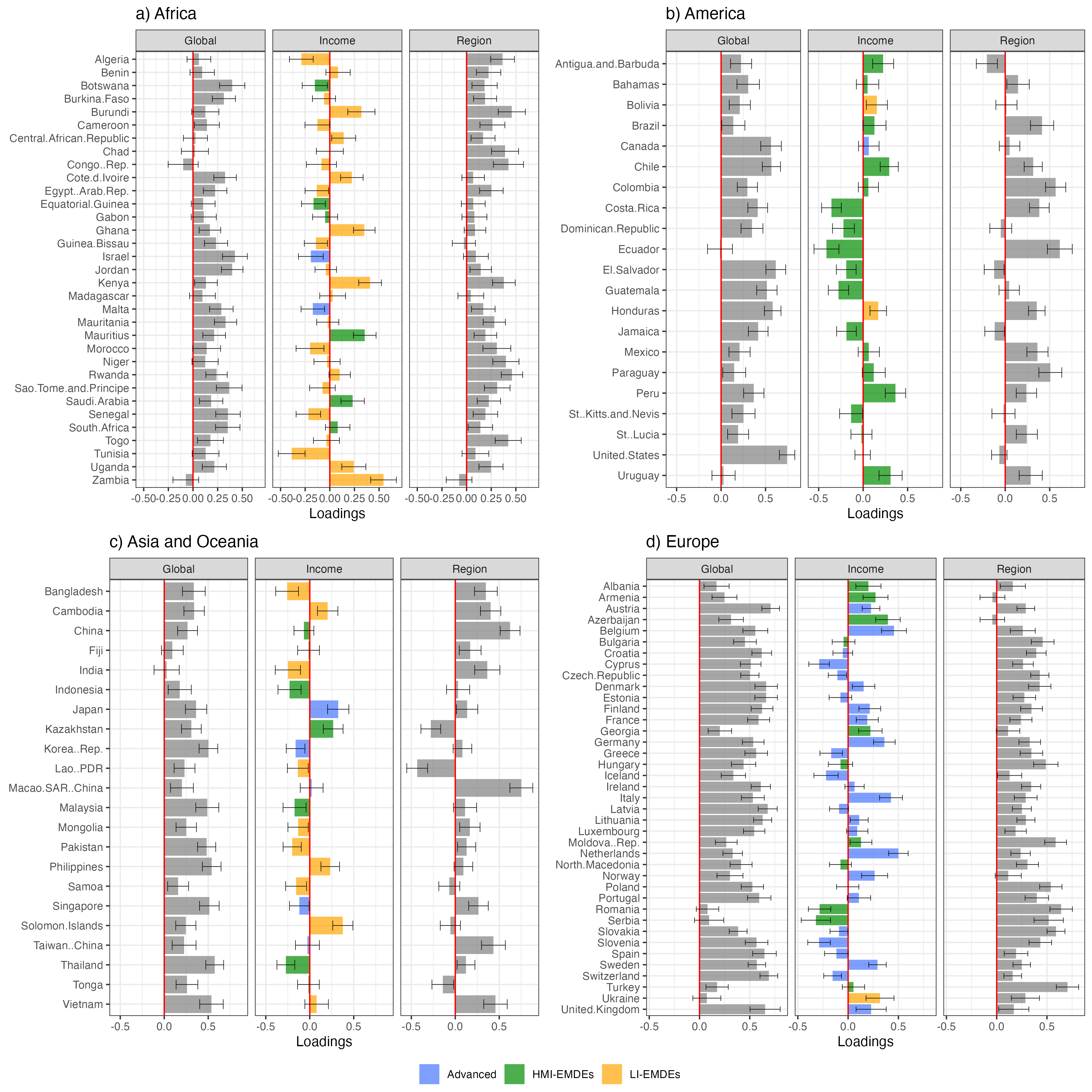}
\end{center}
\end{figure}

The estimated factors and loadings of model (\ref{eq:ML-DFM}) are plotted in Figures \ref{fig:Estimated_factors} and \ref{fig:Estimated_loadings}, respectively, together with their $95\%$ confidence bounds. We can observe that the global international factor, $F_{gt}$, loads in almost all inflations around the world (approximately $84\%$). Most economies with non-significant loadings on $F_{gt}$ are in Africa (Algeria, Benin, Burundi, Central African Rep., Chad, Congo Rep., Madagascar, Tunisia, Zambia, Equatorial Guinea and Gabon) and only two in America (Ecuador and Uruguay), two in Asia (Fiji and India), and three in Europe (Romania, Serbia and Ukraine). All these countries are LI-EMD but Ecuador, Uruguay, Fiji, Romania, Serbia, Equatorial Guinea and Gabon, which are MHI-EMDE economies. %However, approximately $34\%$ (11 out of 33) economies in Africa have not significant loadings on $F_{gt}$, with 9 of those countries being LI-EMD economies ) and the other two () being HMI-EMD economies.
Consequently, in concordance with the related literature, we find a strong global international inflation factor linking the inflation of most economies around the world. It loads in all ADV economies, $83\%$ MHI-EMD economies and $71\%$ LI-EMD economies. When looking at the distribution at different geographical regions, the global factor loads in $92.5\%$ European economies, $90\%$ economies in America and Asia, and $66\%$ economies in Africa.

The international global factor of inflation, $F_{gt}$ moves around an approximately constant level after the global financial crisis, and shows a gradual increase since the COVID pandemic, which becomes sharp after the Russian invasion of Ukraine. Note that, as expected, the global inflation factor registers sharp movements around global recessions; see also \cite{ha2022} for a description of global international inflation extracted from a similar group of countries.

We can also observe that international group-specific factors also play important roles in summarizing the variability of inflation around the world. First, the specific-regional factor in Europe, $F_{Eut}$, is very strong, with its loadings being significant and positive in $90\%$ of countries. The loadings are significant and positive in $73\%$ of countries in Africa and $67\%$ of countries in America and Asia. Note that in these two latter groups of countries, when significant, the loadings are negative in few economies. The regional factor in Asia, $F_{ASt}$, is highly persistent, while $F_{Aft}$ is highly volatile. The European factor, $F_{Eut}$ is the most affected by the Russian invasion while the factors of Africa, America and Asia only have very mild impacts.

Finally, we can observe that the loadings associated to the level of economic development factors, are significant in $68\%$ of LI-EMD economies, $67\%$ of MHI-EMD economies and $74\%$ of ADV economies. The effect on inflation of the Russian invasion is exacerbated in $F_{Adt}$. However, $F_{Mit}$ and $F_{Lit}$, although not affected by the Russian invasion, are highly persistent, with the former showing a smooth increase since the global financial crisis.

\subsection{Estimation of inflation densities}

Using the international factors estimated in the previous subsection, we fit the FA-QR model in (\ref{eq:Quantile_regression}) to each of the 115 economies in our data base, for $\tau=0.05, 0.25, 0.50, 0.75, 0.95$ quantiles. In this section, we describe in detail estimation results for two particular groups of economies and $\tau=0.05, 0.50, 0.95$.\footnote{Results for all other countries and quantiles are available upon request.} In particular, we consider the economies within the Group of Seven (G7), namely, Canada, France, Germany, Italy, Japan, UK, and the US, which are all ADV and located in three different geographical regions, America, Asia and Europe. Therefore, in their corresponding FA-QRs, the parameters $\beta_2=\beta_7=\beta_8=0$ . We also consider the economies within the Group of Five (G5), namely, Brazil, China, India, Mexico, and South Africa, which are all MHI-EMD but India, which belongs to the LI-EMD block. The G5 countries are located in 3 different geographical areas, namely, America, Asia and Africa. Therefore, $\beta_5=\beta_6=0$ for all of them. 

Consider first the estimated parameters of the FA-QR model for the G7 economies, which are reported in Table \ref{tab:estimated_parameters_1}, together with their corresponding $p$-values.\footnote{The standard errors are calculated using the option \textit{ker} of the R software \textit{quantreg} available at https://cran.r-project.org/web/packages/quantreg/index.html.} The global factor, $F_{gt}$, is significant and positive for all G7 economies and quantiles but in Canada and the US. In these last two countries, $F_{gt}$, is only significant for the upper $95\%$ quantile. In any case, the parameter of $F_{gt}$ is larger for the $95\%$ quantile in all G7 economies, implying that the global international factor is more relevant to explain the risk of high inflation in these economies. Furthermore, regardless of the quantile, the parameters of the global factor are larger in Italy than in Germany; see \cite{ciccarelli2010}, who conclude that countries with stronger commitment to price stability (such as Germany) are less affected by the global factor than those with weaker inflation discipline (such as Italy).

With respect to the regional-specific factors, we can observe that they are significant for at least one quantile of the distribution of inflation in all G7 economies. In Canada, the American factor, $F_{Amt}$, is only significant and positive for the $95\%$ quantile, while in the US, it is significant, although negative, for the lower $5\%$ quantile. It seems that the American factor is more relevant in Central and South American countries than in Canada and the US. In Japan, the regional factor, $F_{Ast}$, is significant for the upper quantile $95\%$, while, $F_{Eut}$ is significant in at least one quantile in all European countries. Finally, the factor of ADV economies, $F_{Adt}$, is significant for at least one quantile, in all G7 economies but Italy. %Summarizing, we can conclude that international inflation factors play an important role in explaining inflation densities in the G7 countries, mainly when looking at the upper quantiles.

Table \ref{tab:estimated_parameters_1}, which also reports the $R^1$ coefficients, shows that they are always larger for the upper quantile $95\%$. They are largest in UK with $R^1=0.17, 0.14$ and $0.35$ for $\tau=0.05, 0.50$ and $0.95$, respectively. Finally, we can also observe that the differences between the AICs of the FA-QR and AR-QR models are generally larger when looking at the 95\% quantile and smaller in the 50\% quantile. It seems that international factors are not so important for mean domestic inflation as they are for the upper quantiles of the inflation distribution. 

\begin{landscape}
\begin{table}[]
\resizebox{25cm}{!}{
\centering
\begin{tabular}{@{}lrrrrrrrrrrrrrrrrrrrrrrrrrrr@{}}
\\
& \multicolumn{3}{c}{Canada} && \multicolumn{3}{c}{France} && \multicolumn{3}{c}{US} && \multicolumn{3}{c}{Italy} && \multicolumn{3}{c}{Germany} && \multicolumn{3}{c}{UK} && \multicolumn{3}{c}{Japan}\\
& 0.05 & 0.50 & 0.95 && 0.05 & 0.50 & 0.95 && 0.05 & 0.50 & 0.95 && 0.05 & 0.50 & 0.95 && 0.05 & 0.50 & 0.95 && 0.05 & 0.50 & 0.95 && 0.05 & 0.50 & 0.95 \\
\cline{2-28}
\\
$\mu$ & \textbf{-4.261} & \textbf{2.232} & \textbf{8.216} && \textbf{-2.524} & \textbf{2.107}& \textbf{6.407} && \textbf{-3.834} & \textbf{1.995} & \textbf{6.761} && \textbf{-1.074} & \textbf{2.293} & \textbf{6.200} && \textbf{-1.696} & \textbf{1.541} & \textbf{5.426} && \textbf{-0.609} & \textbf{2.367}  & \textbf{6.692} && \textbf{-3.386} & 0.234 & \textbf{3.841} \\
 & (0.00) & (0.00) & (0.00) && (0.00) & (0.00) & (0.00) && (0.00) & (0.00) & (0.00) && (0.00) & (0.00) & (0.00) && (0.00) & (0.00) & (0.00) && (0.07) & (0.00) & (0.00) && (0.00) & (0.25) & (0.00) \\
$\phi$ & 0.075 & 0.062 & \textbf{-0.198} && -0.011 & \textbf{-0.453} & \textbf{-0.554} && \textbf{0.407} & \textbf{0.299} & \textbf{0.268} && \textbf{-0.229} & -0.166 & \textbf{-0.231} && -0.109 & 0.047 & 0.196 && -0.176 & -0.044 & -0.133 && 0.061 & 0.117 & 0.057 \\
 & (0.52) & (0.55) & (0.07) && (0.93) & (0.00) & (0.00) && (0.02) & (0.02) & (0.02) && (0.07) & (0.25) & (0.01) && (0.29) & (0.65) & (0.21) && (0.26) & (0.66) & (0.40) && (0.66) & (0.11) & (0.37)\\
$\beta_{1}$ & 0.699 & 0.340 & \textbf{1.961} && \textbf{1.077} & \textbf{1.807} & \textbf{1.992} && -0.093 & 0.288 & \textbf{1.310} && \textbf{1.787} & \textbf{1.442} & \textbf{2.063} && \textbf{1.274} & \textbf{0.723} & \textbf{1.232} && \textbf{1.443} & \textbf{1.438} & \textbf{2.222} && \textbf{0.879} & \textbf{0.731} & \textbf{1.139} \\
 & (0.13) & (0.46) & (0.00) && (0.00) & (0.00) & (0.00) && (0.88) & (0.57) & (0.00) && (0.00) & (0.00) & (0.00) && (0.00) & (0.00) & (0.00) && (0.00) & (0.00) & (0.00) && (0.00) & (0.00) & (0.00)\\
$\beta_{2}$ & - & - & - && - & - & - && - & - & - && - & - & - && - & - & - && - & - & - && - & - & -\\
& \\
$\beta_{3}$ & -0.676 & 0.214 & \textbf{0.496} && & & && \textbf{-0.769} & 0.308 & 0.044 &&  &  & &&  &  &  &&  &  &  &&  &  &  \\
 & (0.14) & (0.55) & (0.07) && & & && (0.10) & (0.28) & (0.86) &&  & & && & & && & & && & &\\
$\beta_{4}$ & & & && & & && & & &&  &  & &&  &  &  &&  &  &  && 0.252 & 0.307 & \textbf{0.475} \\
 & & & && & & && & & && & & && & & && & & && (0.27) & (0.11) & (0.07)\\
$\beta_{5}$ & & & && 0.324 & \textbf{0.815} & \textbf{1.537} && & & && \textbf{0.622} & \textbf{0.711} & \textbf{2.070} && \textbf{1.065} & \textbf{0.663} & 0.292 && -0.081 & 0.325 & \textbf{1.667} &&  &  &  \\
 & & & && (0.28) & (0.00) & (0.00) && & & && (0.02) & (0.01) & (0.00) && (0.00) & (0.01) & (0.25) && (0.74) & (0.11) & (0.00) && & &\\
$\beta_{6}$ & \textbf{0.752} & 0.052 & 0.030 && -0.090 & \textbf{0.609} & -0.014 && \textbf{0.518} & -0.120 & \textbf{-0.587} && 0.380 & 0.139 & 0.486 && \textbf{0.660} & 0.237 & 0.651 && 0.417 & \textbf{0.462} & \textbf{0.984} && \textbf{0.892} & \textbf{0.368} & \textbf{0.427} \\
 & (0.00) & (0.87) & (0.92) && (0.79) & (0.01) & (0.94) && (0.02) & (0.65) & (0.01) && (0.22) & (0.65) & (0.13) && (0.00) & (0.32) & (0.13) && (0.12) & (0.03) & (0.00) && (0.00) & (0.08) & (0.03)\\
$\beta_{7}$ & - & - & - && - & - & - && - & - & - && - & - & - && - & - & - && - & - & - && - & - & -\\
& \\
$\beta_{8}$ & - & - & - && - & - & - && - & - & - && - & - & - && - & - & - && - & - & - && - & - & -\\
& \\
 \cline{2-28}
 $R^1$ & 0.05 & 0.02 & 0.10 && 0.05 & 0.12 & 0.28 && 0.15 & 0.08 & 0.23 && 0.07 & 0.12 & 0.31 && 0.12 & 0.09 & 0.23 && 0.17 & 0.14 & 0.35 && 0.09 & 0.10 & 0.13\\
 \cline{2-28}
 AIC & \textbf{1826.5} & 1604.2 & \textbf{1741.4} && \textbf{1548.0} & \textbf{1409.6} & \textbf{1558.5} && \textbf{1775.2} & 1557.6 & \textbf{1674.5} && \textbf{1527.3} & \textbf{1398.9} & \textbf{1683.3} && \textbf{1492.4} & {1366.2} & \textbf{1634.4} && \textbf{1392.2} & \textbf{1286.0} & \textbf{1516.6} && \textbf{1529.8} & \textbf{1335.4} & \textbf{1640.32}\\
 AIC-AR & 1849.3 & 1599.2 & 1791.1 && 1568.9 & 1432.1 & 1732.2 && 1788.9 & \textbf{1508.6} & 1695.5 && 1552.2 & 1417.5 & 1828.0 && 1535.5 & 1369.9 & 1674.5 && 1483.7 & 1297.2 & 1635.4 && 1559.8 & 1350.3 & 1705.3\\
% BIC & & & && & & && \textbf{1786.7} & 1524.8 & \textbf{1683.0} && \textbf{1543.4} & \textbf{1371.5} & \textbf{1693.0} && \textbf{1511.5} & \textbf{1354.4} & \textbf{1651.2} && & & && \textbf{1553.9} & \textbf{1340.8} & \textbf{1662.6}\\
% BIC-AR & & & && & & && 1797.0 & \textbf{1509.4} & 1704.5 && 1561.0 & 1416.9 & 1853.3 && 1544.9 & 1373.9 & 1683.6 && & & && 1569.8 & 1356.6 & 1715.5 \\
 \cline{2-28}
\end{tabular}
}
\caption{Estimated parameters of FA-QR model, for $\tau=0.05, 0.50$ and $0.95$, and for G7 economies. $p$-values in parenthesis. In bold, estimates significant at the $10\%$ level. The table also reports the $R^1$ coefficient and the values of the AIC together with the corresponding values of the AR-QR model (AIC-AR). In bold the smallest value when the difference is greater than 10.}
\label{tab:estimated_parameters_1} 
\end{table}
\end{landscape}

Consider now the results reported in Table \ref{tab:estimated_parameters_2} for the G5 economies. We can observe that the influence of the international factors on domestic inflation densities is, in general, weaker than that observed for the G7 economies. Other works also point out the heterogeneity of the response of average domestic inflation of advanced and emerging economies; see \cite{ha2019a}, \cite{ha2019b}, \cite{forbes2020},  \cite{jorda2020},  and \cite{banerjee2024}. We extend this conclusion to other quantiles of the conditional density of domestic inflation. Remarkably, at the $10\%$ significance level, there is not any significance influence of the international factors on any of the quantiles of the inflation density in Brazil. However, the inflation density in Mexico depends on the global and American factors for $\tau=0.05$ and on the American and MHI-EMD factors for $\tau=0.95$. Once more, we can observe that the maximum $R^1$ is achieved for $\tau=0.95$, but for China that, in any case, has $R^1$ close to zero. Similar conclusions are achieved when looking at the AIC criteria. The AIC of the FA-QR model is smaller than that of the AR-QR model in the $5\%$ and $95\%$ quantiles in Mexico. The AIC of the FA-QR model is also smaller in the $5\%$ quantile in South Africa and China, and in the $95\%$ quantile in India. In all other cases, the AIC of the AR-QR model is either smaller or approximately equal to that of the FA-QR model.

\begin{landscape}
\begin{table}[]
{\scriptsize	 %  
\centering
\begin{tabular}{@{}lrrrrrrrrrrrrrrrrrrrr@{}}
\\
 & \multicolumn{3}{c}{Brazil} && \multicolumn{3}{c}{Mexico} && \multicolumn{3}{c}{South Africa} && \multicolumn{3}{c}{China} && \multicolumn{3}{c}{India}\\
 & 0.05 & 0.50 & 0.95 && 0.05 & 0.50 & 0.95 && 0.05 & 0.50 & 0.95 && 0.05 & 0.50 & 0.95 && 0.05 & 0.50 & 0.95 \\
\cline{2-20}
\\
$\mu$ & \textbf{-1.667} & \textbf{2.191} & \textbf{6.980} && \textbf{-1.535} & \textbf{2.908} & \textbf{7.898} && \textbf{-5.136} & \textbf{3.672} & \textbf{10.001} && \textbf{-5.854} & \textbf{1.715} & \textbf{10.383} && \textbf{-3.511} & \textbf{4.193} & \textbf{12.692} \\
 & (0.00) & (0.00) & (0.00) && (0.00) & (0.00) & (0.00) && (0.00) & (0.00) & (0.00) && (0.00) & (0.00) & (0.00) && (0.00) & (0.00) & (0.00) \\
$\phi$ & \textbf{0.504} & \textbf{0.612} & \textbf{0.837} && \textbf{0.242} & \textbf{0.359} & \textbf{0.360} && \textbf{0.540} & \textbf{0.272} & \textbf{0.437} && 0.108 & 0.083 & 0.156 && 0.029 & \textbf{0.270} & \textbf{0.396} \\
 & (0.00)  & (0.00) & (0.00) && (0.03) & (0.00) & (0.00) && (0.00) & (0.00) & (0.00) && (0.38) & (0.50) & (0.38) && (0.77) & (0.00) & (0.00)\\
$\beta_{1}$ & 0.382 & 0.038 & 0.102 && \textbf{0.859} & 0.196 & -0.144 && \textbf{1.238} & \textbf{0.939} & \textbf{0.873} && 0.597 & \textbf{0.936} & 0.366 && 0.825 & 0.014 & 0.510 \\
 & (0.11) & (0.90) & (0.82) && (0.00) & (0.43) & (0.61) && (0.02) & (0.02) & (0.02) && (0.20) & (0.03) & (0.63) && (0.11) & (0.98) & (0.32)\\
 $\beta_{2}$ & & & && & & && -0.670 & -0.409 & \textbf{-1.303} &&  &  &  &&  &  &  \\
 & & & && & & && (0.17) & (0.33) & (0.00) && & & && & &\\
 $\beta_{3}$ & -0.057 & 0.000 & -0.198 && \textbf{1.330} & 0.445 & \textbf{0.185} &&  &  &  &&  &  &  &&  &  &  \\
 & (0.84) & (1.00) & (0.54) && (0.00) & (0.13) & (0.00) && & & && & & && & &\\
 $\beta_{4}$ & & & && & & && & & && 0.688 & 0.105 & 0.370 && 0.652 & \textbf{1.478} & \textbf{1.879}  \\
 & & & && & & && & & && (0.21) & (0.86) & (0.71) && (0.28) & (0.01) & (0.00)\\
 $\beta_{5}$ & & & && & & && & & && & & && & & \\
 & & & && & &  &&  & &  &&  & & && & & \\
 $\beta_{6}$ & &  &  &&  &  & &&  &  &  &&  &  &  &&  &  &  \\
 & & & && & & && & & && &  & && &  &\\
 $\beta_{7}$ & 0.312 & 0.245 & 0.573 && 0.366 & 0.012 & \textbf{-0.592} && 0.678 & -0.127 & -0.191 && \textbf{0.687} & 0.304 & -0.795 &&  & & \\
 & (0.14) & (0.42) & (0.17) && (0.13) & (0.97) & (0.08) && (0.15) & (0.76) & (0.62) && (0.06) & (0.56) & (0.29) &&  &  &\\
 $\beta_{8}$ & &  &  &&  &  & &&  &  &  &&  &  &  && -0.626 & -0.576 & 0.273  \\
 & & & && & &  &&  &  & && & &  && (0.20) & (0.25) & (0.64) \\
 \cline{2-20}
 $R^1$ & 0.23 & 0.20 & 0.36 && 0.13 & 0.13 & 0.21 && 0.14 & 0.09 & 0.19 && 0.06 & 0.02 & 0.01 && 0.04 & 0.08 & 0.14\\
 \cline{2-20}
AIC & 1662.0 & 1579.4 & 1787.0 && \textbf{1656.7} & 1489.6 & \textbf{1721.5} && \textbf{1949.4} & 1689.7 & 1853.0 && \textbf{1895.7} & 1745.3 & 2028.8 && 2029.1 & 1871.8 & \textbf{2135.9}\\
AIC-AR & 1660.6 & \textbf{1532.6} & \textbf{1584.8} && 1705.0 & \textbf{1461.1} & 1784.3 && 1964.4 & \textbf{1666.7} & 1861.3 && 1915.0 & \textbf{1731.7} & 2026.7 && 2038.9 & \textbf{1849.3} & 2155.9 \\
% BIC & \textbf{1661.7} & \textbf{1509.7} & \textbf{1780.4} && \textbf{1674.8} & 1470.9 & \textbf{1736.1} && \textbf{1964.8} & 1668.9 & \textbf{1862.5} && \textbf{1917.1} & 1749.0 & 2048.7 && 2052.5 & \textbf{1850.3} & \textbf{2147.5}\\
% BIC-AE & 1669.0 & 1517.0 & 1787.7 && 1714.0 & \textbf{1461.8} & 1793.0 && 1972.7 & \textbf{1668.1} & 1869.9 && 1925.3 & \textbf{1738.8} & \textbf{2036.9} && \textbf{2049.6} & 1853.7 & 2165.3 \\
 \cline{2-20} 
\end{tabular}
\caption{Estimated parameters of FA-QR model, for $\tau=0.05, 0.50$ and $0.95$, and for G5 economies. $p$-values in parenthesis. In bold significant estimates at the $10\%$ level. The table also reports the $R^1$ coefficient and the values of the AIC together with the corresponding values of the AR-QR model (AIC-AR). In bold the smallest value when the difference is greater than 10.}
\label{tab:estimated_parameters_2}
}%  
\end{table}
\end{landscape}

\begin{figure}[]
\begin{center}
\includegraphics[trim=0mm 10mm 0mm 0mm, clip, width=1\textwidth]{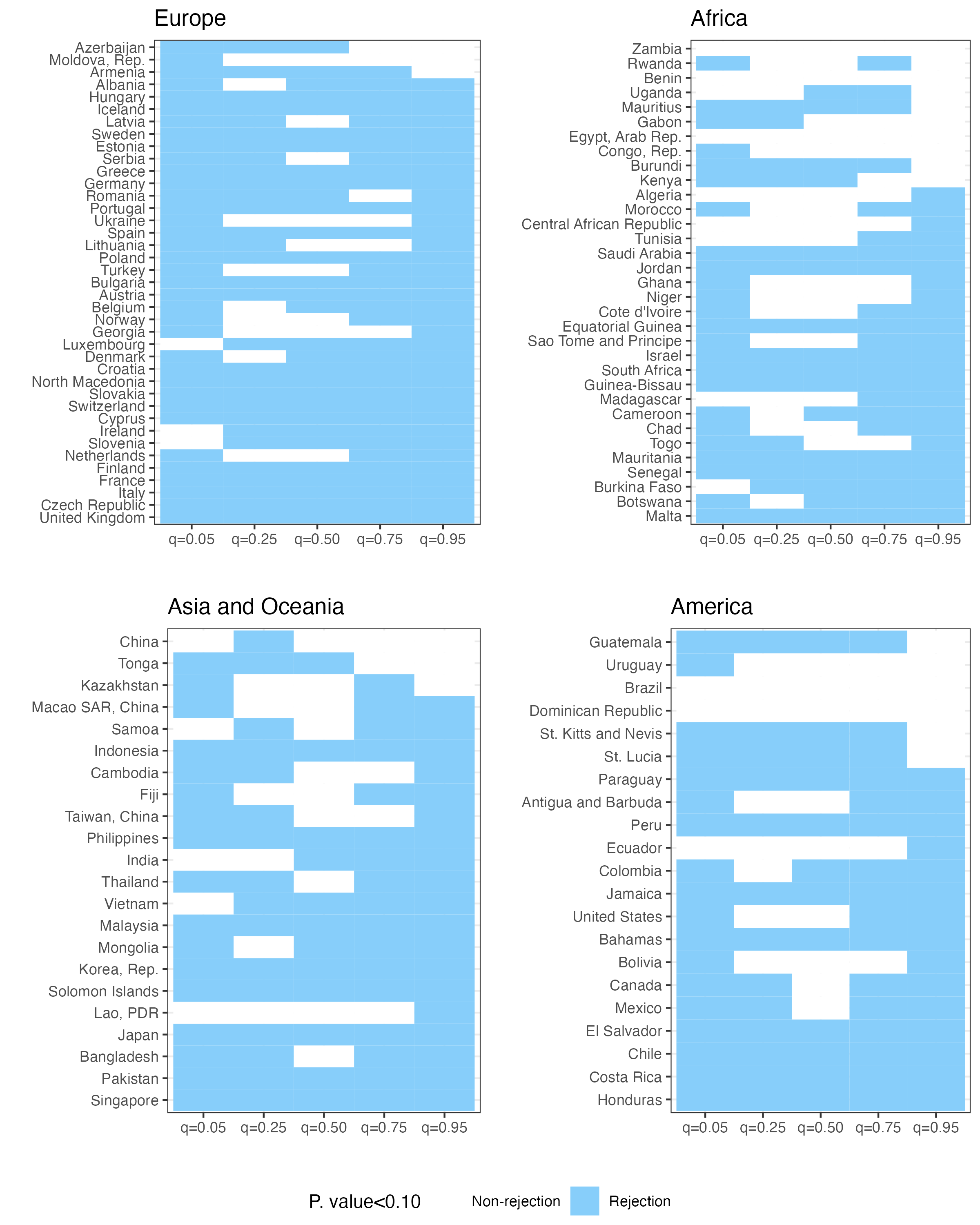}
\caption{Tests of joint significance of the parameters of the FA-QR models associated to international factors. In blue rejection of the null when $p$-value is smaller than 0.10.}
\label{fig:R1}
\end{center}
\end{figure}

For a big picture overview of the results for all $N=115$ inflations and $\tau=0.05, 0.25, 0.50, 0.75$ and 0.95, Figure \ref{fig:R1} represents the combinations of countries and quantiles for which the $F$-test for the joint significance of the factors is rejected. We can observe that the international factors are jointly significant for at least one quantile in all economies in Europe and Asia. This is also the case in America but for Brazil and Dominican Republic, and in Africa but for Zambia, Benin and Egypt. When looking at the results for different quantiles, we can observe that the international factors are not significant in a larger proportion of economies when $\tau=0.5$. However, when $\tau$ is either $0.05$ or $0.95$, we can observe that the international factors are significant in a large number of economies. All in all, we find that international factors are important when modelling the distribution of inflation all around the world. 

\begin{figure}[]
\caption{Predicted domestic inflation densities in the US, Germany, Mexico, and China obtained using the FA-QR model. The bullets represent the observed inflation.}
\label{fig:densities_1}
\begin{center}
\includegraphics[width=0.45\textwidth]{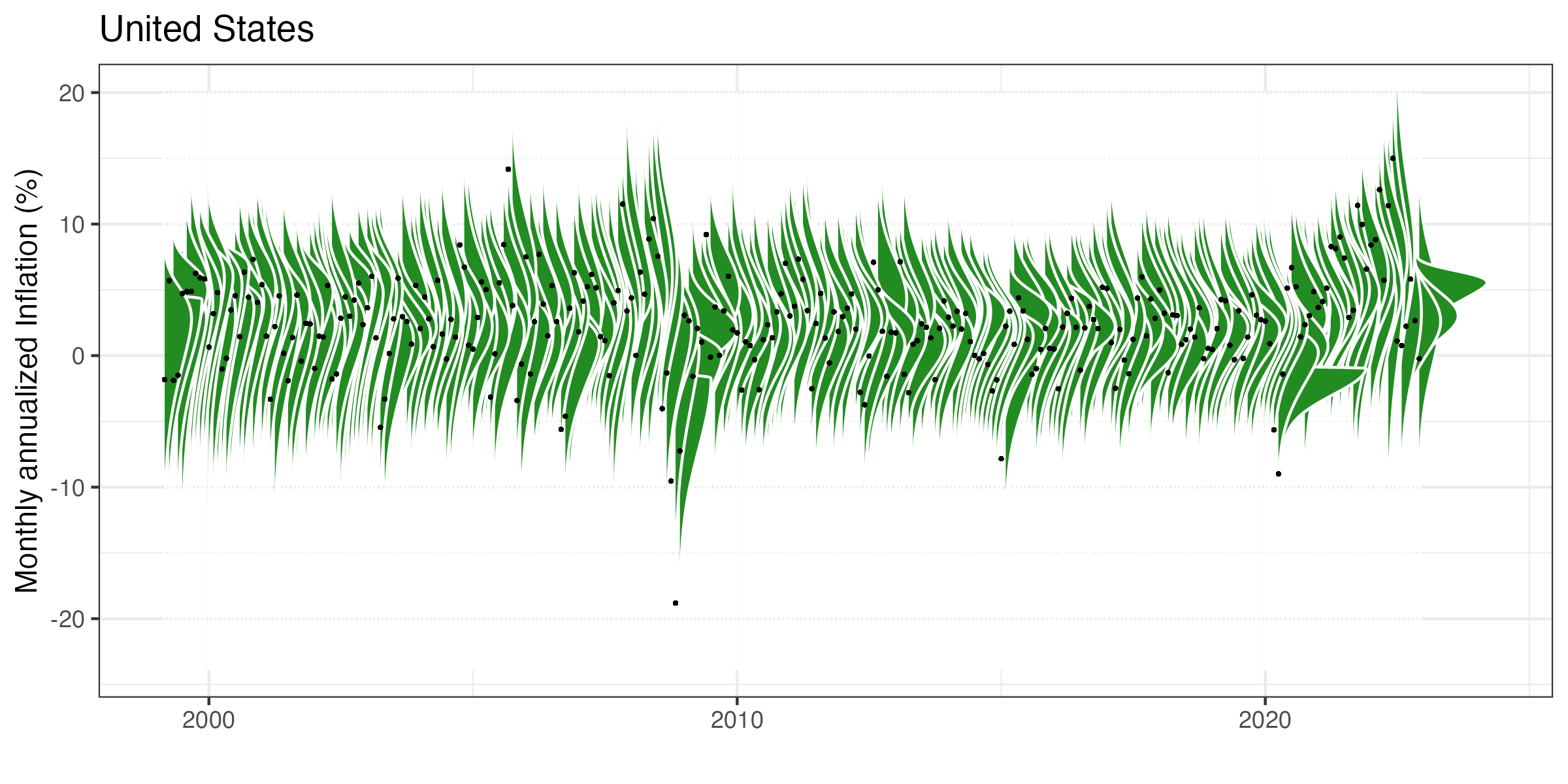}
\includegraphics[width=0.45\textwidth]{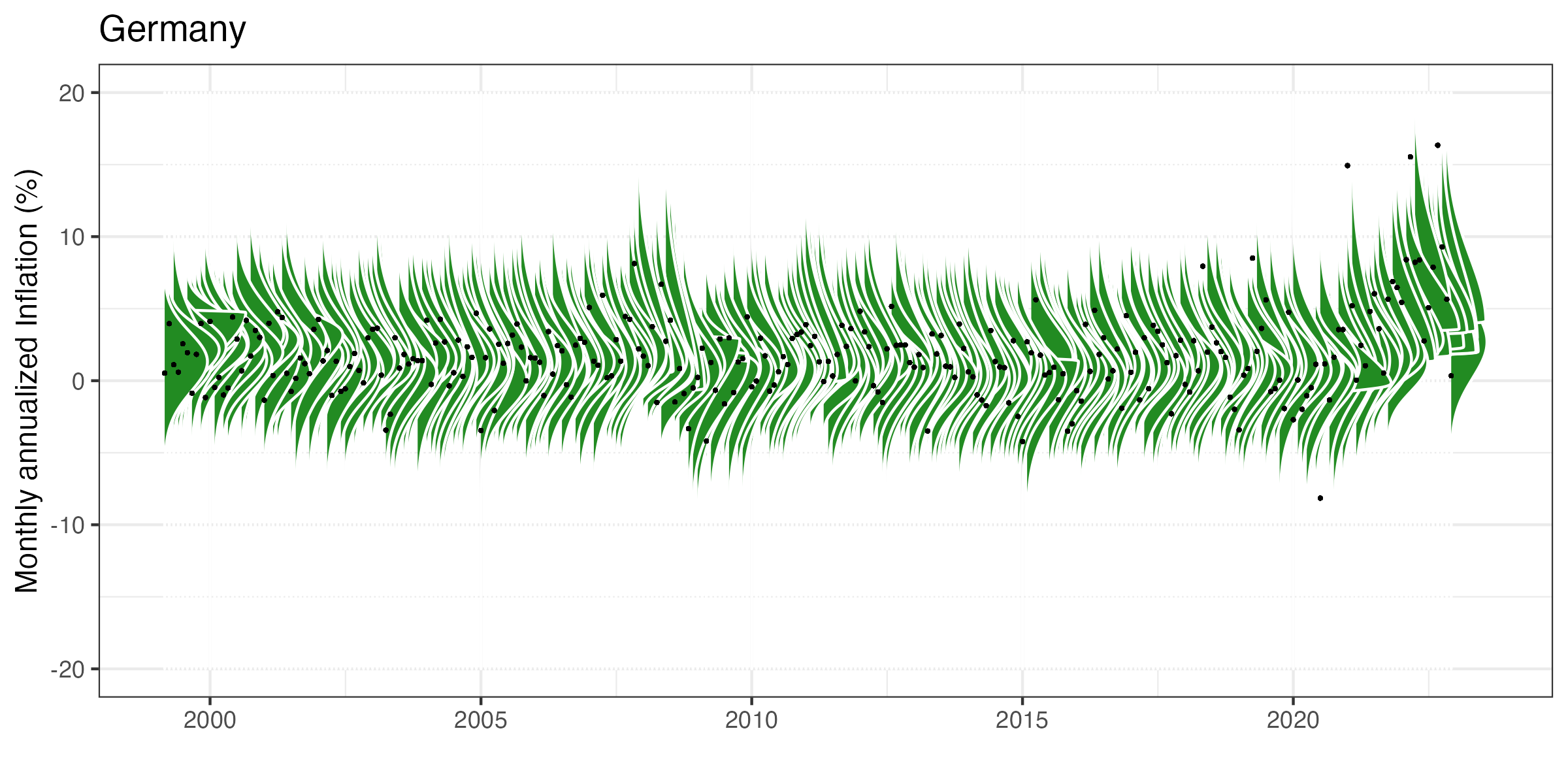}
\includegraphics[width=0.45\textwidth]{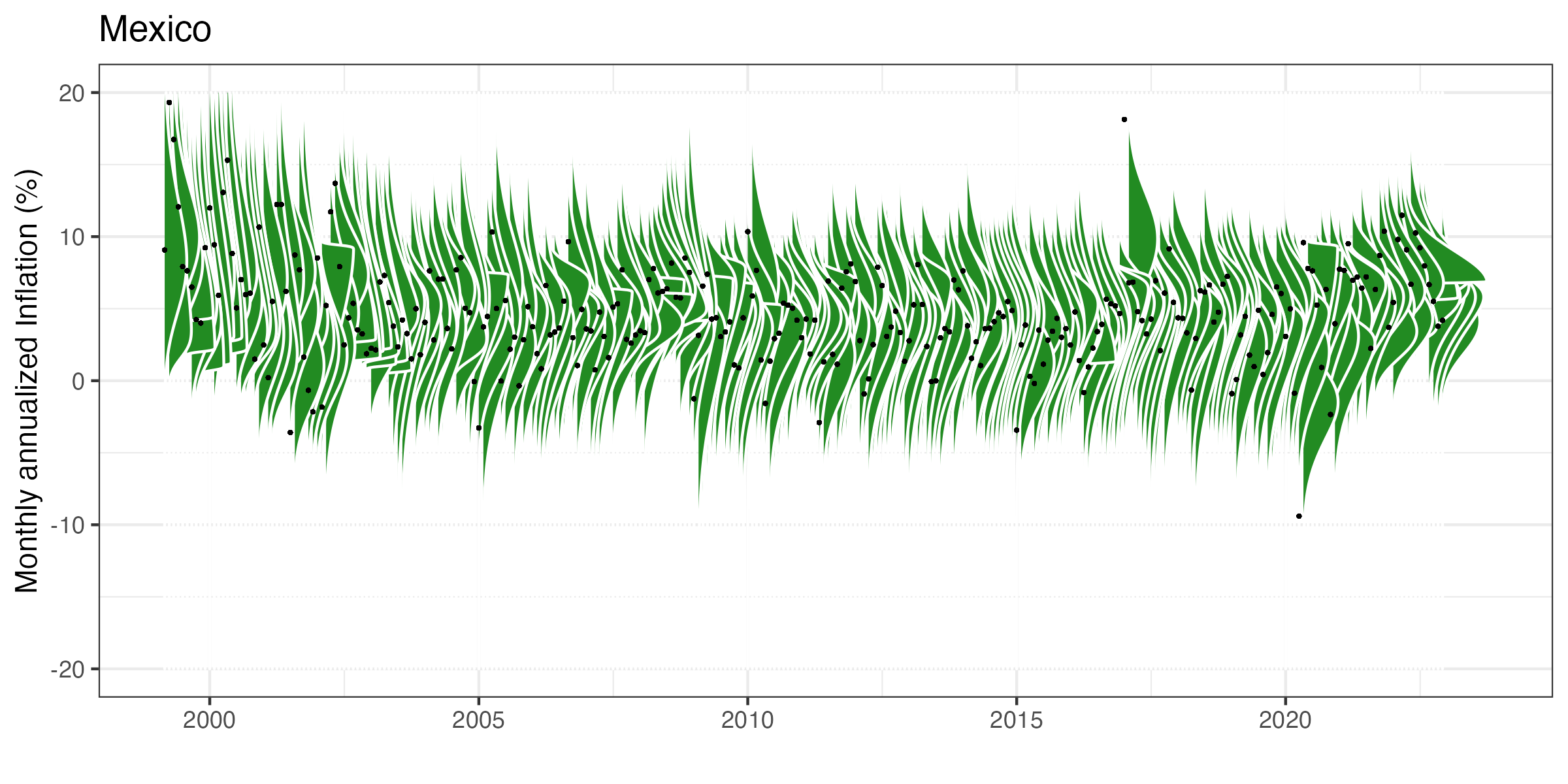}
\includegraphics[width=0.45\textwidth]{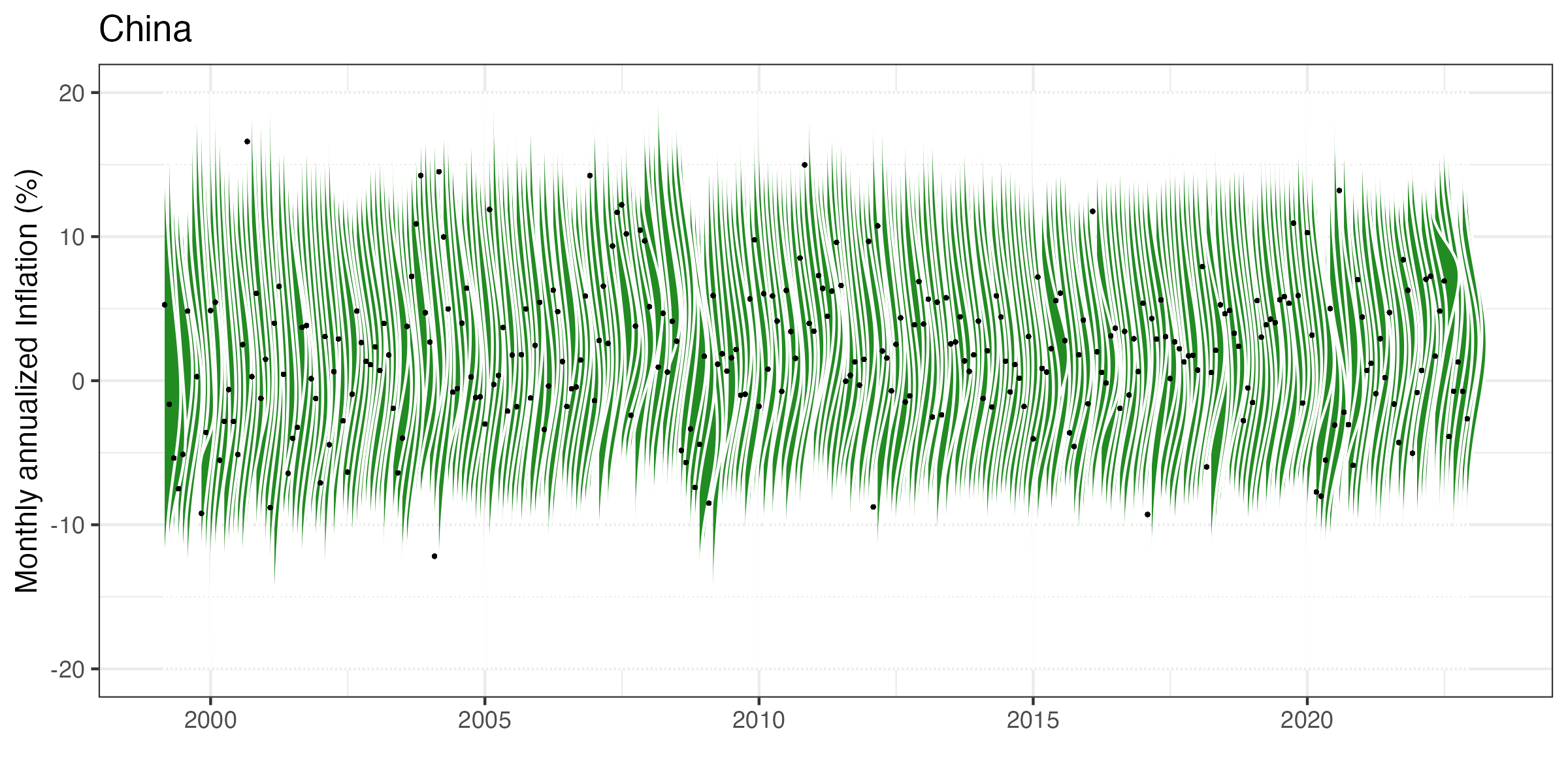}
\end{center}
\end{figure}

\begin{figure}[]
\caption{Estimates of $IaR_{it+1}(3)$ (left) and $DaR_{it+1}(0)$ (right), for all economies in Africa (first row), America (second row), Asia (third row), and Europe (fourth row).}
\label{fig:Iar}
\begin{center}
\includegraphics[width=0.75\textwidth]{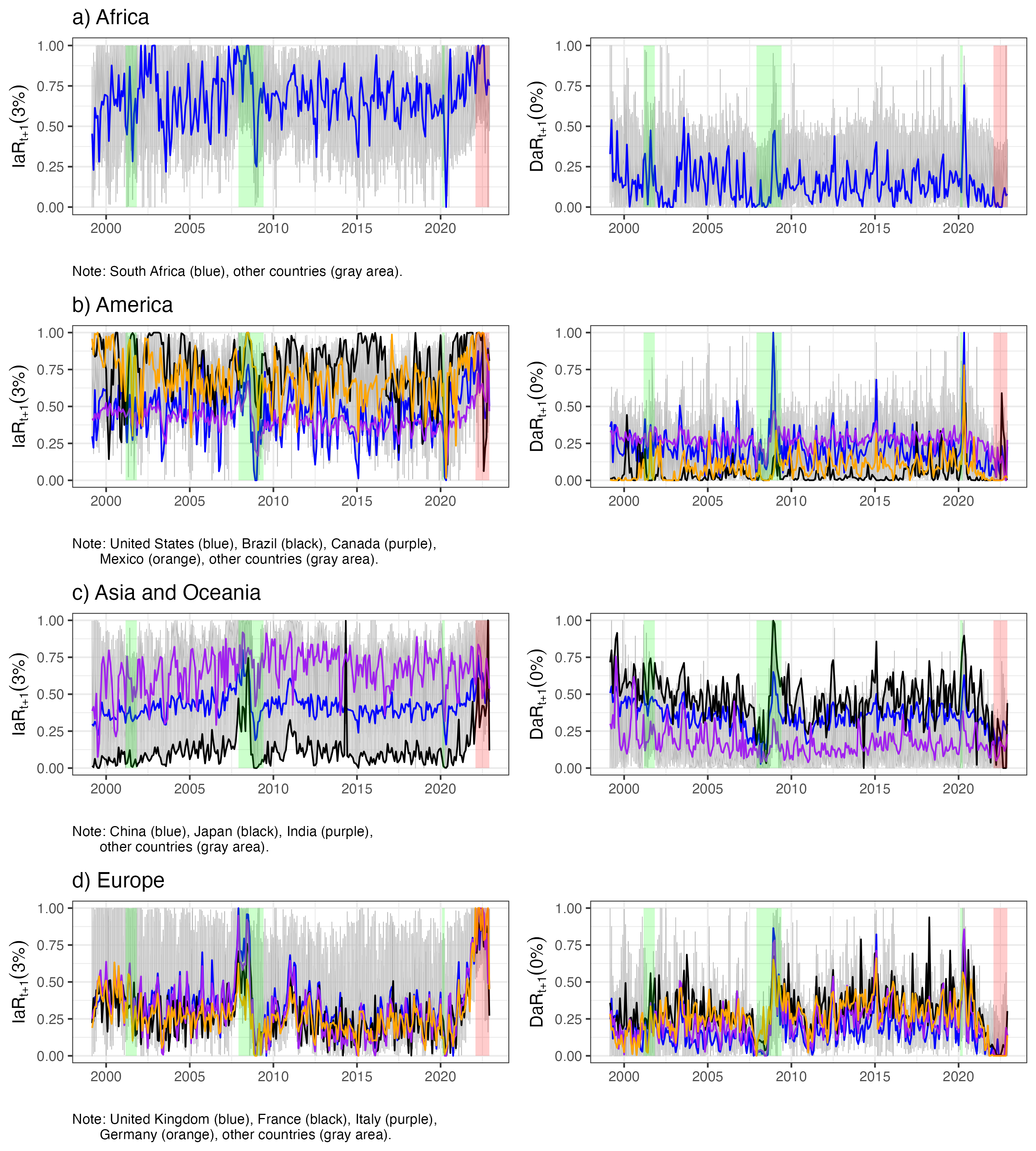}
\end{center}
\end{figure}

Finally, using the estimated quantiles reported in Tables \ref{tab:estimated_parameters_1} and \ref{tab:estimated_parameters_2}, for the G7 and G5 economies, we estimate the conditional distribution of domestic inflation in each country, $\tilde{k}_{it+1}$. Figure \ref{fig:densities_1}, which plots them for China, Germany, Mexico and the US, shows that the Russian invasion has a very strong effect on Germany while inflation densities are very stable in China. Based on $\tilde{k}_{it+1}$, we also estimate $IaR_{it+1}(3)$ (probability of inflation being larger than $3\%$) and $DaR_{it+1}(0)$ (probability of inflation being lower than $0\%$), in each country. Figure \ref{fig:Iar} plots these estimates for all $N=115$ economies, highlighting those corresponding to the G7 and G5 economies. It is remarkable that, while the Russian invasion increases the probability of high inflation all over the world, the effect in Europe is strongest, moving from probabilities between $1\%$ and $40\%$ to probabilities over $90\%$ that have been only observed after the global financial crisis. On the other hand, in some countries as India, there is not any impact of the Russian invasion. Figure \ref{fig:Iar} also shows that the probability of deflation is usually low with a pick after the COVID19 crisis.

\section{Out-of-sample forecasts}

In this section, we analyse the out-of-sample forecast accuracy of the estimated FA-QR models. %; see \cite{inoue2005} about the robustness of measuring the predictive accuracy in-sample instead of out-of-sample. 
To do so, the sample size is divided into an estimation period from February 1999 to December 2011 ($R=155$ observations) and an out-of-sample period from January 2012 to December 2022 ($P=132$ observations); see \cite{hansen2012} for rules to chose $R$ and $P$. Using a rolling-window scheme, we compute the scoring function QS($i,\tau$) for $i=1,...,115$ and $\tau=0.05, 0.25, 0.5, 0.75 $ and $0.95$, and the $CRPS(i)$ statistic, for each of the three weighting schemes: CRPS-L, CRPS-R, and CRPS-E.\footnote{The time series of estimated parameters in the out-of-sample period together with the factors extracted in each rolling window are reported in the On-line Appendix A.} We also compute the corresponding ratios $\frac{QS^{FA-QR}}{QS^{AR-QR}}$ and $\frac{CRPS^{FA-QR}}{CRPS^{AR-QR}}$ with respect to the benchmark AR-QR model. Next, we describe the main results.

\subsection{Out-of-sample forecast evaluation}

\begin{figure}[]
\begin{center}
\includegraphics[width=1\textwidth]{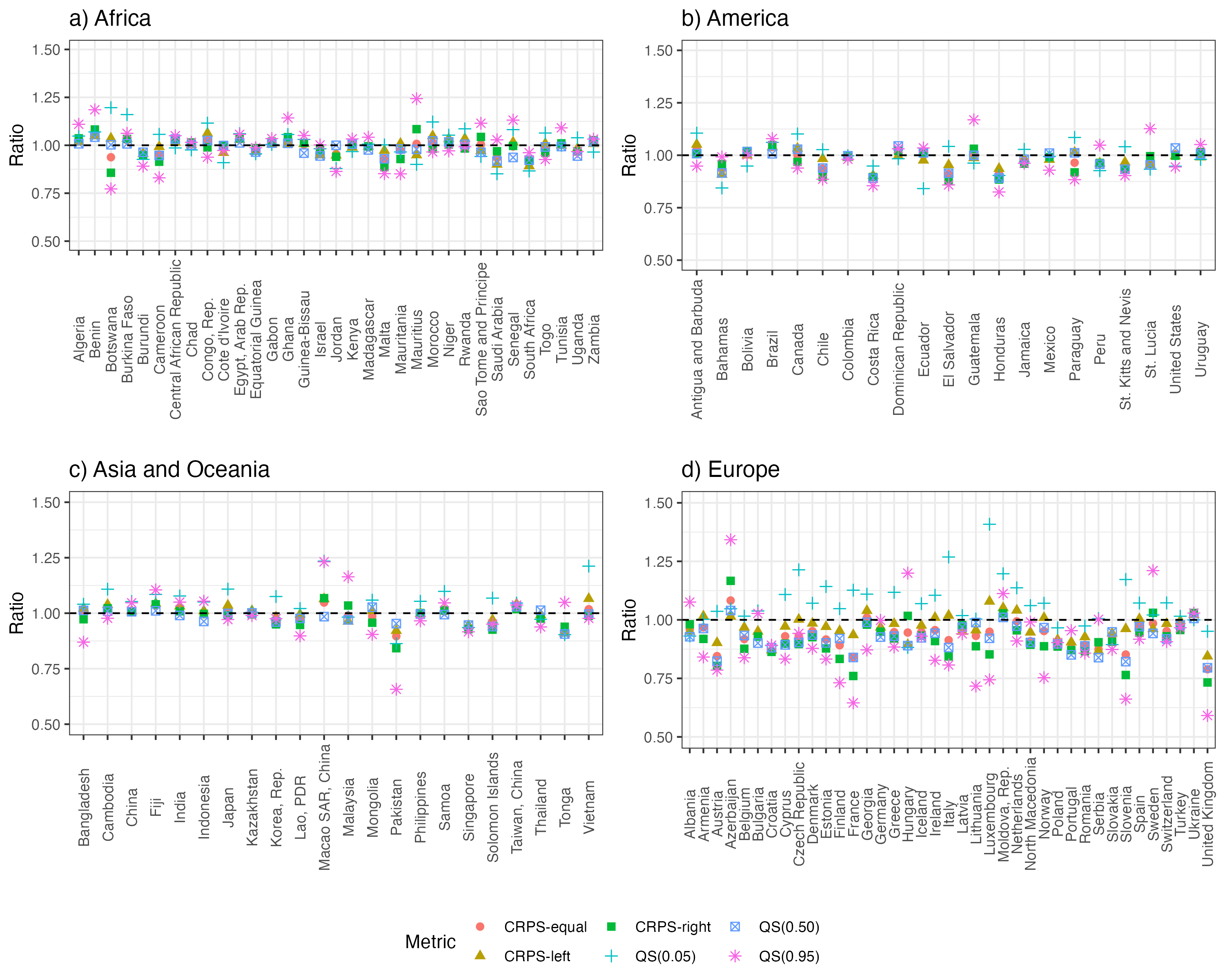}
\caption{QS ratios for $\tau=0.05, 0.50$ and 0.95, and CRPS-E, CRPS-R, and CRPS-L losses between the FA-QR and AR-QR models.}
\label{fig:Ratio}
\end{center}
\end{figure}

Figure \ref{fig:Ratio} plots the ratios  $\frac{QS^{FA-QR}}{QS^{AR-QR}}$ , for $\tau=0.05, 0.50$ and 0.95, and $\frac{CRPS^{FA-QR}}{CRPS^{AR-QR}}$, for CRPS-E, CRPS-R and CRPS-L losses.\footnote{The ratios corresponding to FA-QR models in which only one of the three factors is included are reported in the On-line Appendix B.} We can observe that, regardless of the loss used, all ratios (except QS(0.05)) are below one in most European countries. This implies a very strong power of international factors to forecast the density of inflation in European economies. The results are not so clear when looking at all other continents, in which the analysis has to be done separately for each country. 

To assess the significance of the difference between losses of the FA-QR and AR-QR models, we implement the one-sided DM and AM tests for equal predictive accuracy against the alternative of the forecasts obtained with the FA-QR model being better than those of the AR-FA model and, consequently, the international factors having forecast power. The results, reported in Table \ref{tab:DM} for the G7 and G5 economies, clearly point out towards the relevance of considering international movements in inflation when forecasting the quantiles of domestic inflation. Within the G7 countries, we can observe that, regardless of the loss function considered, the FA-QR model generates smaller losses when forecasting the quantiles of the density of domestic inflation in UK, France, and Italy. Within these countries, the only exception is the score associated with the left 0.05 quantile. Somehow different is the behaviour of domestic inflation in Germany as, in this case, the international factors seem relevant to forecast the mean of the distribution but not the tails. Within the G5 countries, we can observe that, in South Africa, international factors improve forecasts of all quantiles of domestic inflation but for the right quantile, QS(0.95). However, in Mexico, international factors are relevant to forecast only this quantile.

\begin{table}[h!]
\begin{center}
\begin{tabular}{|c|ccc|ccc|}
\hline
\multicolumn{1}{|c|}{} & \textbf{QS(0.05)} & \textbf{QS(0.50)} & \textbf{QS(0.95)} & \textbf{CRPS-E} & \textbf{CRPS-L} & \textbf{CRPS-R} \\
\hline
\textbf{US} & 0.95 & 1.03 & 0.94 & 1.00 & 1.00 & 1.00 \\
\textbf{Canada} & 1.10 & 1.03 & 0.94 & 1.00 & 1.03 & 0.97 \\
\textbf{Japan} & 1.11 & 1.00 & 0.97 & 1.01 & 1.04 & 0.99 \\
\textbf{UK} & 0.95 & $0.80^{***(**)}$ & $0.59^{**(**)}$ & $0.79^{***}$ & $0.85^{***}$ & $0.73^{***}$ \\
\textbf{France} & 1.13 & $0.84^{***}$ & $0.65^{**}$ & $0.84^{**}$ & 0.94 & $0.76^{***}$\\
\textbf{Germany} & 1.00 & $0.93^{**(*)}$ & 1.00 & $0.95^*$ & 0.97 & 0.95 \\
\textbf{Italy} & 1.27 & $0.88^{**(*)}$ & $0.81^{***}$ & $0.91^*$ & 1.02 & $0.84^{***}$ \\
\textbf{India} & 1.08 & 0.99 & 1.05 & 1.01 & 1.03 & 1.01 \\
\textbf{Brazil} & 1.06 & 1.01 & 1.08 & 1.04 & 1.04 & 1.05 \\
\textbf{China} & 1.05 & 1.01 & 1.05 & 1.01 & 1.01 & 1.01 \\
\textbf{South Africa} & $0.87^{*}$ & $0.92^{**(**)}$ & 0.96 & $0.91^{**}$ & $0.89^{**}$ & $0.92^{**}$ \\
\textbf{Mexico} & 0.99 & 1.01 & $0.93^*$ & 0.99 & 0.98 & 0.99 \\ 
\hline
$\%$ total & 12.20 & 40.90 & 31.30 & 34.80 & 19.10 & 41.70 \\
$\%$ total in Africa & 12.10 & 24.20 & 21.20 & 21.20 &12.10 & 24.20  \\
$\%$ total in America & 14.30 & 28.60 & 33.30 & 33.30 & 19.00 & 28.60 \\
$\%$ total in Asia  & 13.60 & 22.70 & 22.70 & 22.70 & 18.20 & 27.30\\
$\%$ total in Europe & 10.30 & 71.80 & 43.60 & 53.80 & 25.60 & 71.80\\
$\%$ total in ADV & 8.60 & 68.60 & 48.60 & 48.60 & 17.10 & 68.60\\
$\%$ total in HMI & 19.00 & 35.70 & 21.40 & 38.10 & 31.00 & 35.70\\
$\%$ total in LI & 7.90 & 21.10 & 26.30 & 18.40 & 7.90 & 23.70\\
\hline
\end{tabular}
\end{center}
\caption{Ratios of forecast losses obtained with the FA-QR and AR-QR models. The ratios market with stars are those for which the loss of forecasts obtained with the FA-QR model are significantly smaller than those of the AR-QR model according to the DM test of equal predictive accuracy. * $10\%$, ** $5\%$, and *** $1\%$ significance levels. The ratios significant according to the AM tests are marked in parenthesis. The lower part of the Table reports the percentage of countries with significant DM tests.}
\label{tab:DM}
\end{table}

To summarize the results for all $N=115$ countries, Table \ref{tab:DM} also reports the percentages of countries in which the null of the DM test is rejected for the different losses considered. Focusing on the results for the CRPS-R loss, we can observe that international factors are relevant for $41.70\%$ of the total number of countries. This percentage increases to $71.80\%$ among European countries and to $68.60\%$ among advanced economies. It is also remarkable that the CRPS-R obtained when using the international factors to forecast the quantiles of the inflation distribution is significantly smaller in $23.70\%$ of low-income emerging economies; see \cite{ha2019c} for the influence of international shocks in the inflation of low-income economies. 

\begin{figure}[]
\caption{Fluctuation test for equal predictive ability of the FA-QR model with respect to the benchmark AR-QR model: G7 countries (left column), and G5 countries (right column).}
\label{fig:GR}
\begin{center}
\includegraphics[width=0.4\textwidth]{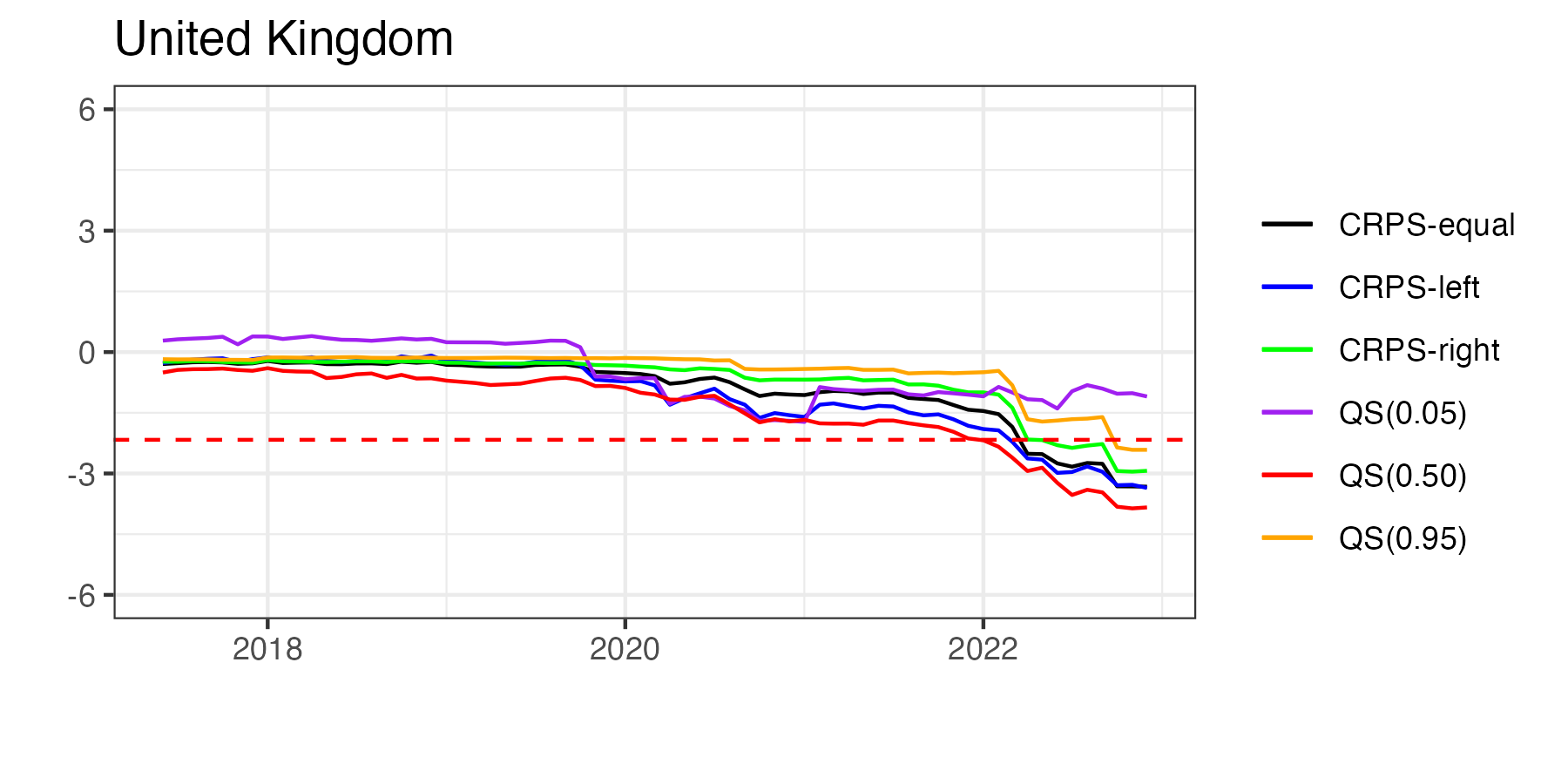}
\includegraphics[width=0.4\textwidth]{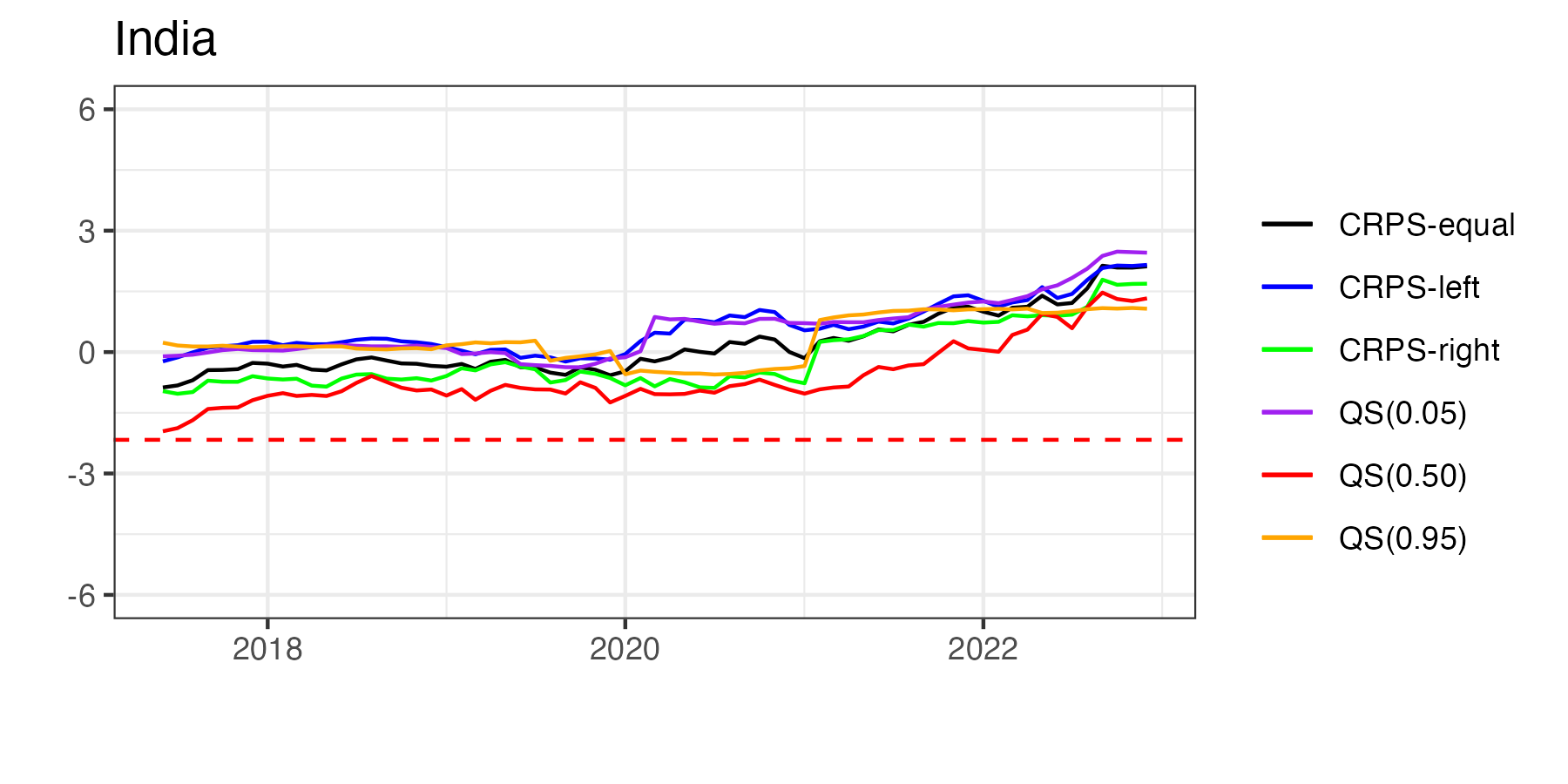}\\
\includegraphics[width=0.4\textwidth]{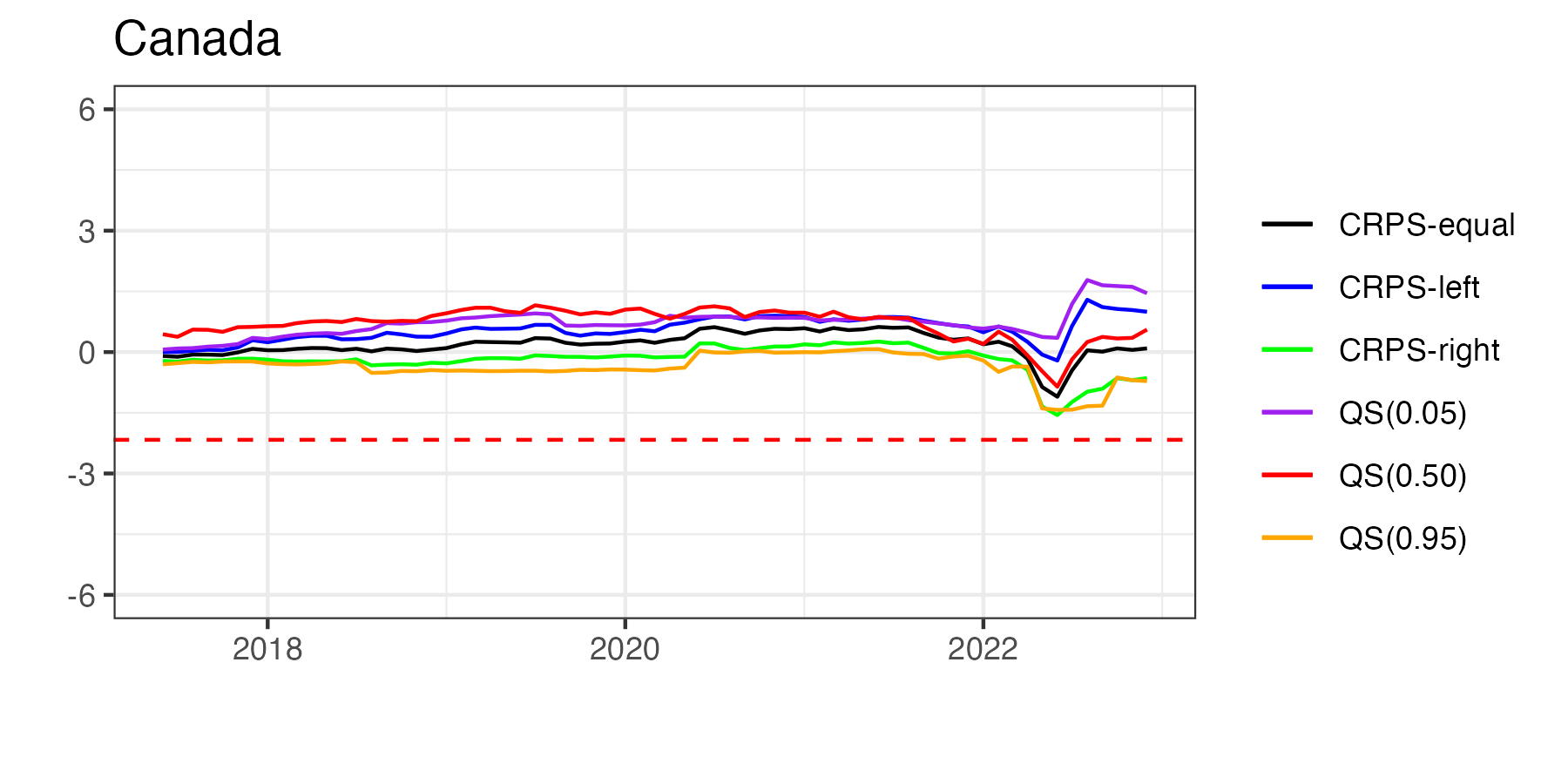}
\includegraphics[width=0.4\textwidth]{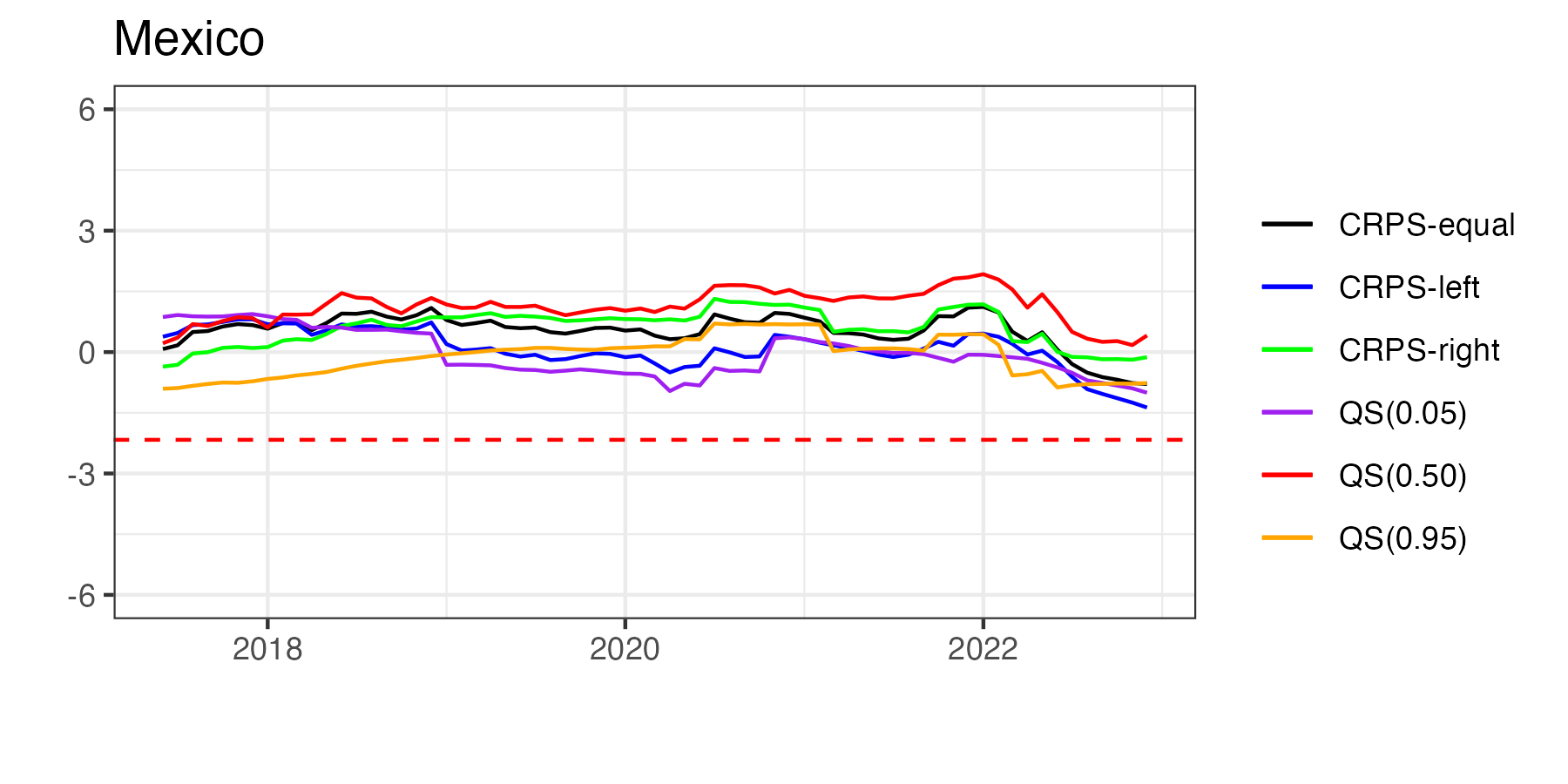}\\
\includegraphics[width=0.4\textwidth]{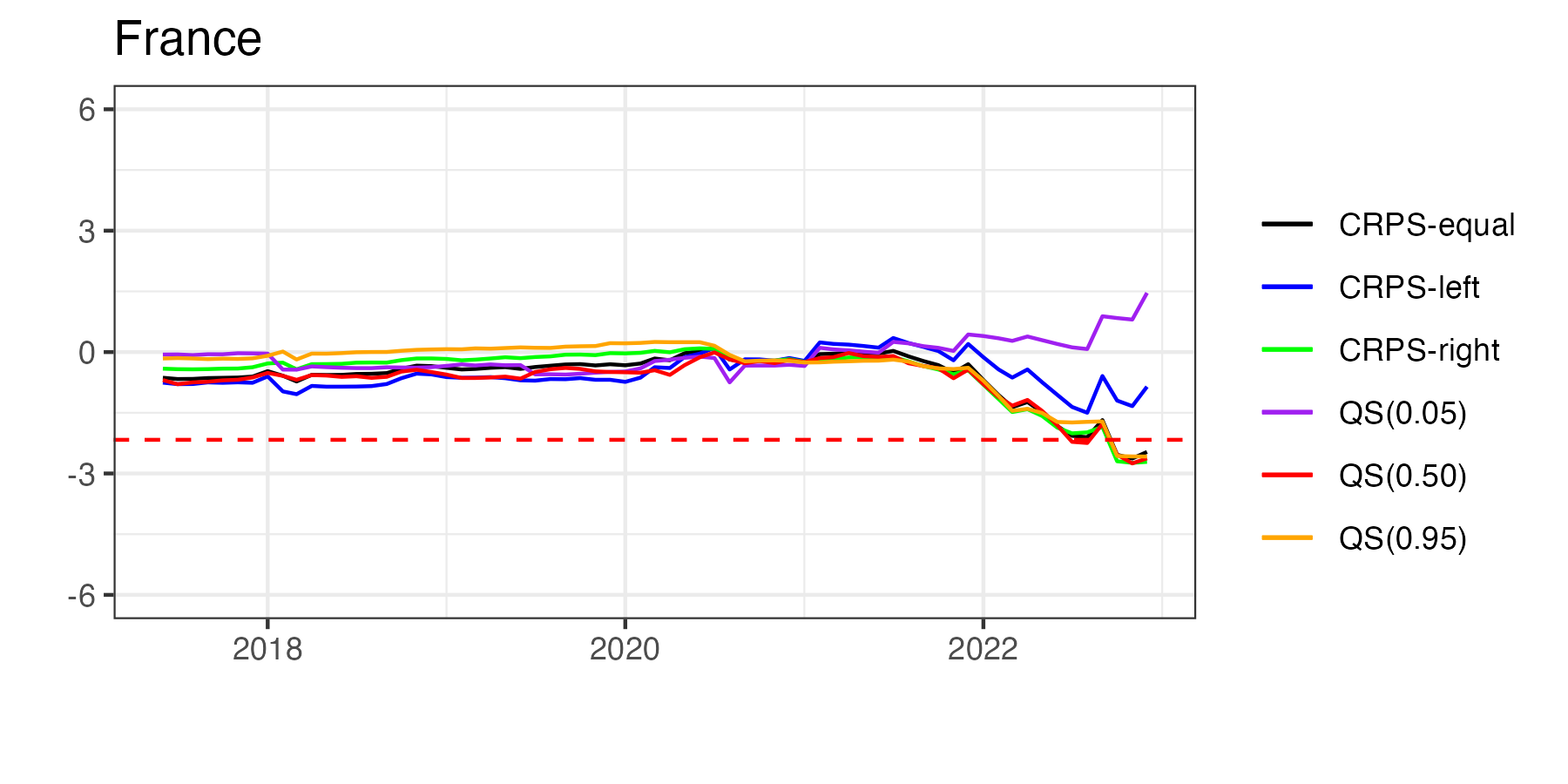}
\includegraphics[width=0.4\textwidth]{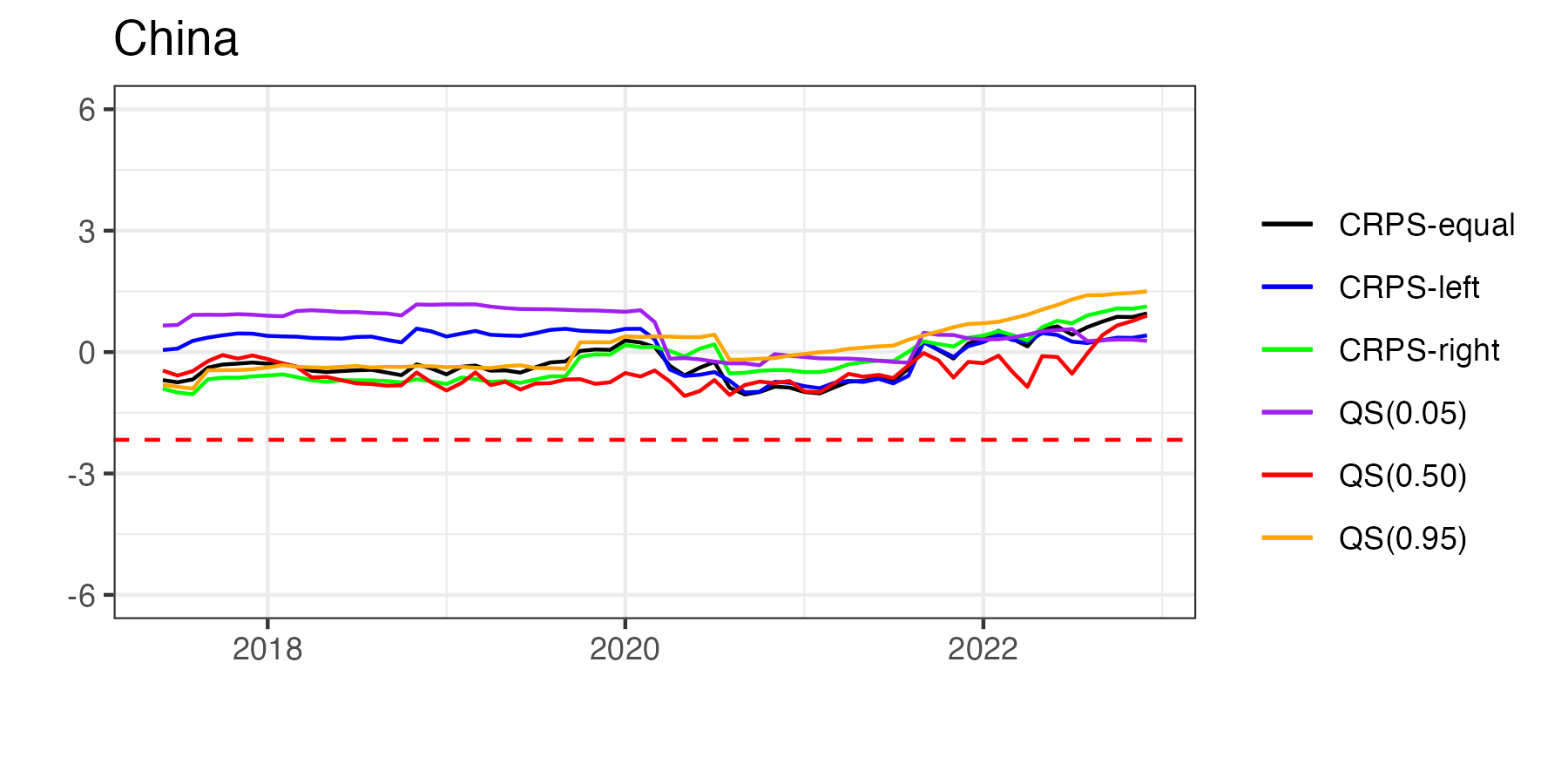}\\
\includegraphics[width=0.4\textwidth]{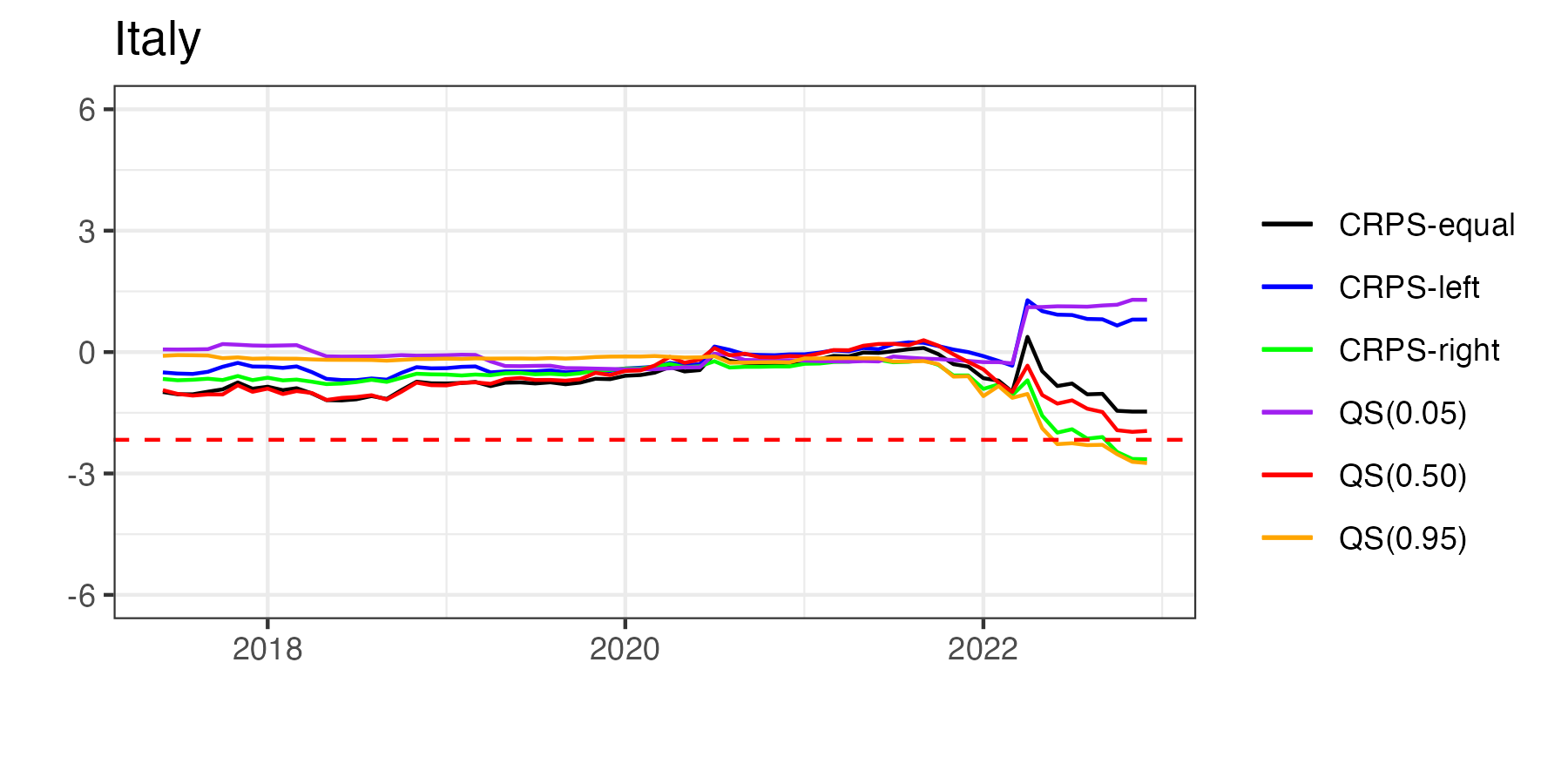}
\includegraphics[width=0.4\textwidth]{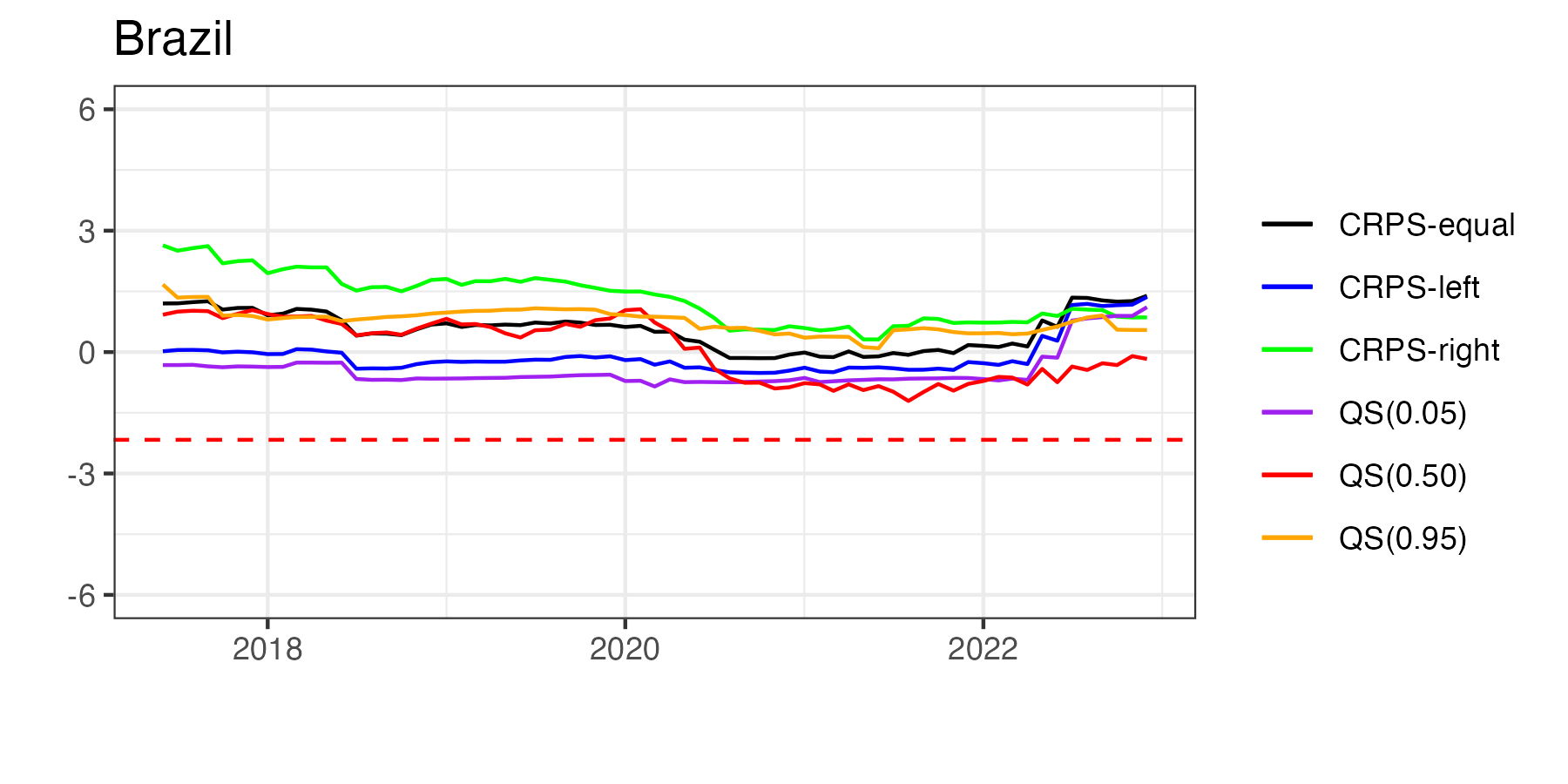}\\
\includegraphics[width=0.4\textwidth]{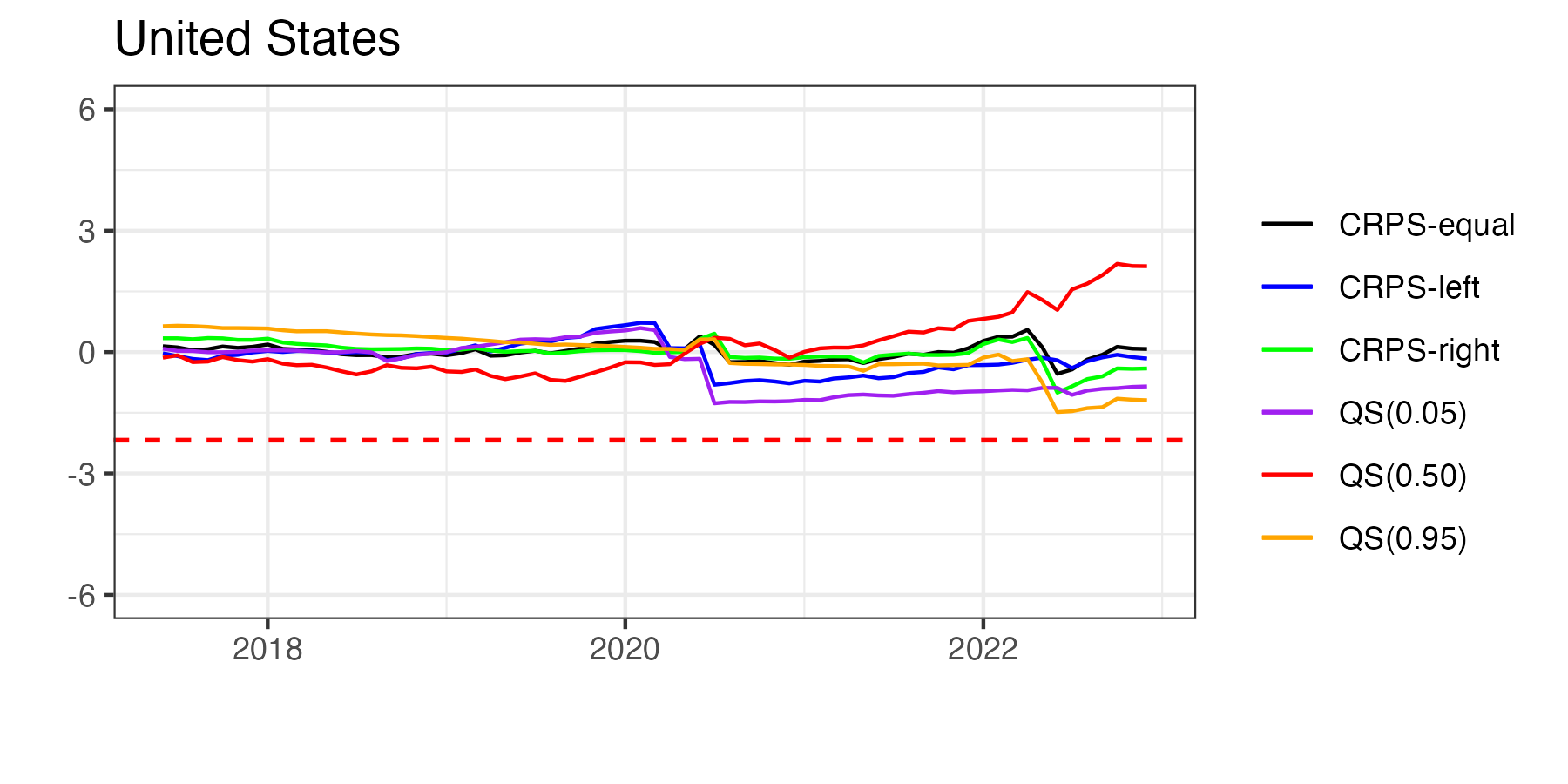}
\includegraphics[width=0.4\textwidth]{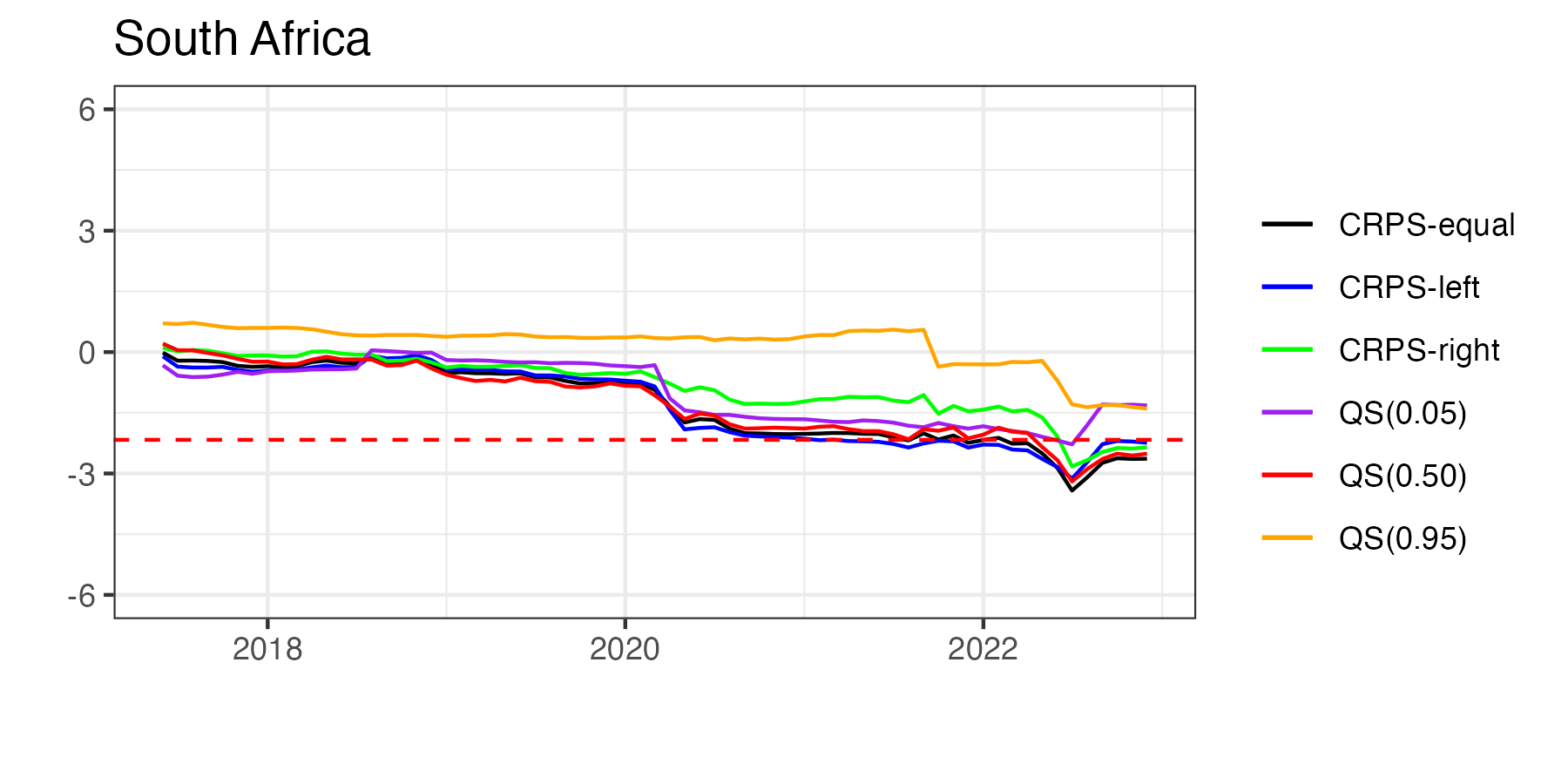}\\
\end{center}
\includegraphics[width=0.4\textwidth]{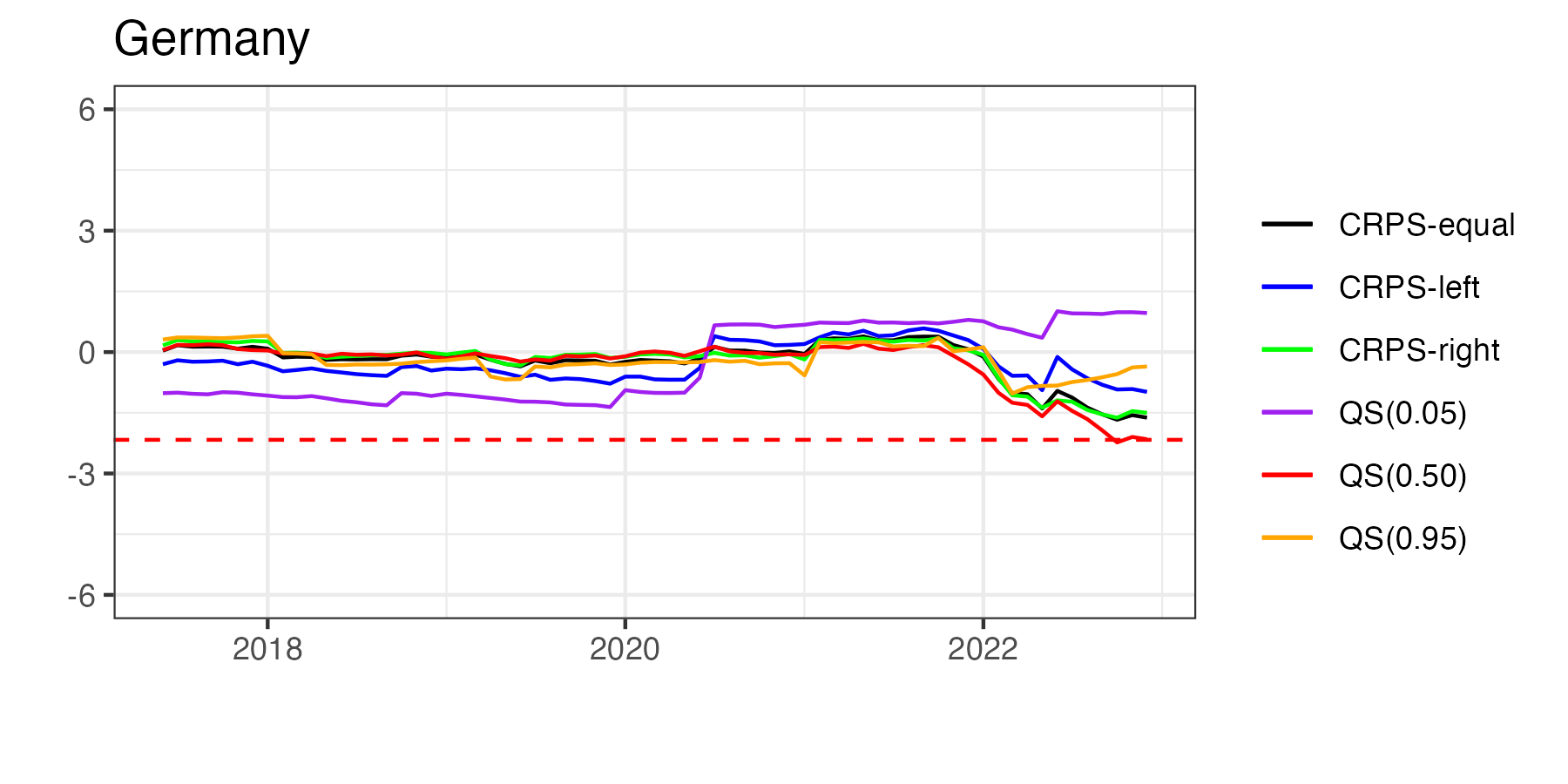}\\
\includegraphics[width=0.4\textwidth]{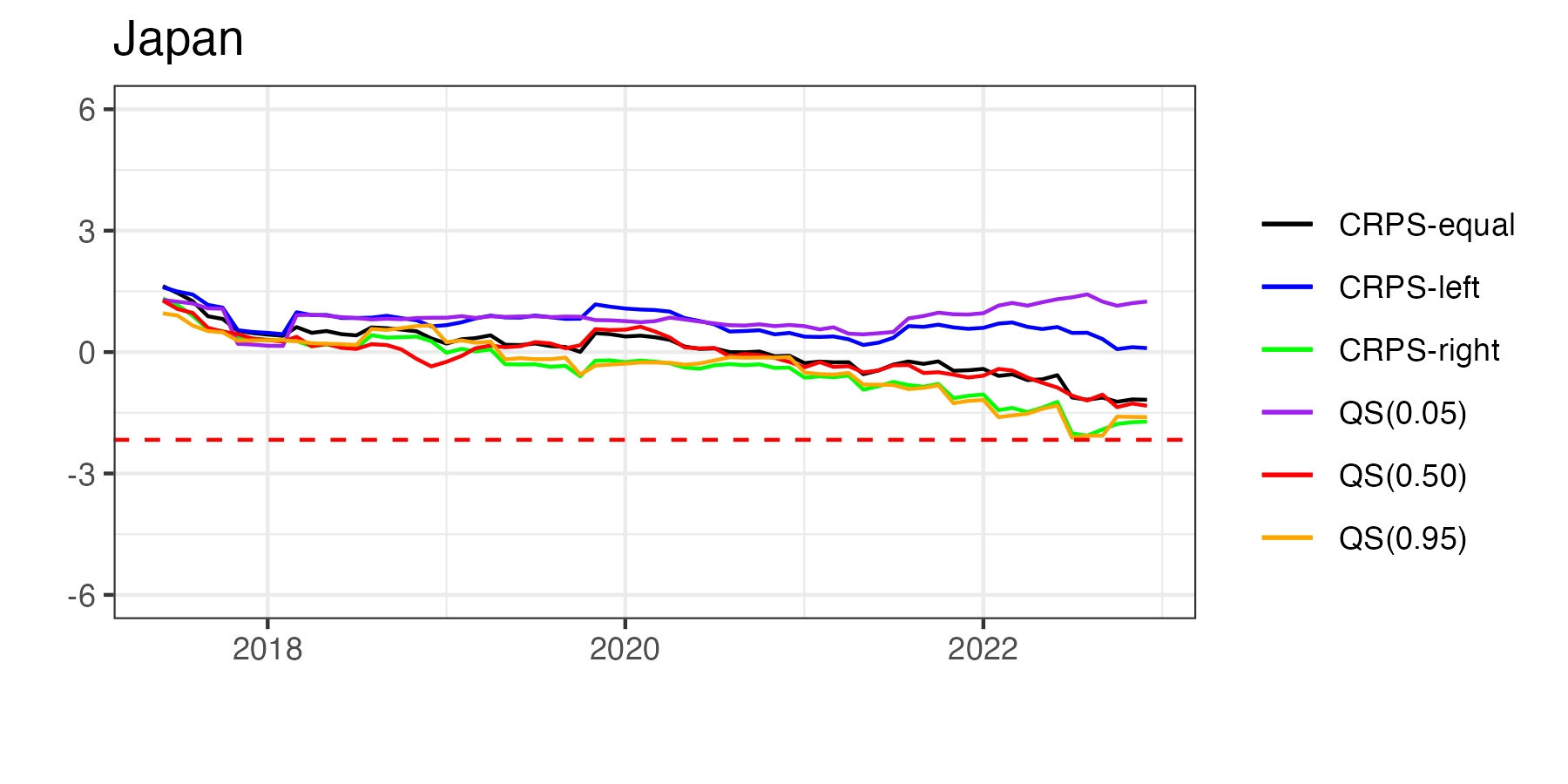}\\
\end{figure}

Finally, in order to detect "pockets of predictability", we run the fluctuation test of Giacomini and Rossi (2010) to compare through time equal predictive accuracy between the FA-QR model and the benchmark AR-QR model. The results are plotted in Figure \ref{fig:GR} for the economies within the G7 and G5. The null of equal predictive accuracy is rejected in the more recent years in the G7 economies but in the US and Canada. The forecasting power of the international factors is increasing through time being stronger after 2022 and it is mostly related with losses based on the right tail of the distribution of inflation. Consider, for example, the case of domestic inflation in the UK, with the forecasting accuracy of international factors increasing over the out-of-sample period and being significant after 2022 for all losses but QS(0.05). The picture is somehow different when looking at the G5 economies. We observe that, in China, Mexico and Brazil, the forecasting power of international factors has been approximately constant over the out-of-sample period, while, in India, they have been loosing power. However, in South Africa, the pattern is similar to that observed in the G7 economies but with the international factors being more importnat to explain the left tail than the right tail of the distribution of domestic inflation. 

\section{Final considerations}
\label{section:final}

% Podemos analizar si la inflación está relacionada con el precio de commodities como, por ejemplo, el oil price: \cite{cicarelli2010} check that global inflation is not simply a stand-in for typical common shocks such us commodity prices.

In this paper, we put together for the first time international factors and the vulnerability of domestic inflation. To this end, we use monthly data and employ FA-QRs and ML-DFMs to analyse inflation dynamics in a balanced panel of inflation observed in 115 countries over the period January 1999 to December 2022.

We show that the influence of international factors, measured as common components of a large set of inflations around the world, is relevant to understand the distribution of domestic inflation in a large proportion of countries. These factors can be global (with weights for most countries), regional, factors common to economies in a given region, or associated to the level of economic development. Looking at the in-sample fit, we show that, in all but 5 of the 115 economies considered, the international factors (either the global, the regional or the development level factors) are significant for at least one quantile (usually the upper quantiles in advanced economies and the lower quantiles for developing economies). We also show that the relevance of international factors is stronger in advanced than in low-income economies. We also analyse the out-of-sample power of the international factors with similar conclusions. The observed heterogeneity among countries can be explained by inflation also depending on domestic drivers within each country; see, for example, \cite{busetti2021OBES} and \cite{binici2024}, who analyse and disentangle the confluence of domestic and external factors in explaining the evolution of inflation dynamics in Europe, where post-pandemic inflation reached the highest level in four decades.

Our results can help to understand better some of the puzzles about inflation modelling and forecasting, which are often related to domestic models; see, for example, the very recent essay by \cite{mankiw2024}. We have seen that international factors should be considered for an appropriate understanding of domestic inflation in a large proportion of countries.  

The increasing influence of international factors on domestic inflation in advanced economies points towards the coordination of corrective actions by central banks. It seems important that central banks in developed countries coordinate their strategies to control the risk of high domestic inflation. Furthermore, if inflation is largely determined abroad, a central bank could be less concerned about inflation exceeding its target, priorizing job creation. Alternatively, it may also need to make larger adjustments to interest rates to stabilize inflation; see \cite{forbes2020}. Furthermore,  due to the increasingly influential international factors, \cite{ha2022} point out that central banks in EMDEs face more challenges than those in advanced eonomies. They recommend for EMDE policies a careful calibration, credible formulation and clear communication of policies to make these econmies more resilient to international factors of inflation.

\section*{Acknowledgements}

All authors acknowledge financial support from the Spanish National Research Agency (Ministry of Science and Technology) Project PID2022-139614NB-C22/AIE/10.13039/501100011033 (MINECO/FEDER). We are also very grateful to Aaron Amburgey and Mike McCracken for sharing codes for their test. Comments from participants at the $XIV_t$ Workshop in Time Series Econometrics (April 2024, Zaragoza, Spain), at the Women in Econometrics Conference (June 2024, Bologna, Italy), at the International Symposium of Forecasting (July 2024, Dijon, France), and at a seminar at the Economics department of the University of Salamanca (October 2024), are also acknowledged. This research was partly developed while the third author was visiting the department of Economics of the University of Bologna. Thanks are also due to its hospitality and to Salvador Madariaga grant PRX22/00195 that financially support her visit. Any remaining errors are obviously our responsibility.

\bibliographystyle{apalike}
\bibliography{References_GiS}

\newpage

\appendix

\onehalfspacing

\begin{center}
\section*{\large{Online Appendix\\International vulnerability of inflation}}
\end{center}

\addcontentsline{toc}{section}{Supplementary material}
\renewcommand{\thesubsection}{\Alph{subsection}}

\vspace{0.5cm}

\subsection{Results of rolling-window estimation.}
\label{appendix:rw}

\renewcommand{\theequation}{A.\arabic{equation}}
\renewcommand{\thetable}{A.\arabic{table}}
\renewcommand{\thefigure}{A.\arabic{figure}}

\setcounter{equation}{0}
\setcounter{table}{0}
\setcounter{figure}{0}

\begin{figure}[]
\begin{center}
\subfigure[]{\includegraphics[width=0.30\textwidth]{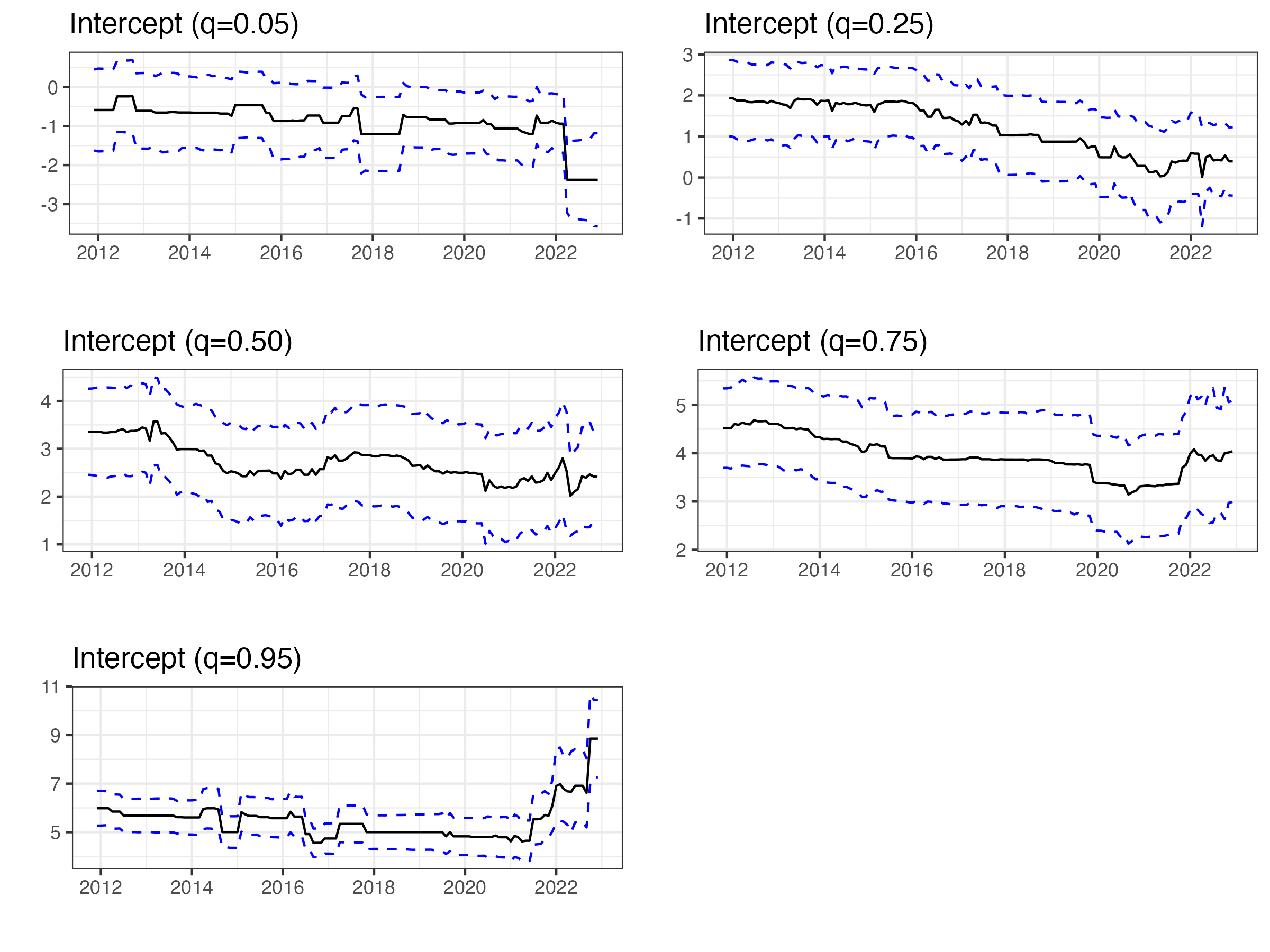}}
\subfigure[]{\includegraphics[width=0.30\textwidth]{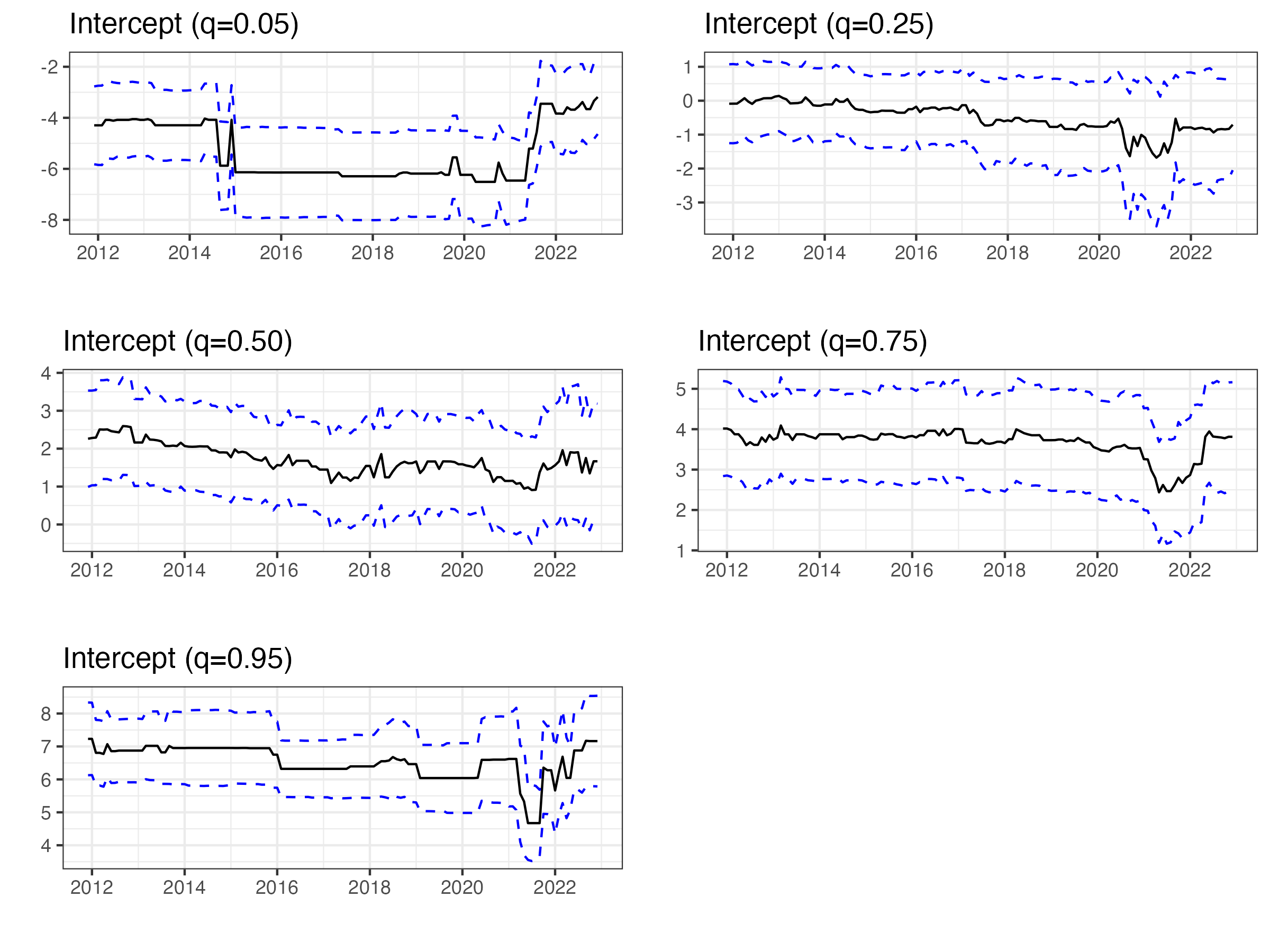}}
\subfigure[]{\includegraphics[width=0.30\textwidth]{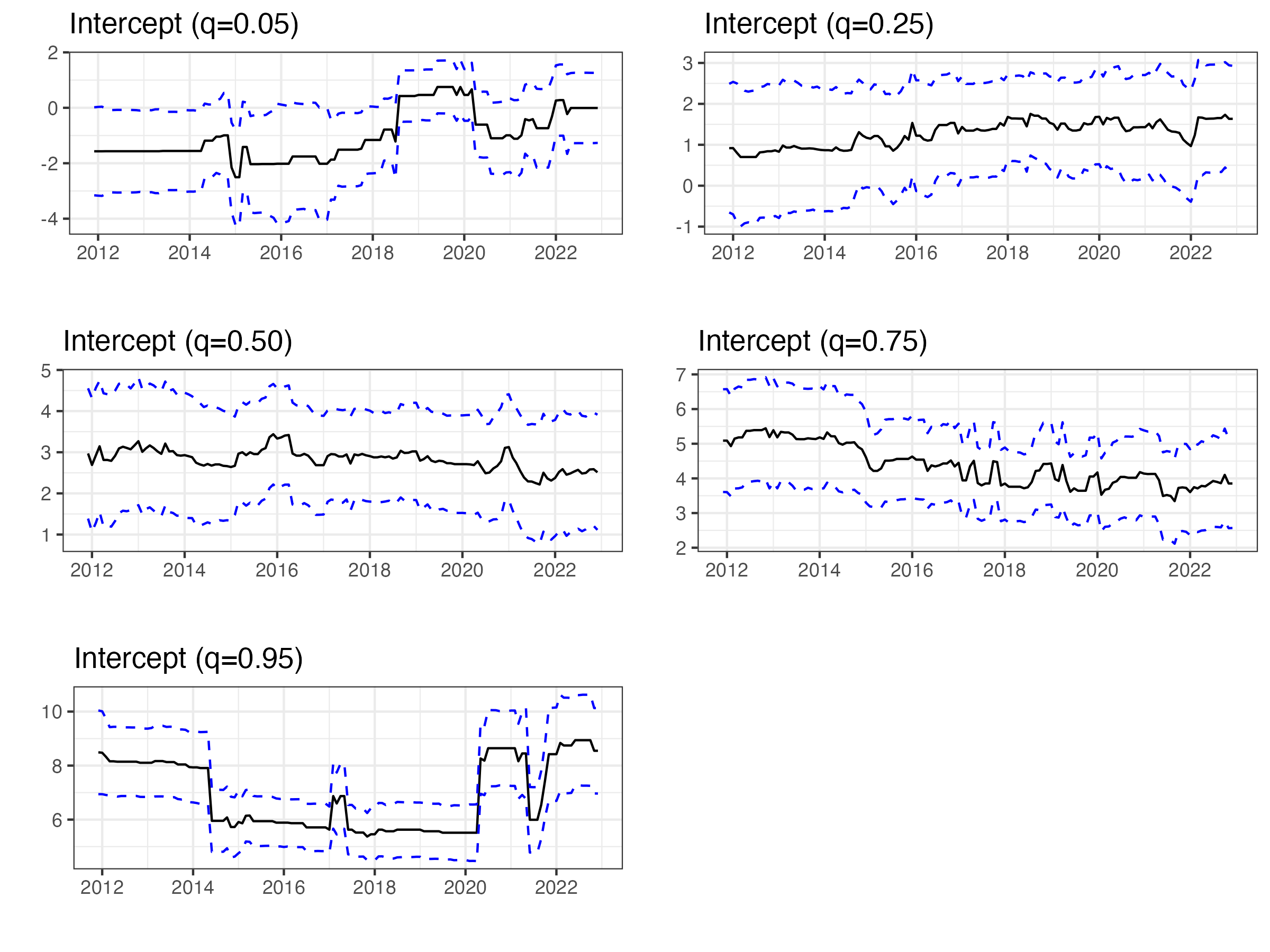}}\\
\subfigure[]{\includegraphics[width=0.30\textwidth]{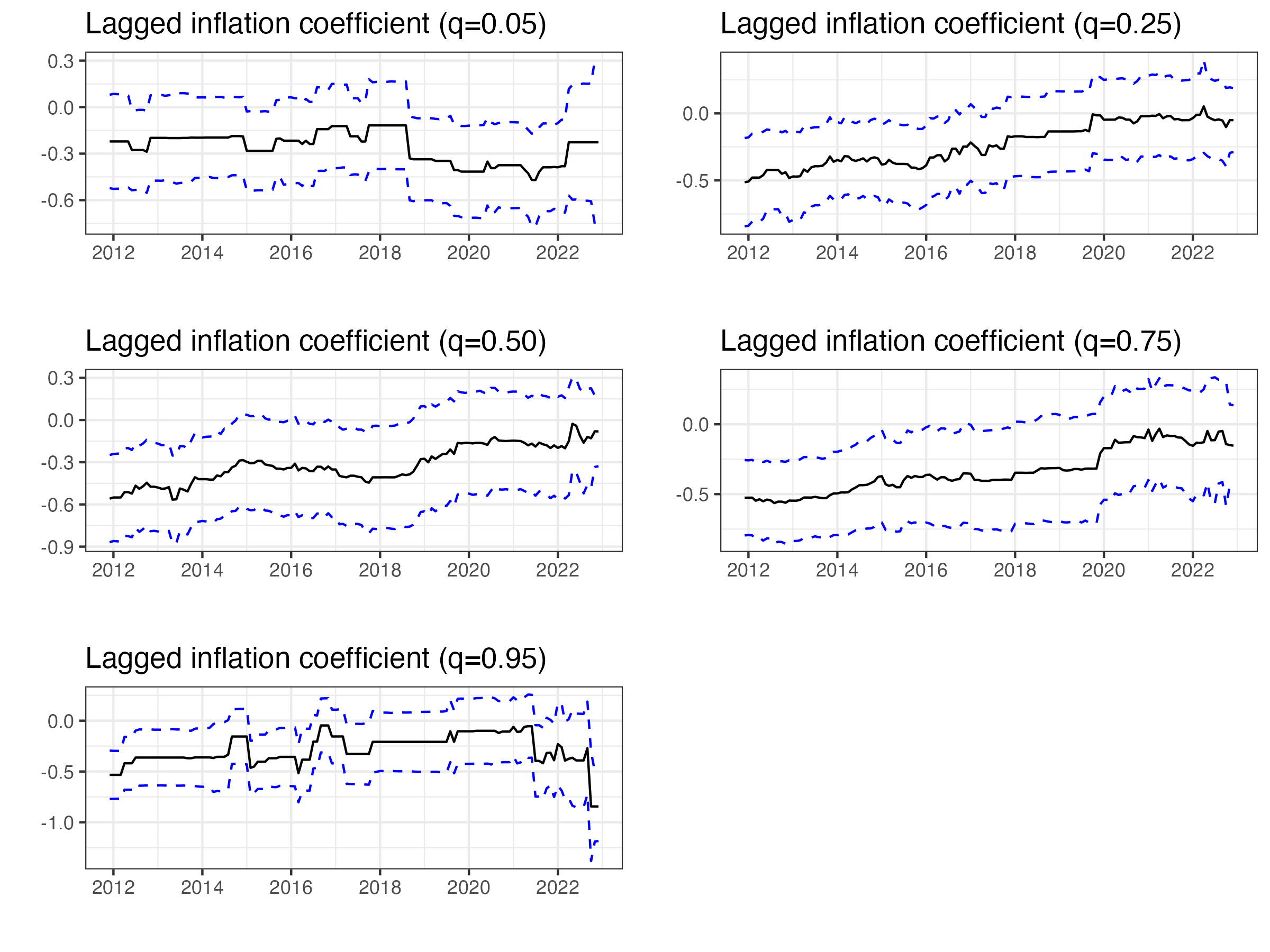}}
\subfigure[]{\includegraphics[width=0.30\textwidth]{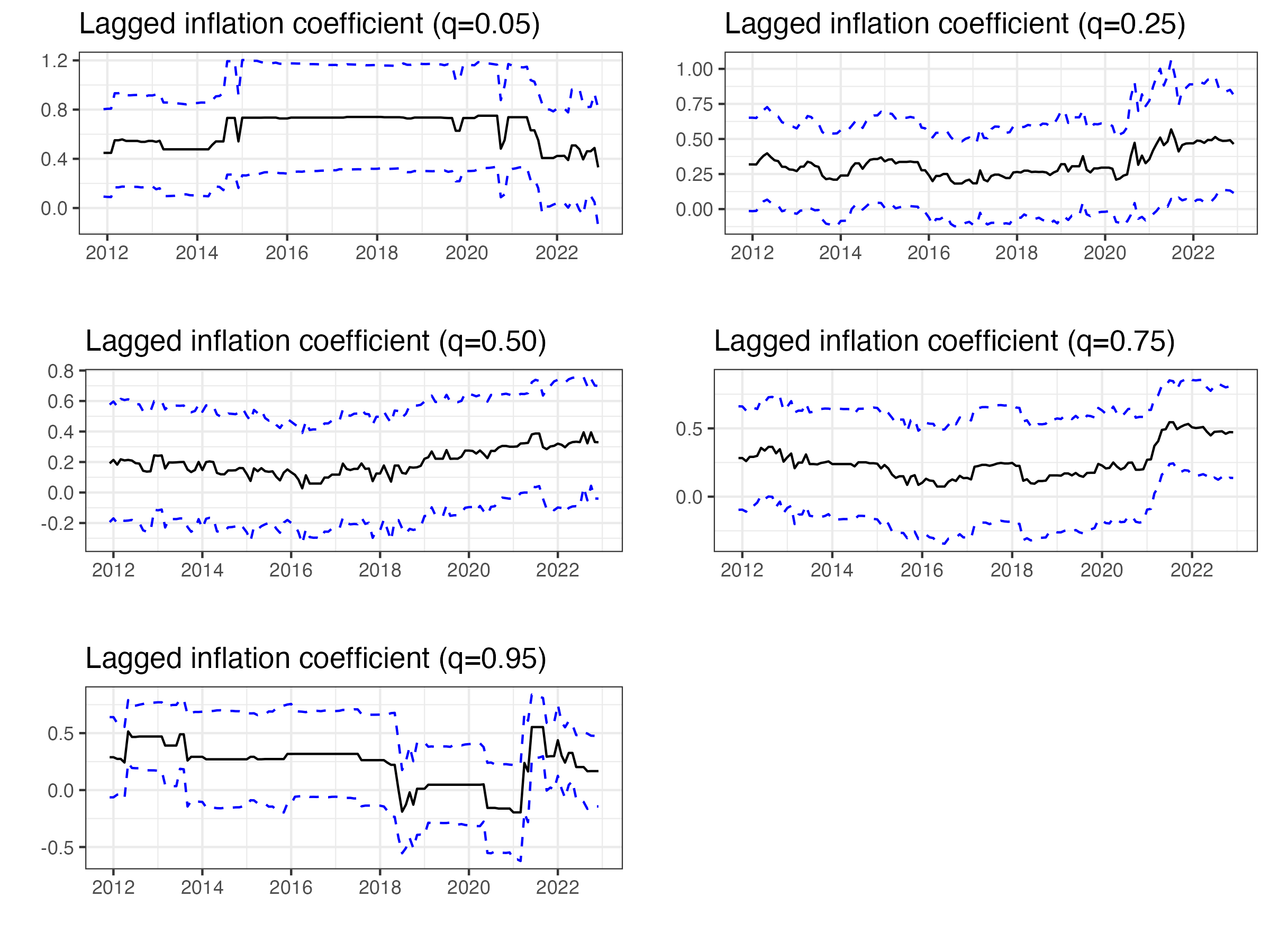}}
\subfigure[]{\includegraphics[width=0.30\textwidth]{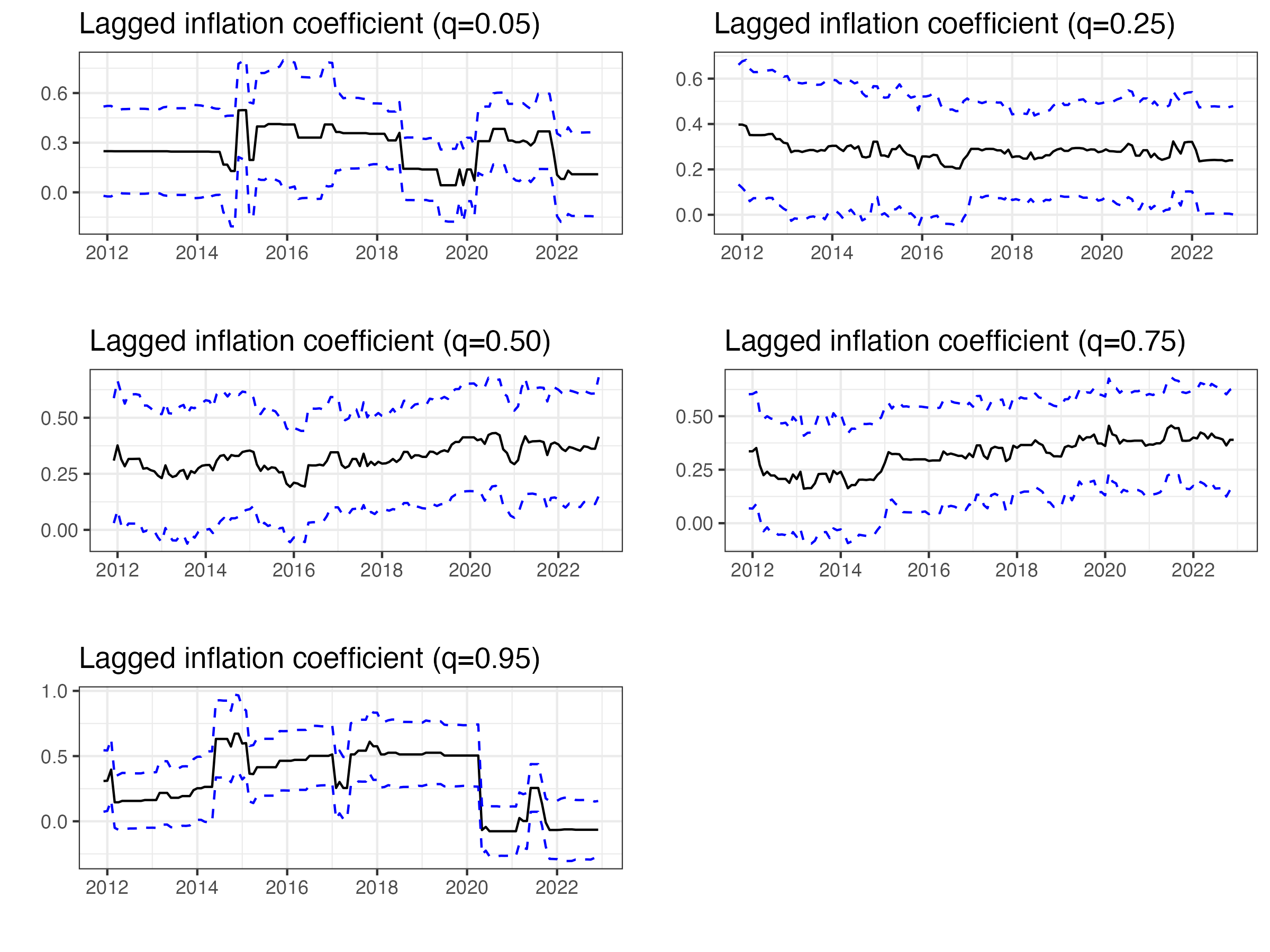}}\\
\subfigure[]{\includegraphics[width=0.30\textwidth]{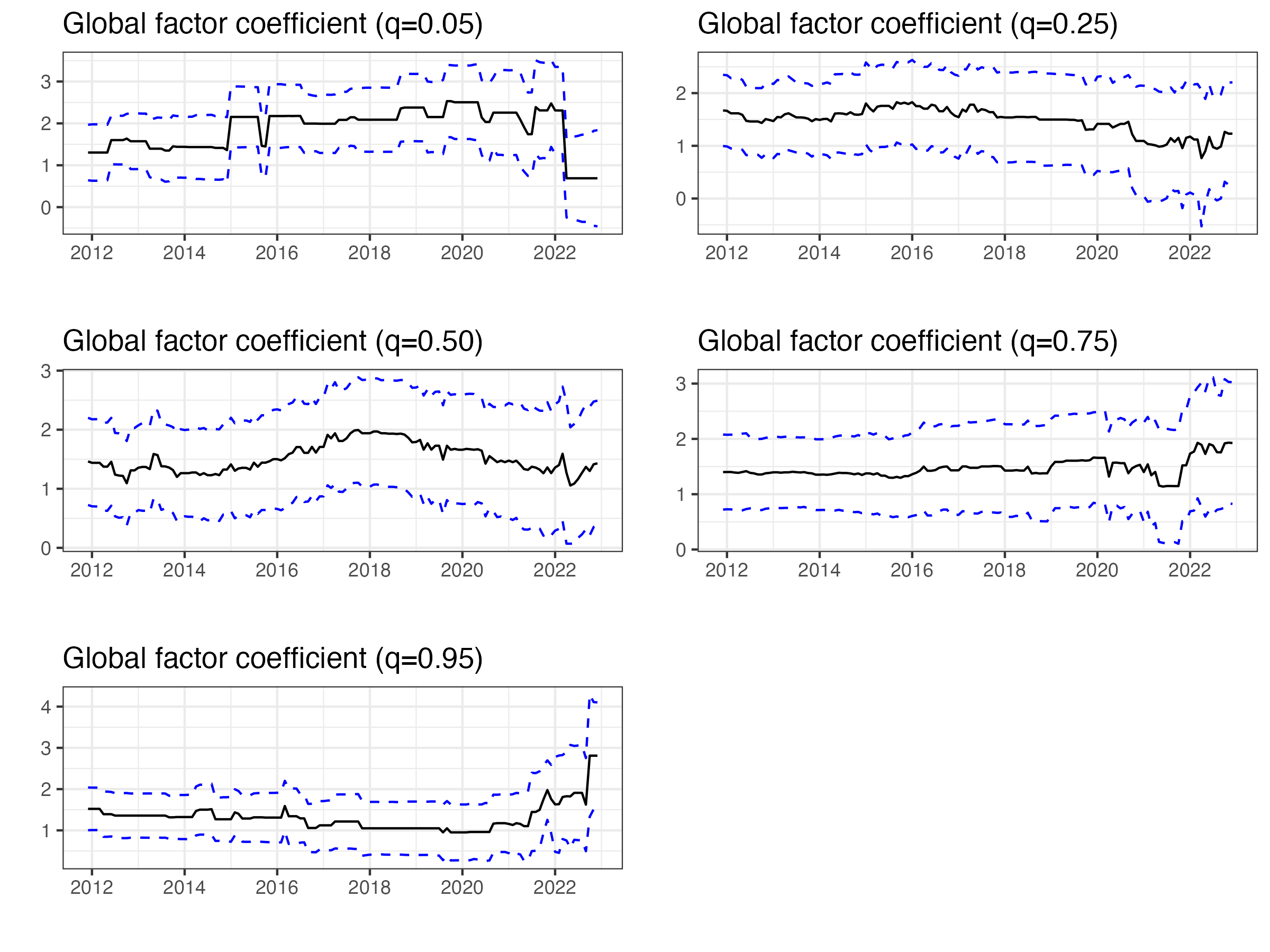}}
\subfigure[]{\includegraphics[width=0.30\textwidth]{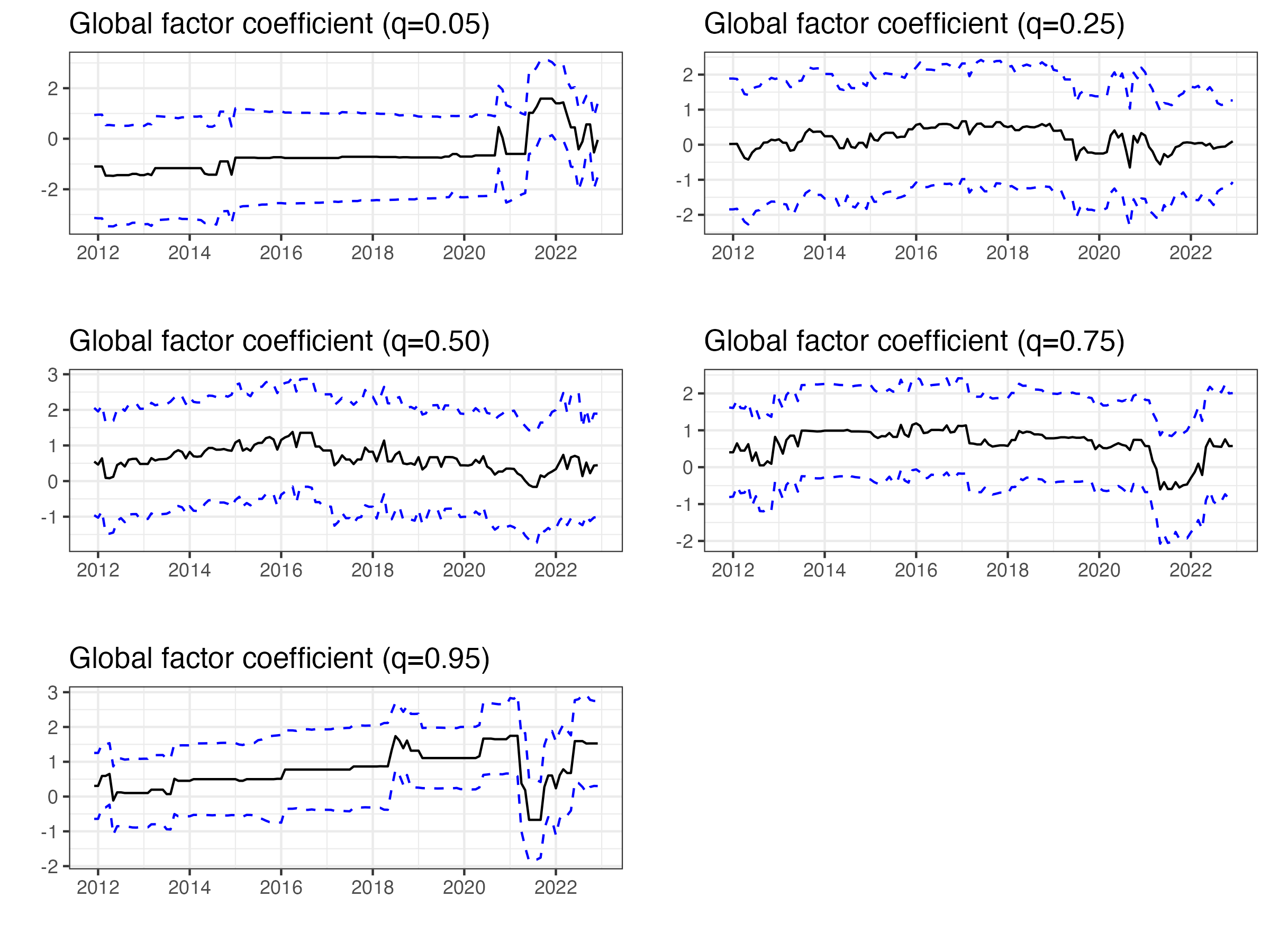}}
\subfigure[]{\includegraphics[width=0.30\textwidth]{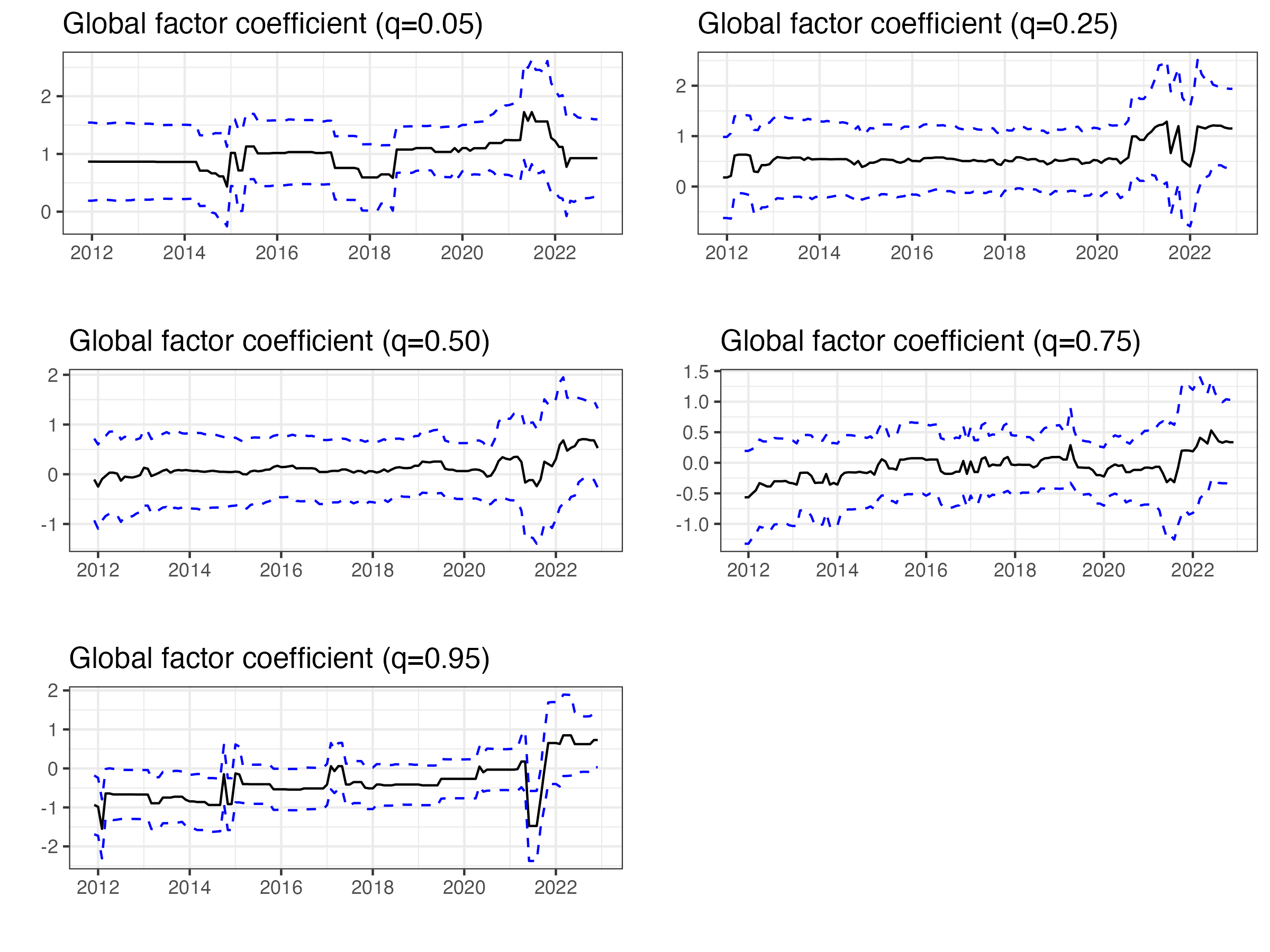}}\\
\subfigure[]{\includegraphics[width=0.3\textwidth]{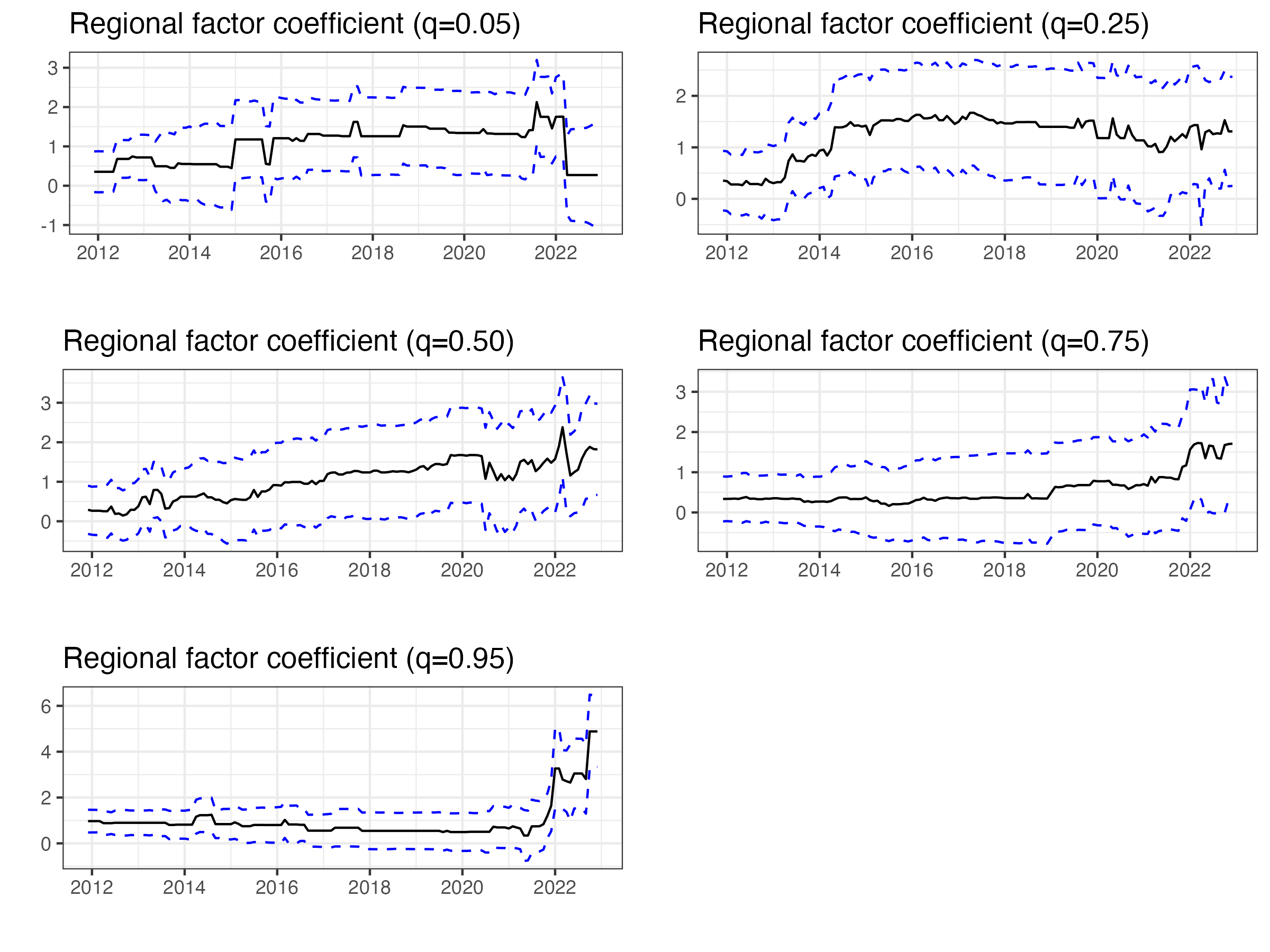}}
\subfigure[]{\includegraphics[width=0.3\textwidth]{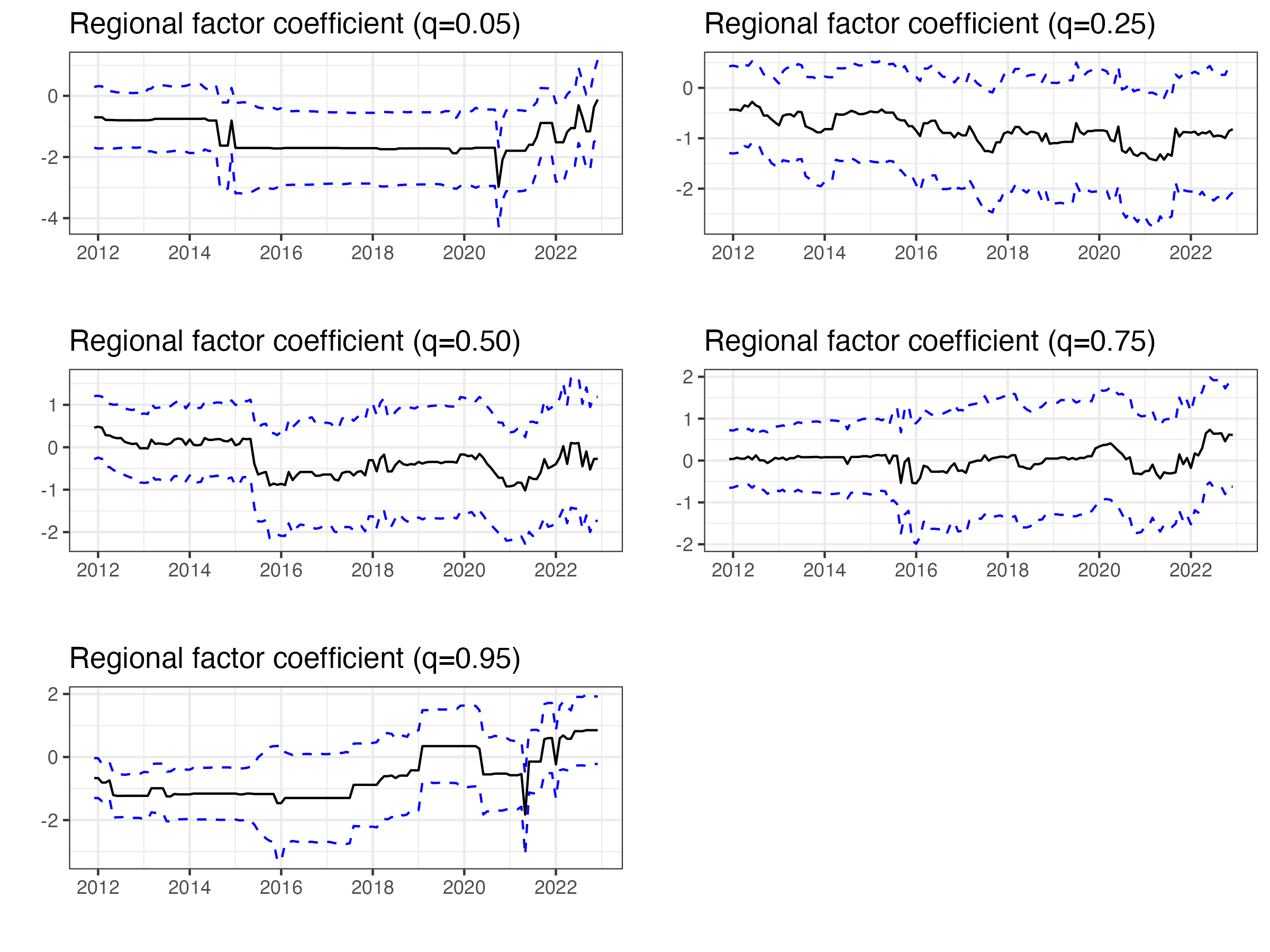}}
\subfigure[]{\includegraphics[width=0.3\textwidth]{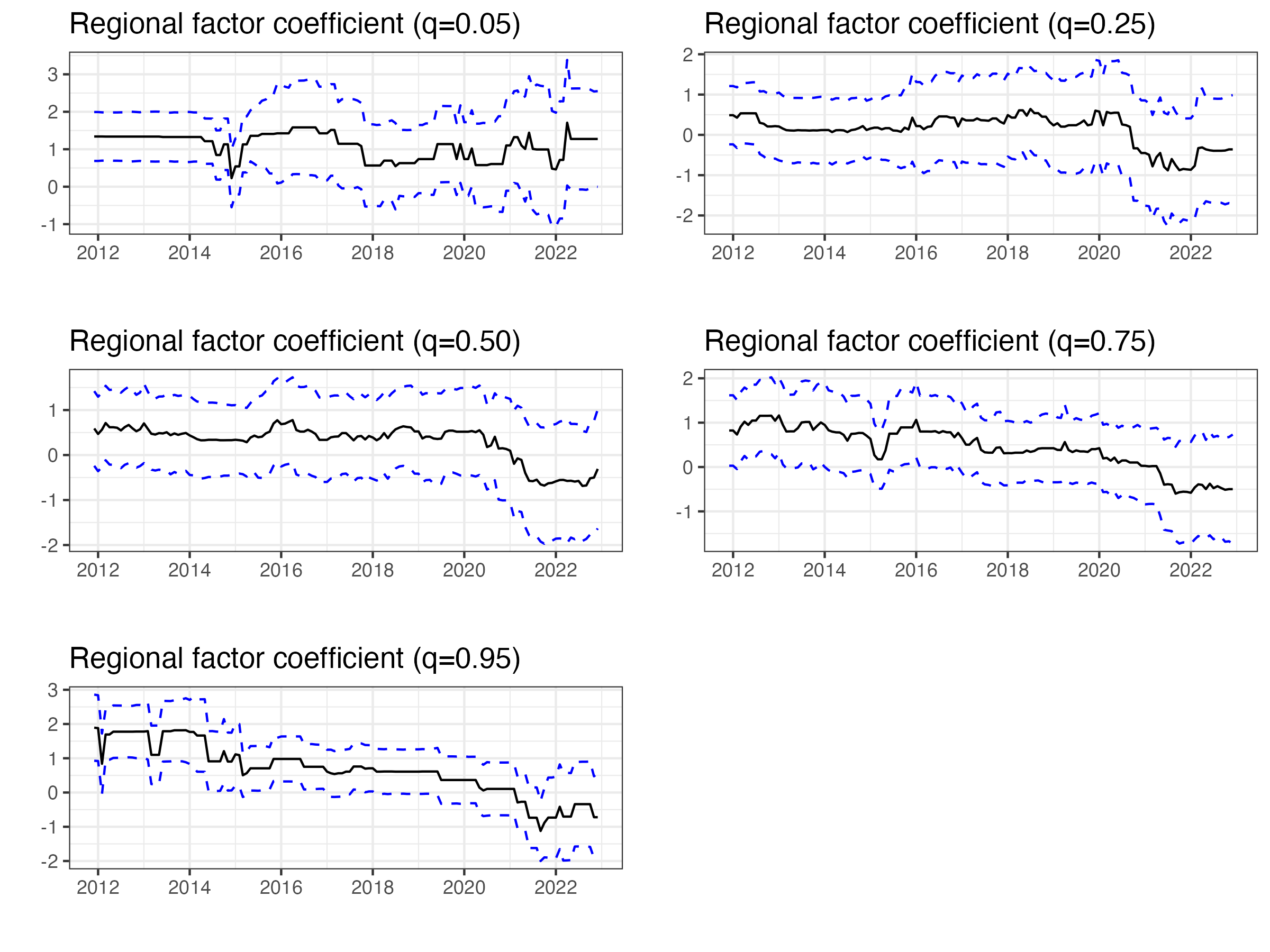}}\\
\subfigure[]{\includegraphics[width=0.30\textwidth]{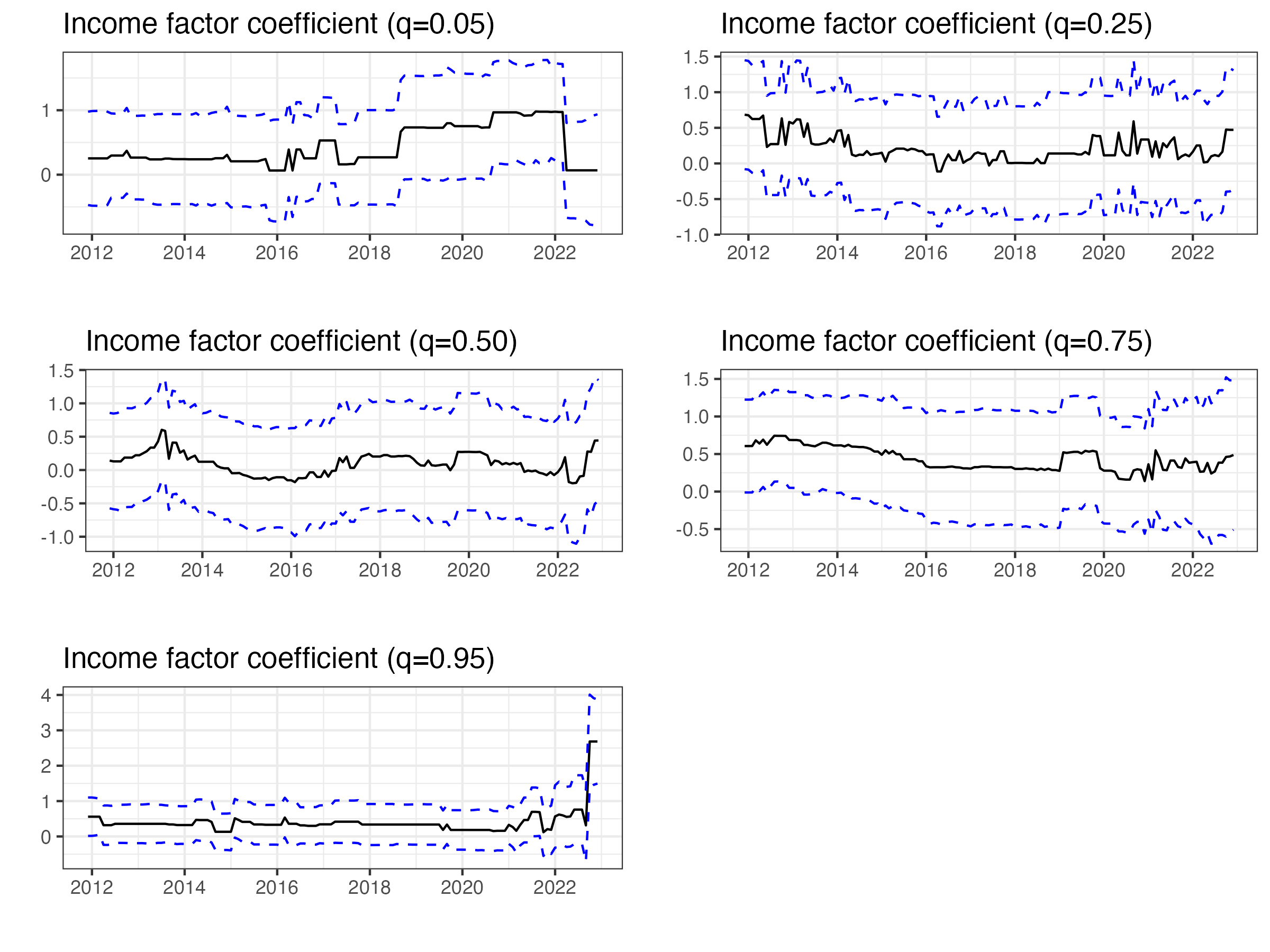}}
\subfigure[]{\includegraphics[width=0.3\textwidth]{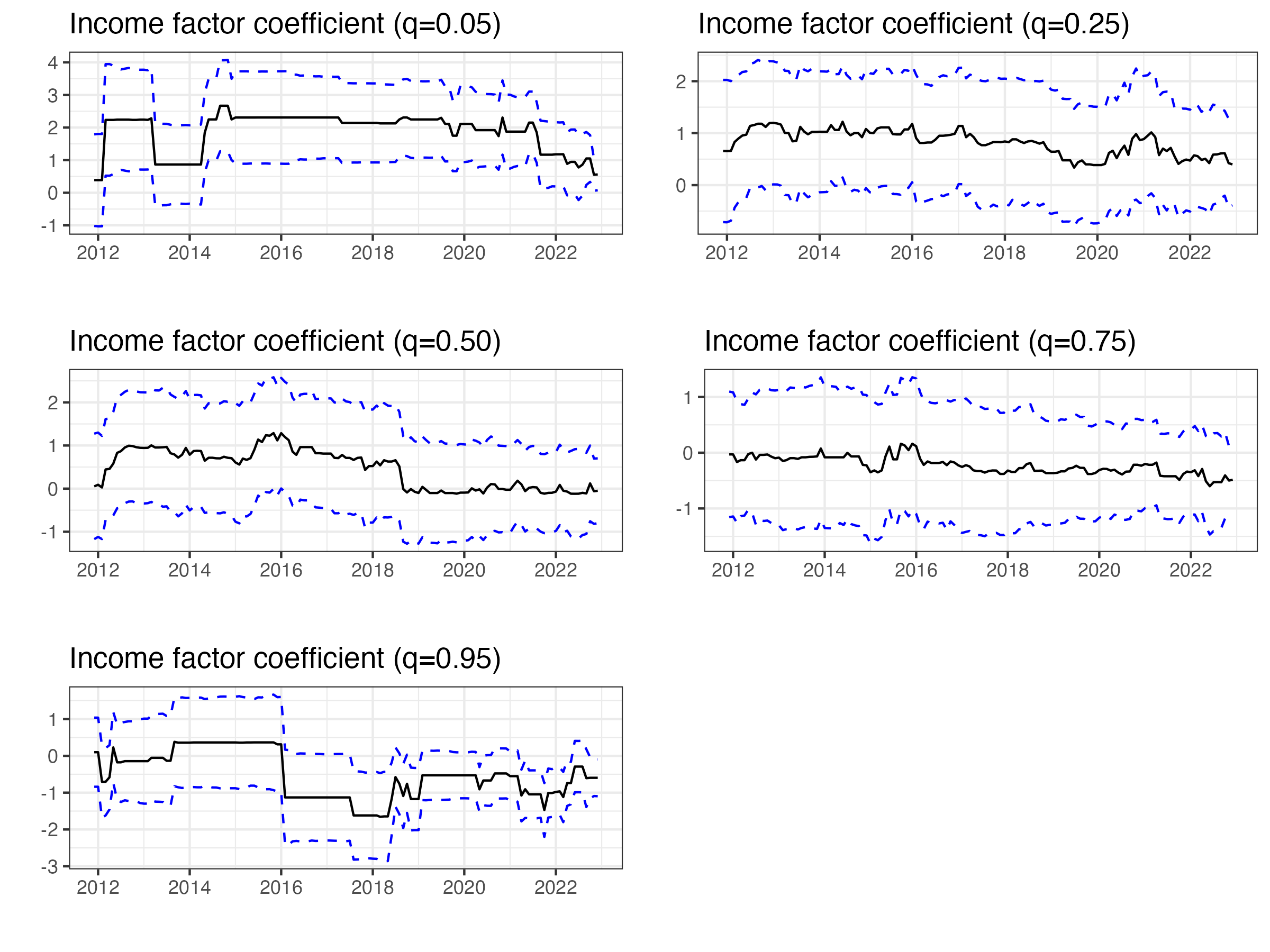}}
\subfigure[]{\includegraphics[width=0.30\textwidth]{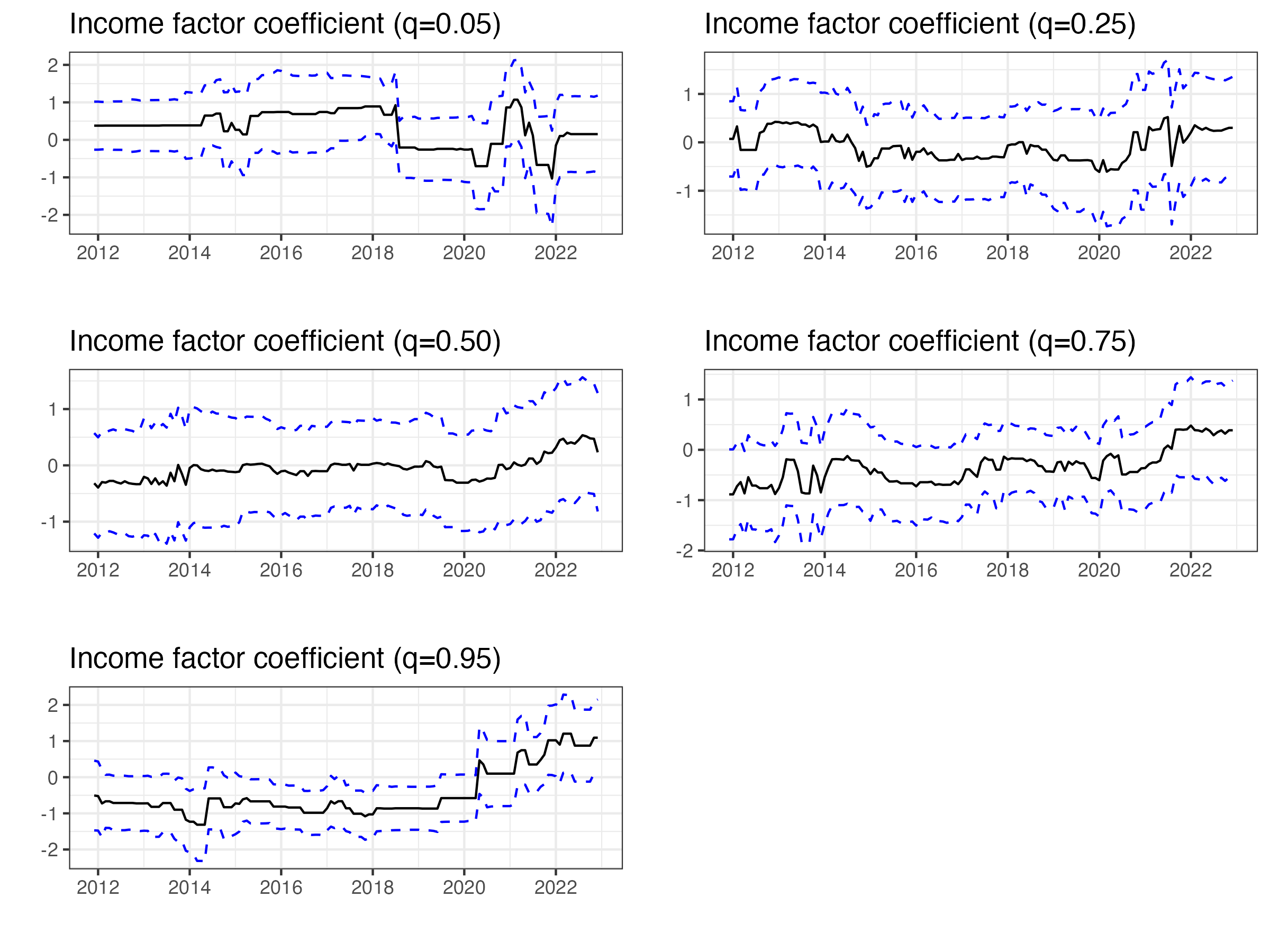}}
\caption{Estimated parameters of FA-QR models for $\tau=0.05, 0.25, 0.50, 0.75$ and 0.95, obtained in the rolling-window: Intercept (first row), autoregressive (second row), global factor (third row), regional factor (fourth row), and economic development factor (fith row). Results for Italy (first column), the US (second column), Mexico (third colmun).}
\label{fig:tv_estimates}
\end{center}
\end{figure}

In this Appendix, we report the estimated parameters of the FA-QR models for $\tau=0.05, 0.25, 0.50, 0.75$ and 0.95, obtained in the rolling-window scheme and used to obtain the out-of-sample forecasts of the inflation densities. The estimated parameters, plotted in  Figure \ref{fig:tv_estimates} together with their 95\% confidence bounds, correspond to FA-QR models for the quantiles of inflation in Italy, the US and Mexico. The first rolling-window vintage is based on observations from February 1999 to December 2011. Analysing the possibility of time-varying parameters of the FA-QR models, we can observe that all parameters are mostly constant for all quantiles and econmies except when looking at the 95\% quantile. It is remarkable that, in this case, the parameter associated to the global factor is increasing through time, but for the very last years associated to the Ukraine invasion in the case of the US and Mexico. Furthermore, the parameter associated to the regional factor is increasing through time in Italy and decreasing in Mexico. %Finally, it is also important to note the increasing parameter of the factor related with the level of economic development in the 50\%, 75\% and 95\% quantiles in Mexico; see also \cite{forbes2020}, who conclude that the role of global factors mesaured by commodity prices has been increasing over the last decade.

\begin{figure}[]
\label{fig:vintage_factors}
\begin{center}
\subfigure[Global factor]{\includegraphics[width=0.4\textwidth]{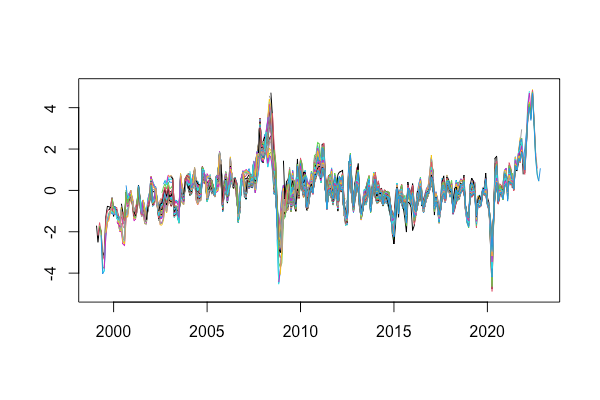}}\\
\subfigure[African factor]{\includegraphics[width=0.4\textwidth]{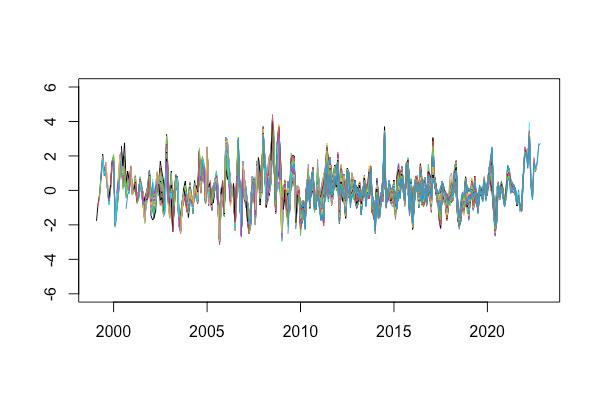}}
\subfigure[Advanced economies factor]{\includegraphics[width=0.4\textwidth]{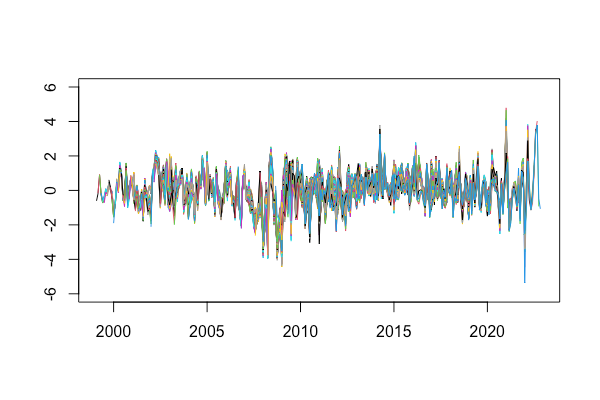}}
\subfigure[American factor]{\includegraphics[width=0.4\textwidth]{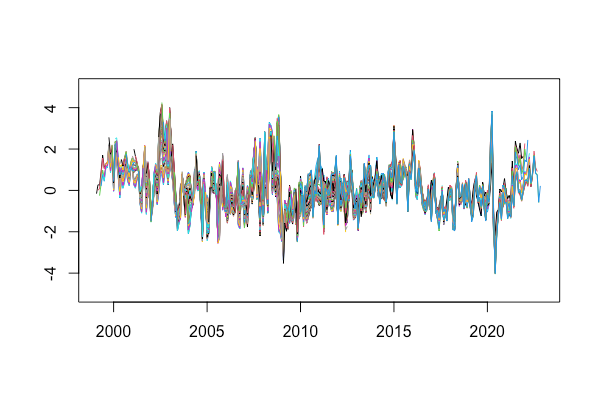}}
\subfigure[HMI economies factor]{\includegraphics[width=0.4\textwidth]{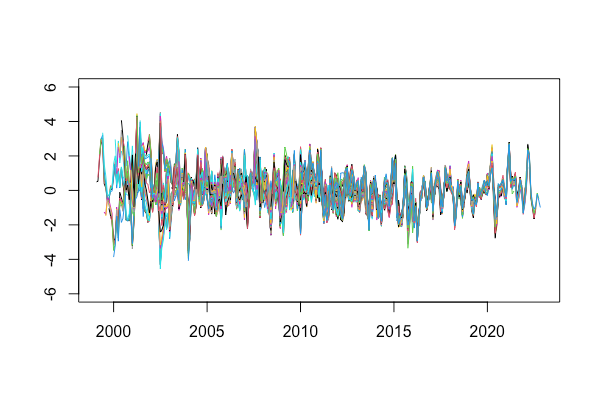}}
\subfigure[Asian factor]{\includegraphics[width=0.4\textwidth]{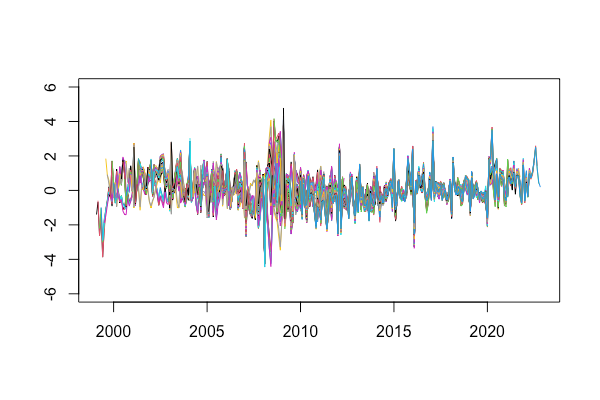}}
\subfigure[LI economies factor]{\includegraphics[width=0.4\textwidth]{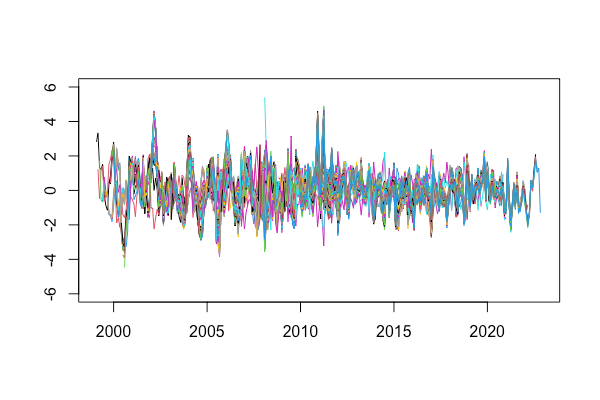}}
\end{center}
\subfigure[European factor]{\includegraphics[width=0.40\textwidth]{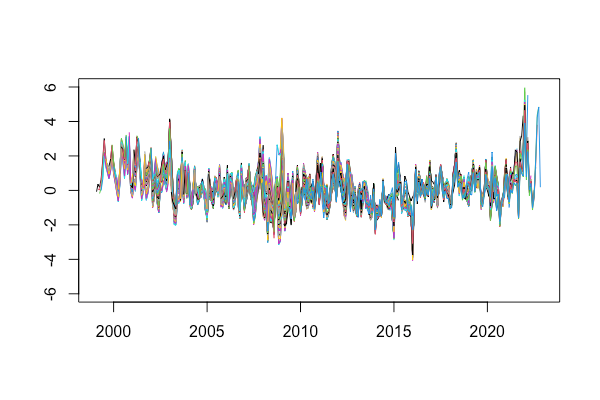}}
\caption{Vintages of factors extracted in each estimation period in the rolling-window.}
\end{figure}

As recomended by \cite{amburgey2022}, in each rolling-window vintage, we extract the three international factors used to estimate the FA-QR models. Figure \ref{fig:vintage_factors} plots the extracted factors obtained in each rolling-window vintage. This figure shows that the factors extracted are rather consistent through vintages. 
   
\newpage

\subsection{Out-of-sample results for restricted FA-QR models.}
\label{appendix:oos}

\renewcommand{\theequation}{B.\arabic{equation}}
\renewcommand{\thetable}{B.\arabic{table}}
\renewcommand{\thefigure}{B.\arabic{figure}}

\setcounter{equation}{0}
\setcounter{table}{0}
\setcounter{figure}{0}

\begin{figure}[]
\begin{center}
\includegraphics[width=1\textwidth]{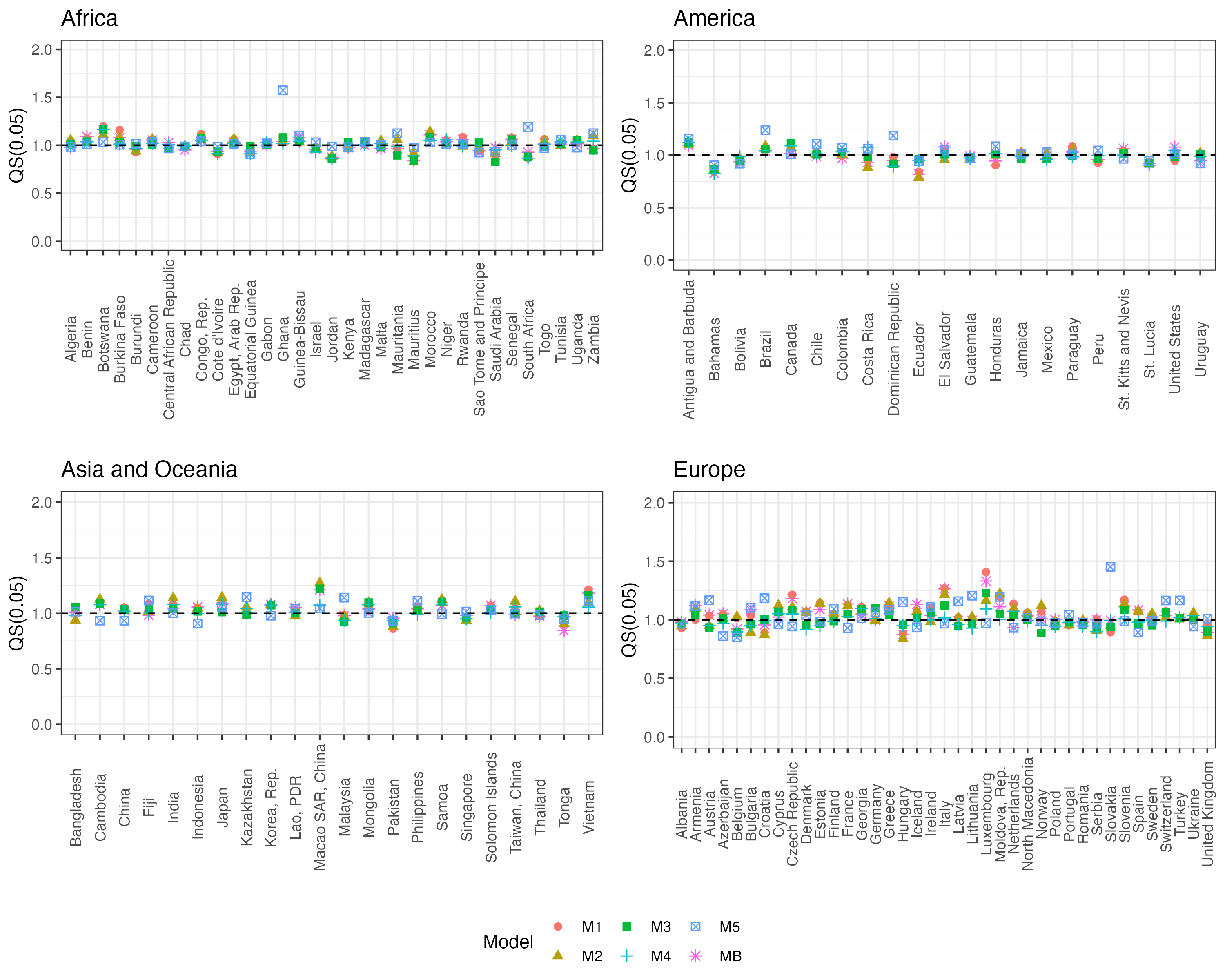}
\caption{Ratio of QS for $\tau=0.05$ losses between the FA-QR and AR-QR models.}
\label{fig:Ratio_1}
\end{center}
\end{figure}

\begin{figure}[]
\begin{center}
\includegraphics[width=1\textwidth]{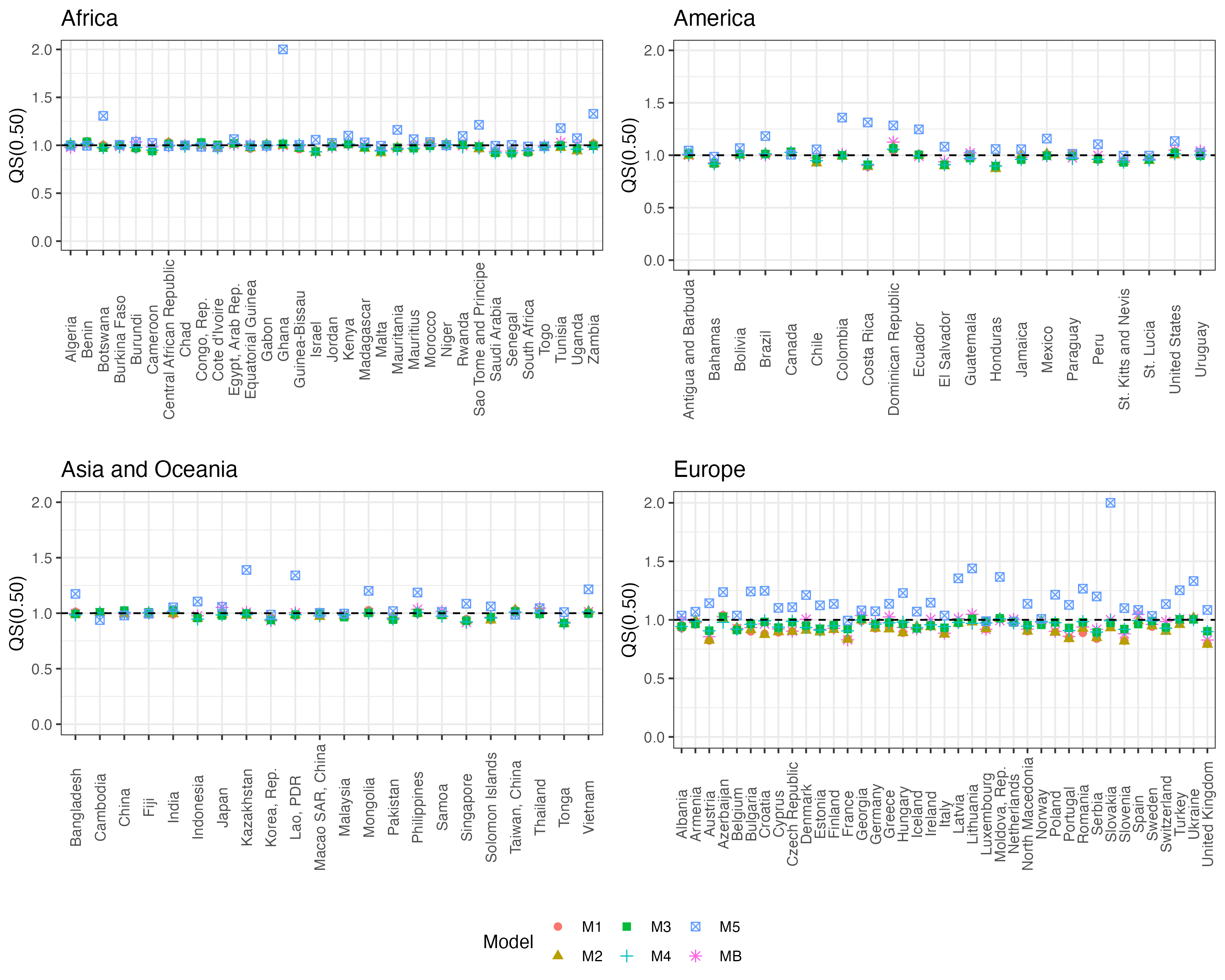}
\caption{Ratio of QS for $\tau=0.50$ losses between the FA-QR and AR-QR models.}
\label{fig:Ratio_2}
\end{center}
\end{figure}

\begin{figure}[]
\begin{center}
\includegraphics[width=1\textwidth]{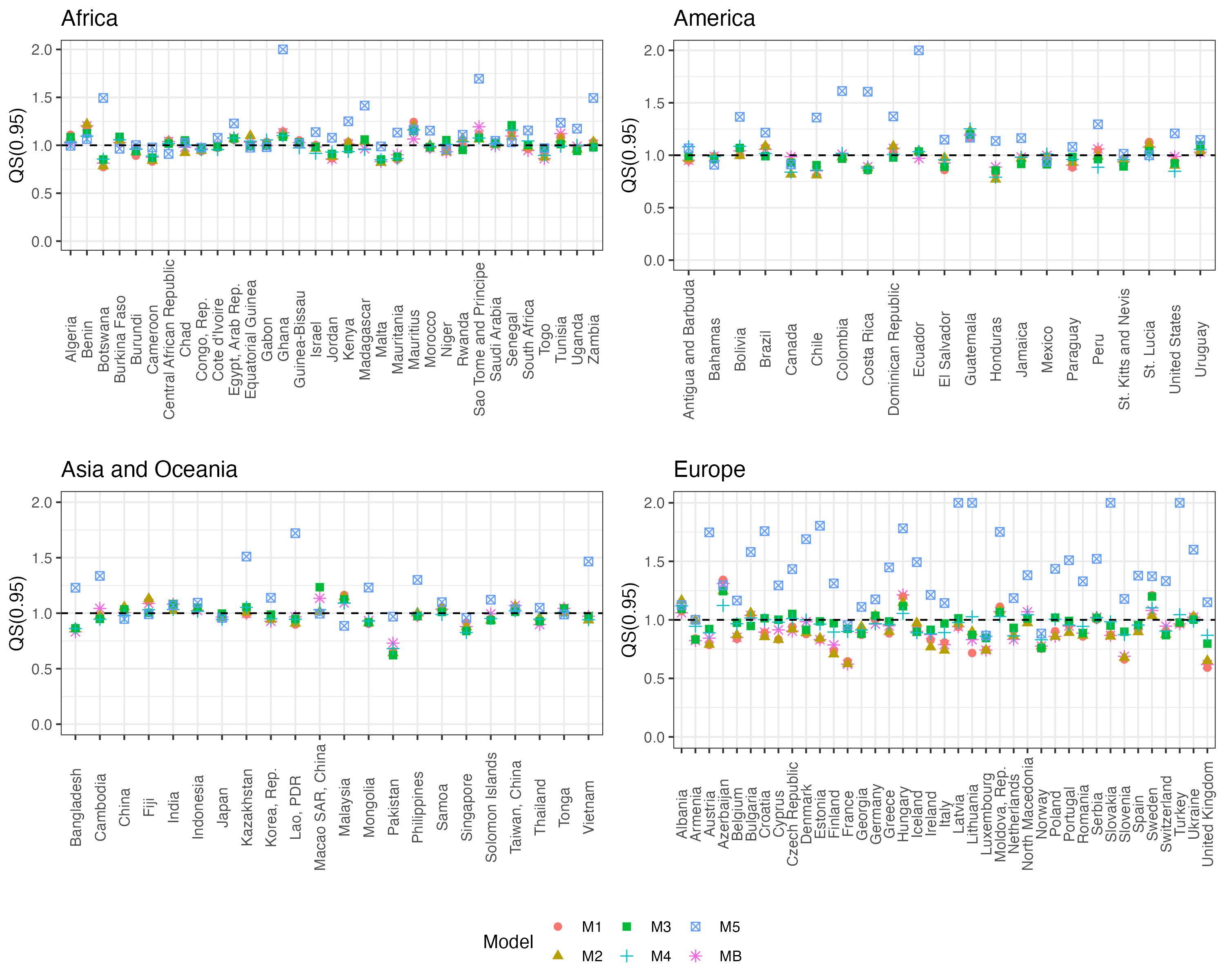}
\caption{Ratio of QS for $\tau=0.95$ losses between the FA-QR and AR-QR models.}

\label{fig:Ratio_3}
\end{center}
\end{figure}

\begin{figure}[]
\begin{center}
\includegraphics[width=1\textwidth]{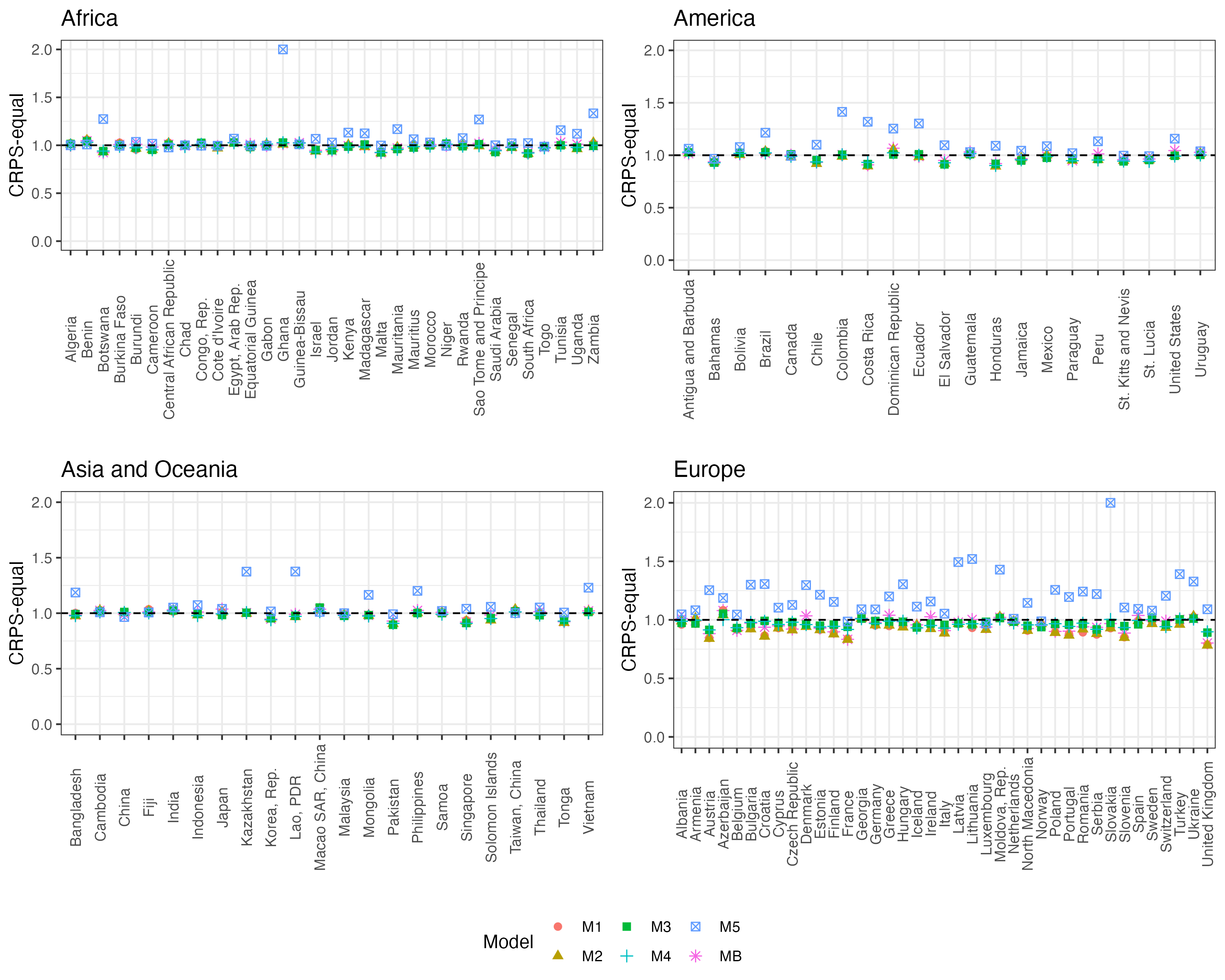}
\caption{Ratio of CRPS-E loss with equal weights between the FA-QR and AR-QR models.}
\label{fig:Ratio_4}
\end{center}
\end{figure}

\begin{figure}[]
\begin{center}
\includegraphics[width=1\textwidth]{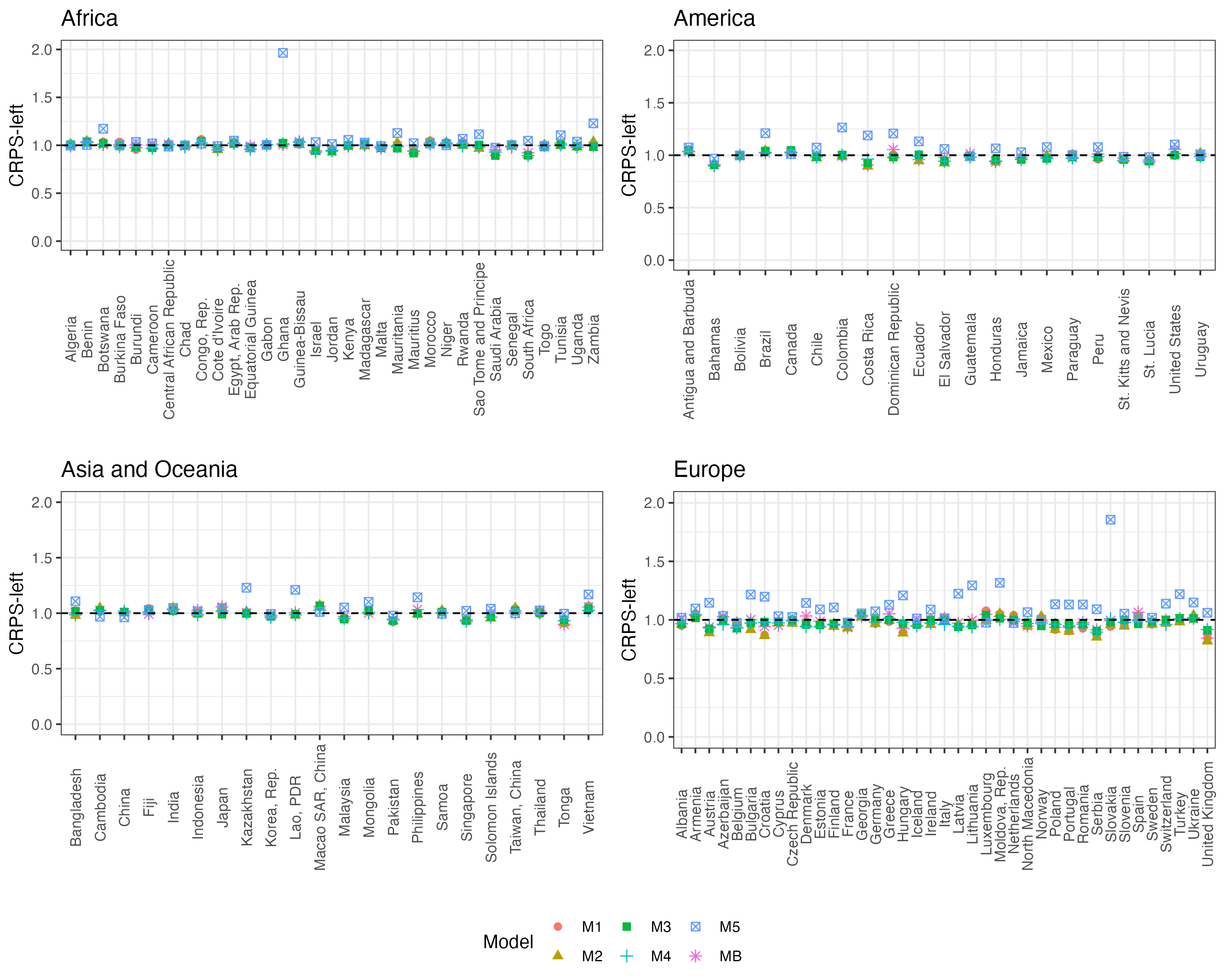}
\caption{Ratio of CRPS-L loss with left weights between the FA-QR and AR-QR models.}
\label{fig:Ratio_5}
\end{center}
\end{figure}

\begin{figure}[]
\begin{center}
\includegraphics[width=1\textwidth]{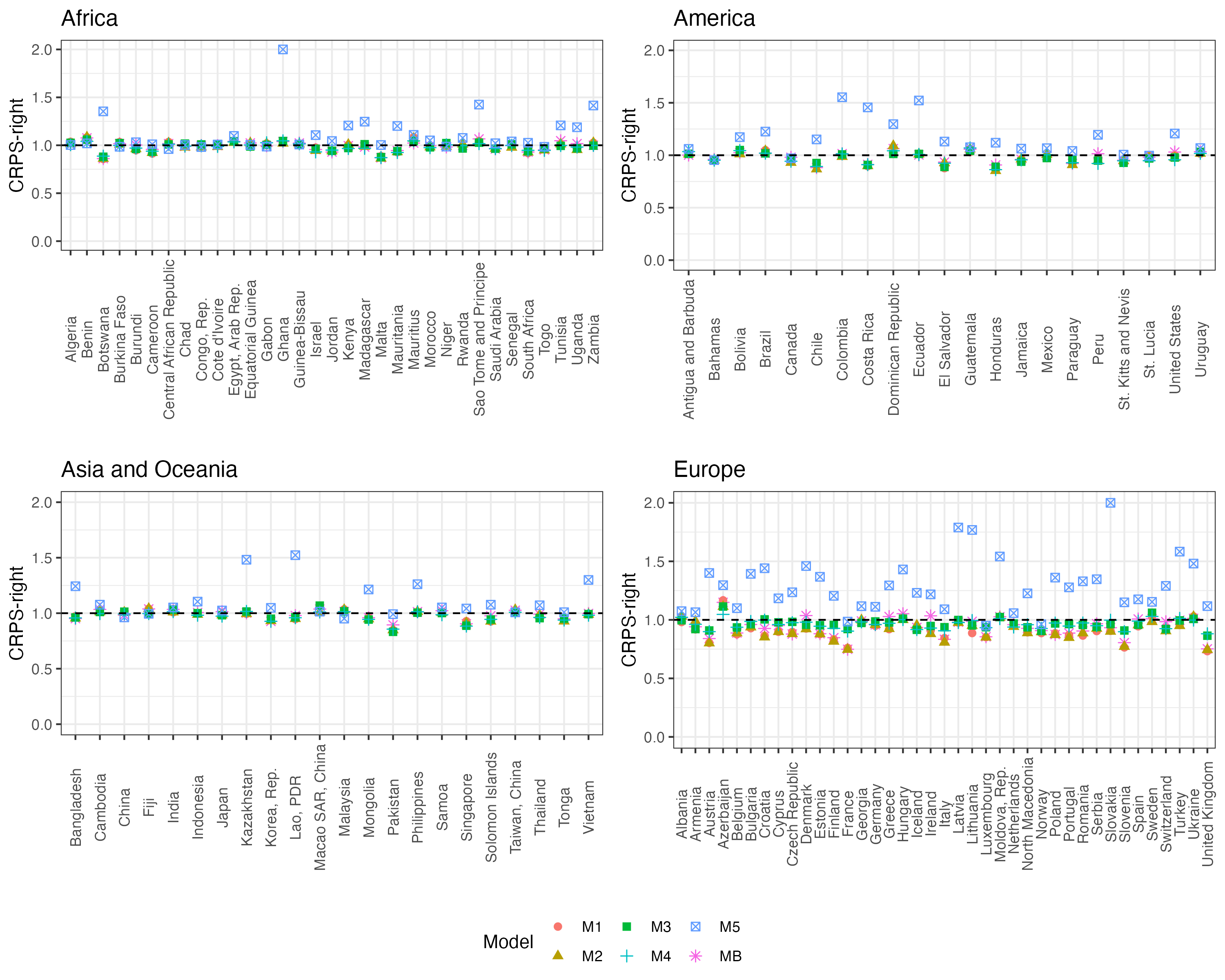}
\caption{Ratio of CRPS-R loss with right weights between the FA-QR and AR-QR models.}
\label{fig:Ratio_6}
\end{center}
\end{figure}

In this Appendix, we report the out-of-sample ratios $\frac{QS^{FA-QR}}{QS^{AR-QR}}$ and $\frac{CRPS^{FA-QR}}{CRPS^{AR-QR}}$ corresponding to restricted versions of the FA-QR model with respect to the benchmark AR-QR model. The FA-QR model with all three international factors is denoted as M!, while the model with only the global and reginal factors is M" and the model with the global and economic development factor is M3. M4 denotes the model that only has the global international factor. As a reference, model M5 stands for the model that only has a constant, i.e. the forecasts are based on the empirical quantile. Finally, MB stands for the model in which only the significant parameters are used to forecast the quantiles. Figure \ref{fig:Ratio_1} to \ref{fig:Ratio_3} plot the ratios obtained for the  $\frac{QS^{FA-QR}}{QS^{AR-QR}}$ for $\tau=0.05, 0.50$ and 0.95, respectively. We can observe that, as expected, M5 works worse than the AR-QR model, specially for the $\tau=0.50$ and 0.95 quantiles. For $\tau=0.05$ and 0.50, all models have ratios close to one. However, when $\tau=9.95$, the M1 model seems to be the winner in European ecnomies, closely followed by M2. Finally,  Figures \ref{fig:Ratio_4} to \ref{fig:Ratio_6} plot the $\frac{CRPS^{FA-QR}}{CRPS^{AR-QR}}$ ratios between CRPS-E, CRPS-L and CRPS-R losses, respectively. Once more, we observe that the AR-QR model beats forecasts obtained using the empirical quantiles, with the improvement being larger for the CRPS-R loss. All other FA-QR models considered have ratios close to one for CRPS-E and CRPS-L. However, in concordance with previous results, the forecast power of FA-QR models improve over that of the benchmark AR-QR model when looking at the CRPS-R ratios, mainly in European economies. The performance of the M1 and ;2 models is similar and best among the specifications considered. 

%\section*{References Appendix}

%Acharya, V., Engle, R., and Richardson, M. (2012). Capital shortfall: a new approach to ranking and regulating systemic risk. \textit{American Economic Review}, 102:59--64.

\end{document}